\documentclass[a4paper,fleqn,usenatbib]{mnras}
\pdfoutput=1

\usepackage[T1]{fontenc}
\usepackage{ae,aecompl}
\usepackage{amsmath}
\usepackage[dvipsnames]{xcolor}
\usepackage{graphicx}
\usepackage[colorinlistoftodos]{todonotes}
\usepackage{framed} 
\usepackage[framed]{ntheorem}
\usepackage{rotating}
\usepackage[titletoc]{appendix}
\usepackage{titlesec}
\usepackage{courier}       
\usepackage{titletoc}
\usepackage{rotating}
\usepackage{ntheorem}
\usepackage{esint}
\usepackage{empheq}
\usepackage{listings}
\usepackage{natbib}
\usepackage{pdflscape}
\usepackage{makecell}
\usepackage{scalerel}
\usepackage{tikz}
\usetikzlibrary{svg.path}

\definecolor{orcidlogocol}{HTML}{A6CE39}
\tikzset{
	orcidlogo/.pic={
		\fill[orcidlogocol] svg{M256,128c0,70.7-57.3,128-128,128C57.3,256,0,198.7,0,128C0,57.3,57.3,0,128,0C198.7,0,256,57.3,256,128z};
		\fill[white] svg{M86.3,186.2H70.9V79.1h15.4v48.4V186.2z}
		svg{M108.9,79.1h41.6c39.6,0,57,28.3,57,53.6c0,27.5-21.5,53.6-56.8,53.6h-41.8V79.1z M124.3,172.4h24.5c34.9,0,42.9-26.5,42.9-39.7c0-21.5-13.7-39.7-43.7-39.7h-23.7V172.4z}
		svg{M88.7,56.8c0,5.5-4.5,10.1-10.1,10.1c-5.6,0-10.1-4.6-10.1-10.1c0-5.6,4.5-10.1,10.1-10.1C84.2,46.7,88.7,51.3,88.7,56.8z};
	}
}

\newcommand\orcidicon[1]{\href{https://orcid.org/#1}{\mbox{\scalerel*{
				\begin{tikzpicture}[yscale=-1,transform shape]
				\pic{orcidlogo};
				\end{tikzpicture}
			}{|}}}}

\usepackage{hyperref} 
\usepackage{txfontsb}

\title[\texttt{SSE} and \texttt{BSE} in \texttt{Nbody6++GPU}, \texttt{MOCCA} \& \texttt{McLuster}]{Preparing the next gravitational million-body simulations: Evolution of single and binary stars in \texttt{Nbody6++GPU}, \texttt{MOCCA} and \texttt{McLuster}}

\author[A. W. H. Kamlah et al. (2021)]{
	A. W. H. Kamlah$^{1,2}$\thanks{E-mail: albrecht.kamlah@stud.uni-heidelberg.de},
	A. Leveque$^{6}$,
	R. Spurzem$^{2,3,4}$,
	M. Arca Sedda$^{2}$,
	\newauthor
	A. Askar$^{7}$,
	S. Banerjee$^{8,9}$,
	P. Berczik$^{3,5}$,
	M. Giersz$^{6}$,
	\newauthor
	J. Hurley$^{10,15}$,
	D. Belloni$^{13,14}$,
	L. K\"uhmichel$^{2}$,
	and L. Wang$^{11,12}$
	\\
	$^{1}$Max-Planck-Institut f\"ur Astronomie, K\"onigstuhl 17, 69117 Heidelberg, Germany \\
	$^{2}$Astronomisches Rechen-Institut, Zentrum f\"ur Astronomie, University of Heidelberg,
	M\"onchhofstrasse 12-14, 69120, Heidelberg, Germany \\
	$^{3}$National Astronomical Observatories and Key Laboratory of Computational Astrophysics, Chinese Academy of Sciences, 20A Datun Rd.,\\
	Chaoyang District, Beijing 100101, China \\
	$^{4}$Kavli Institute for Astronomy and Astrophysics, Peking University, Yiheyuan Lu 5, Haidian Qu, 100871, Beijing, China 20A Datun Rd., \\
	Chaoyang District, 100012, Beijing, China \\
	$^{5}$Main Astronomical Observatory, National Academy of Sciences of Ukraine, 27 Akademika Zabolotnoho St., 03680, Kyiv, Ukraine \\
	$^{6}$Nicolaus Copernicus Astronomical Centre, Polish Academy of Sciences, ul. Bartycka 18, 00-716 Warsaw, Poland \\
	$^{7}$Lund Observatory, Department of Astronomy and Theoretical Physics, Lund University, Box 43, SE-221 00 Lund, Sweden \\
	$^{8}$Helmholtz-Instituts f\"ur Strahlen- und Kernphysik (HISKP), Nussallee 14-16, D-53115 Bonn, Germany \\
	$^{9}$Argelander-Institut f\"ur Astronomie (AIfA), Auf dem H\"ugel 71, D-53121, Bonn, Germany \\
	$^{10}$Centre for Astrophysics and Supercomputing, Swinburne University of Technology, Hawthorn VIC 3122, Australia \\
	$^{11}$Department of Astronomy, School of Science, The University of Tokyo, 7-3-1 Hongo, Bunkyo-ku, Tokyo, 113-0033, Japan \\
	$^{12}$RIKEN Center for Computational Science, 7-1-26 Minatojima-minami-machi, Chuo-ku, Kobe, Hyogo 650-0047, Japan \\
	$^{13}$National Institute for Space Research, Av. dos Astronautas, 1758, 12227-010, Sao Jose dos Campos, SP, Brazil \\ 
	$^{14}$Departamento de F\'isica, Universidad T\'ecnica Federico Santa Mar\'ia, Av. Espa\~na 1680, Valpara\'iso, Chile \\
	$^{15}$The ARC Centre of Excellence for Gravitational Wave Discovery - OzGrav-Swinburne University of Technology, VIC 3122, Australia
}

\date{Accepted XXX. Received YYY; in original form ZZZ}

\pubyear{2021}

\begin{document}
	\label{firstpage}
	\pagerange{\pageref{firstpage}--\pageref{lastpage}}
	\maketitle
	
	\begin{abstract}
		We present the implementation of updated stellar evolution recipes in the codes \texttt{Nbody6++GPU, MOCCA} and \texttt{McLuster}. We test them through numerical simulations of star clusters containing $1.1\times 10^5$ stars (with $2.0\times 10^4$ in primordial hard binaries) performing high-resolution direct $N$-body (\texttt{Nbody6++GPU}) and Monte-Carlo (\texttt{MOCCA}) simulations to an age of 10~Gyr. We compare models implementing either delayed or core-collapse supernovae mechanisms, a different mass ratio distribution for binaries, and white dwarf natal kicks enabled/disabled. Compared to \texttt{Nbody6++GPU}, the \texttt{MOCCA} models appear to be denser, with a larger scatter in the remnant masses, and a lower binary fraction on average. The \texttt{MOCCA} models produce more black holes (BHs) and helium white dwarfs (WDs), whilst \texttt{Nbody6++GPU} models are characterised by a much larger amount of WD-WD binaries. The remnant kick velocity and escape speed distributions are similar for the BHs and neutron stars (NSs), and some NSs formed via electron-capture supernovae, accretion-induced collapse or merger-induced collapse escape the cluster in all simulations. The escape speed distributions for the WDs, on the other hand, are very dissimilar. We categorise the stellar evolution recipes available in \texttt{Nbody6++GPU}, \texttt{MOCCA} and \texttt{Mcluster} into four levels: the one implemented in previous \texttt{Nbody6++GPU} and \texttt{MOCCA} versions (\texttt{level A}), state-of-the-art prescriptions (\texttt{level B}), some in a testing phase (\texttt{level C}), and those that will be added in future versions of our codes.
	\end{abstract}
	
	
	\section{Introduction}
	The stellar environment in star clusters provides the ideal laboratory for investigating stellar binary evolution as well as gravitational wave (GW) physics. This is because the densities are typically so high that stars can interact in close gravitational encounters or even physically collide with each other. These interactions support the presence of more tightly bound binary stars, which can act as a source of huge amounts of gravitational energy to the cluster. This will result in enhanced mass-segregation: more massive stars and binaries sink to the centre of the system, where they undergo close gravitational encounters and in the case of high densities, stellar collisions, which has been predicted and tested theoretically \citep{Heggie1975,PortegiesZwartMcMillan2002,Khalisietal2007,Gierszetal2015,Wangetal2016,Askaretal2017a,ArcaSedda2019,Rizzutoetal2021a,Rizzutoetal2021b} and verified observationally \citep{LadaLada2003,Cantat-Gaudin2014,Martinazzietal2014,Kamannetal2018b,Giesersetal2018,Giesersetal2019}. 
	\\
	Simulations of such star clusters fundamentally aim to solve the equations of motion describing $N$ bodies moving under the influence of their own self-gravity. For this purpose a variety of computational approaches have been developed beginning in the first half of the last century. The two main methods in the regime of around $10^5-10^7$ particles that stand out today are either related to direct $N$-body simulation or Monte-Carlo modelling  \citep{Aarsethetal1974,AarsethLecar1975,GierszHeggie1994,Spurzem1999}. Direct $N$-body simulation -- orbit integration of the orbits of many particles in a self-gravitating bound star cluster -- is the most suitable method to understand relaxation \citep{Larson1970a,Larson1970b} and evolutionary processes in the regime of star clusters. Here, statistical physics still plays a role and more approximate models may be used. These models are based on the Fokker-Planck equation, which can be solved either directly or by a Monte Carlo Markoff chain method \citep{Henon1975,Cohn1979,Stodolkiewicz1982,Stodolkiewicz1986,Giersz1998,Gierszetal2015,Merritt2015,Askaretal2017a,Kremeretal2020a,Kremeretal2021}.
	\\
	Beyond solving the equations of motion for the $N$ bodies, the complete description of a \textit{realistic} star cluster becomes much more complicated, because the stellar evolution of single and binary stars has an enormous impact on the dynamical evolution of star clusters. Single and binary stars may suffer significant mass loss over the lifetime of the cluster depending on their initial zero-age main sequence (ZAMS) mass and their metallicity. This mass loss changes the potential of the star cluster and subsequently has an effect on the orbits of the stars. In our models of single stars, this mass loss is dominated by stellar winds and outflows \citep{Hurleyetal2000,Tout2008a}. In the models of binary stars, the member stars can interact with each other closely and other astrophysical processes involving dynamical mass transfer, tidal circularisation and stellar spin synchronisation happen \citep{MardlingAarseth2001,Hurleyetal2002b,Tout2008b}. In the case of compact objects like black holes (BHs), neutron stars (NSs), and white dwarfs (WDs) repeated encounters between stars and binaries may lead to sudden orbit shrinking of a binary up to a point when finally a huge proportion of further orbit shrinking is due to the emission of gravitational radiation \citep{Fayeetal2006,Bremetal2013,AntoniniGieles2020,ArcaSedaetal2020b,Mapellietal2020b}. The gravitational waves that accompany these subsequent gravitational inspiral events might be detectable with the (Advanced) Laser Interferometer Gravitational-Wave Observatory (aLIGO) \citep{Aasietal2015, Abbottetal2018,Abbottetal2019b}, (Advanced) Virgo Interferometer (aVirgo) \citep{Acerneseetal2015,Abbottetal2018,Abbottetal2019b} if they emit signals coming from merging NSs \citep{Abbottetal2017a,Abbottetal2017b}, stellar mass BHs \citep{Abbottetal2016a} or the process of core collapse in  supernovae (SNe) \citep{Ott2009}. If, for example, the binary consists of two BHs then this gravitational wave inspiral may lead to the formation of intermediate-mass BHs (IMBHs) as has been confirmed in simulations \citep{Gierszetal2014,Gierszetal2015,ArcaSedda2019,Rizzutoetal2021a,DiCarloetal2019,DiCarloetal2020a,DiCarloetal2020b,DiCarloetal2021,Banerjee2021a,Banerjee2021c}. A recent aLIGO and aVirgo detection of such an IMBH with a total mass of around 142 $\mathrm{M}_{\odot}$ \citep{Abbottetal2020b} invites further simulations focussing on this particular aspect. 
	\\
	A subclass of star clusters that we aim to simulate across cosmic time are globular clusters (GCs). The Milky way hosts over 150 of these \citep{Harris1996,Baumgardtetal2018}. Their old age and relatively large numbers not only in our galaxy, but also in much more massive elliptical galaxies such as M87 \citep{Tamuraetal2006a,Tamuraetal2006b,Doyleetal2019}, and at higher redshifts \citep{Zicketal2018a,Zicketal2018b,Zicketal2020} all suggest that they play an important role as a fundamental building block in a hierarchy of cosmic structure formation \citep{Reina-Campos2019,Reina-Camposetal2020,Reina-Camposetal2021}. Although becoming increasingly sophisticated, observational studies using astrophysical instruments such as Multi Unit Spectroscopic Explorer (MUSE) \citep{Husseretal2016,Giesersetal2018,Giesersetal2019,Kamannetal2018a,Kamannetal2018b,Kamannetal2020a,Kamannetal2020b} and Gaia \citep{Kuhnetal2019,Bianchinietal2018a,Bianchinietal2018c,Bianchinietal2019,deBoeretal2019,HuangKoposov2021} are not sufficient on their own to resolve the complete evolution of GCs across cosmic time, because they effectively only take snapshots of these clusters today. These observations must therefore be supplemented with astrophysical simulations \citep{Krumholzetal2019}. Due to their typical sizes, simulations of GCs over billions of years are at the edge of high-resolution direct $N$-body simulations today, which are computationally possible and feasible. The \texttt{Dragon} simulations were the first, and last to date, direct gravitational million-body simulations of such a GC \citep{Wangetal2016}. Similarly, the last direct million-body simulation of a nuclear star cluster (NSC) (similar particle number as the \texttt{Dragon} simulations, but scaled in a way to resemble a NSC) harbouring a central and accreting SMBH were performed by \citet{Panamarevetal2019}. While \citet{Wangetal2015} made the technical programming advances necessary to perform million-body simulations with \texttt{Nbody6++GPU} in the first place by parallelising the integrations across multiple GPUs accelerating the (\textit{regular}) direct force integrations and the energy checks to an unprecedented degree and while \citet{Panamarevetal2019} expanded the code to include a central and accreting SMBH, the stellar evolution prescriptions in both of these codes were largely unchanged.
	\\
	To this end, we updated the stellar evolution routines in the direct-force integration code \texttt{Nbody6++GPU} \citep{Wangetal2015}, which are the \texttt{SSE} \citep{Hurleyetal2000} and \texttt{BSE} \citep{Hurleyetal2002b} stellar evolution implementations. These updates mirror the updates in \texttt{Nbody7} by \citet{Banerjeeetal2020,Banerjee2021a}. The results are then compared with the Hénon-type Monte-Carlo code \texttt{MOCCA} \citep{HypkiGiersz2013,Gierszetal2013}, which also conveniently models the evolution of single and binary stars with the \texttt{SSE} and \texttt{BSE} routines. This study is therefore also a continuation of the productive collaboration between the teams surrounding these modelling methods \citep{Gierszetal2008,Downingetal2010,Downingetal2011,Gierszetal2013,Wangetal2016,Rizzutoetal2021a}. Finally, in the appendix, we present an updated version of \texttt{McLuster} \citep{Kuepperetal2011a}, which now includes a mirror of the stellar evolution available in \texttt{Nbody6++GPU}.
	
	\section{Methods}
	\subsection{Direct $N$-body simulations with \texttt{Nbody6++GPU}}
	The state-of-the-art direct force integration code \texttt{Nbody6++GPU} is optimised for high performance GPU-accelerated supercomputing \citep{Spurzem1999,NitadoriAarseth2012,Wangetal2015}. This code follows a long-standing tradition in a family of direct force integration codes of gravitational $N$-body problems, which were originally written by Sverre Aarseth (\citet{Aarseth1985,Spurzem1999,Aarseth1999a,Aarseth1999b,Aarseth2003b,Aarseth2008} and sources therein) and now spans a more than 50 year-long history of development. The afore-mentioned code \texttt{Nbody7} \citep{Aarseth2012} also stems from this family, but it is its own serial code using the algorithmic regularization chain method \citep{Mikkolaaarseth1993,MikkolaTanikawa1999a,MikkolaTanikawa1999b,MikkolaMerritt2008,HellstroemMikkola2010}. It is not optimised for massively parallel supercomputers, unlike \texttt{Nbody6++GPU}, which is currently one of the best available high accuracy, massively parallel, direct $N$-body simulation codes. Two very promising alternative and supposedly faster codes have been published during the preparation of this paper; the \texttt{PeTar} \citep{Wangetal2020b,Wangetal2020c,Wangetal2020d} and \texttt{FROST/MSTAR} \citep{Rantalaetal2020,Rantalaetal2021} codes. These two codes are more recently developed and less mature. 
	\\
	The \texttt{Dragon} simulations performed with \texttt{Nbody6++GPU} by \citet{Wangetal2016} are currently still the world-record holder for the largest and most realistic star cluster simulations. The code is optimised for large-scale computing clusters by utilising MPI \citep{Spurzem1999}, OpenMP and GPU \citep{NitadoriAarseth2012,Wangetal2015} parallelisation techniques. In combination with intricate and highly sophisticated algorithms, such as the Kustaanheimo-Stiefel (KS) regularisation \citep{Stiefel1965}, the Hermite scheme with hierarchical block time-steps \citep{McMillan1986,Hutetal1995,Makino1991,Makino1999} and the Ahmad-Cohen (AC) neighbour scheme \citep{AhmadCohen1973}, the code thus allows for star cluster simulations of realistic size without sacrificing astrophysical accuracy by not properly resolving close binary and/or higher-order subsystems of (degenerate) stars. With \texttt{Nbody6++GPU} we can include hard binaries and close encounters (binding energy comparable or larger than the thermal energy of surrounding stars) using two-body and chain regularization \citep{MikkolaTanikawa1999a,MikkolaTanikawa1999b,MikkolaAarseth1998}, which permits the treatment of binaries with periods of days in conjunction and multi-scale coupling with the cluster environment. The AC scheme permits for every star to divide the gravitational forces acting on it into the regular component, originating from distant stars, and an irregular part, originating from nearby stars (``neighbours''). Regular forces, efficiently accelerated on the GPU, are updated in larger regular time steps, while neighbour forces are much more fluctuating and need update in much shorter time intervals. Since neighbour numbers are usually small compared to the total particle number, their implementation on the CPU using OpenMP \citep{Wangetal2015} provides the best overall performance. Post-Newtonian dynamics of relativistic binaries is currently still using the orbit-averaged Peters \& Matthews formalism \citep{PetersMathews1963,Peters1964}, as described e.g. in  \citet{DiCarloetal2019,DiCarloetal2020a,DiCarloetal2020b,DiCarloetal2021,Rizzutoetal2021a,Rizzutoetal2021b,ArcaSeddaetal2021}. In those papers a collisional build-up of massive black holes, over one or even several generations of mergers, was found. The final merger of two massive black holes seen in the simulations is comparable to the most massive one observed by LIGO/Virgo \citep{Abbottetal2020b}.
	
	There is an experimental version of the  \texttt{Nbody6++GPU} code available on request, which uses a full post-Newtonian dynamics up to order PN3.5 including spins of compact objects, spin-orbit coupling to next-to-lowest order and spin-spin coupling to lowest order \citep{Blanchet2014}. It will provide more accurate orbital evolution and better predictions for gravitational waveforms in the final phases before coalescence. An early version of this code variant (only up to PN2.5) has been published in \citet{Kupietal2006,Bremetal2013}.

	\subsection{Monte-Carlo modelling with \texttt{MOCCA}}
	For modelling star clusters there are Monte Carlo methods available that statistically solve the Fokker-Planck equation, which describes gravitational N-body systems \citep{Henon1975}. This method is computationally much less taxing than direct $N$-body \citep{Gierszetal2008, Downingetal2012,HypkiGiersz2013,Gierszetal2013}, but that comes at a cost. It is less realistic in the sense that it can only describe spherical systems. This means that rotation cannot be implemented in these Monte Carlo simulations unlike direct $N$-body simulations \citep{EinselSpurzem1999,Spurzem2001,Ernstetal2007,Kimetal2008,Amaro-Seoaneetal2010,FiestasSpurzem2010,Hongetal2013}. This assumption means that \texttt{MOCCA}, for example, cannot investigate tidal tails \citep{BaumgardtMakino2003,Madridetal2017}. For the Monte-Carlo models of star cluster simulations in this paper we use the MOnte Carlo Cluster SimulAtor \texttt{MOCCA} \citep{HypkiGiersz2013,Gierszetal2013}. This code is based on an improvement of the original Hénon-type Monte-Carlo Fokker-Planck method by \citet{Stodolkiewicz1982,Stodolkiewicz1986} and in a further iteration by \citet{Giersz1998,Giersz2001} and ultimately by \citet{Gierszetal2013}. This approach combines the statistical treatment of the process of relaxation with the particle based approach of direct $N$-body simulations. With this, they are able to model spherically symmetric star clusters over long dynamical times. Three- and four-body interactions in the star cluster simulation are computed separately by the \texttt{FEWBODY} code \citep{Fregeauetal2004}. Furthermore, the escapers from tidally limited star clusters are described by \citet{Fukushigeheggie2000}. Here, the escaping stars stay in the system for some time depending on the excess energy above the escape energy.
	\\ 
	The \texttt{MOCCA} Survey Database I \citep{Askaretal2017a}, which provides a grid of about 2000 GC models, something that is currently unthinkable with direct $N$-body simulations, is a major outcome of the work with \texttt{MOCCA} and is also a testament to the strengths of this modelling approach, which has led to a large number of subsequent studies \citep{Morawskietal2018,Morawskietal2019,ArcaSedda2019, Hongetal2020b,Hongetal2018,Hongetal2020a,Levequeetal2021}. With this database, we can choose appropriate initial conditions for realistic star cluster simulations using direct $N$-body methods. It is important to stress, that despite some important physical simplification of the Monte Carlo method, the results of the MOCCA simulations agree very well with the results of $N$-body simulations for clusters with different initial number of stars (from $10^4$ up to $10^6$) and evolving in different host environments \citep{Gierszetal2013,Gierszetal2016a,HeggieGiersz2014,Wangetal2016,Madridetal2017}. The agreement is not only good for the cluster global properties, but also for properties of the binary population \citep{Gelleretal2019,Rizzutoetal2021a}.
	
	\subsection{Summary: stellar evolution updates (\texttt{SSE \& BSE}) in \texttt{NBODY6++GPU} and \texttt{MOCCA}}
	In this paper we present updates in the \texttt{SSE \& BSE} routines in the two codes \texttt{Nbody6++GPU \& MOCCA}. The details of these updates are shown in Tab.\ref{Nbody6++GPU_Stellar_evolution_levels_literature} and Tab.\ref{MOCCA_Stellar_evolution_levels_literature}, respectively. These updates make \texttt{MOCCA \& Nbody6++GPU} largely competitive in their stellar evolution with other codes that are used to simulate star clusters, such as the Monte-Carlo code \texttt{CMC} \citep{Kremeretal2018a,Kremeretal2019,Kremeretal2020a} with the \texttt{COSMIC} implementation \citep{Breiviketal2020a} or the new, massively parallel direct $N$-body code \texttt{PeTar} \citep{Wangetal2020c}. Furthermore, we are now in a position to model the full evolution of aLIGO/aVirgo gravitational wave sources and their progenitor stars up until the eventual merger according to our best current theoretical understanding. We also implemented the \texttt{SSE \& BSE} version that is shown in Tab.\ref{Nbody6++GPU_Stellar_evolution_levels_literature} into our version of \texttt{McLuster} and we are now able to produce initial star cluster models that have proper evolution of multiple stellar populations (this will be elaborated in a further publication). The details are shown in Appendix B, where also two use-cases are demonstrated to confirm excellent agreement with the \texttt{SSE \& BSE} updates in \texttt{Nbody7} and the results in \citet{Banerjeeetal2020,Banerjee2021a}. 
	\\ 
	The \texttt{SSE \& BSE} implementation within our versions of \texttt{Nbody6++GPU, MOCCA \& McLuster} all contain: 
	\begin{itemize}
		\item updated metallicity dependent stellar winds \citep{Vinketal2001,VinkdeKoter2002,VinkdeKoter2005,Belczynskietal2010},
		\item updated metallicity dependent core-collapse SNe, their remnant masses and fallback \citep{Fryeretal2012,Banerjeeetal2020},
		\item updated electron-capture supernovae (ECSNe), accretion-induced collapse (AIC) and merger-induced collapse (MIC) remnant masses and natal kicks \citep{Nomoto1984,Nomoto1987,NomotoKondo1991,SaioNomoto1985,SaioNomoto2004,Kieletal2008a,GessnerJanka2018}
		\item (P)PISNe remnant masses \citep{Belczynskietal2010,Belczynskietal2016,Woosley2017},
		\item updated fallback-scaled natal kicks for NSs and BHs \citep{Fulleretal2003,Schecketal2004,Fryer2004,FryerKusenko2006, MeakinArnett2006,MeakinArnett2007,FryerYoung2007,Schecketal2008,Fryeretal2012,Banerjeeetal2020},
		\item and BH natal spins (see also \citet{Belczynskietal2020, BelczynskiBanerjee2020}) from
		\begin{itemize}
			\item \texttt{Geneva} model \citep{Eggenbergeretal2008,Ekstroem2012,Belczynskietal2020,Banerjee2021a},
			\item \texttt{MESA} model \citep{Spruit2002,Paxtonetal2011,Paxtonetal2015,Belczynskietal2020,Banerjee2021a},
			\item and the \texttt{Fuller} model \citep{FullerMa2019,Fulleretal2019,Banerjee2021a}.
		\end{itemize}
	\end{itemize} 
	The \texttt{SSE \& BSE} implementation within \texttt{MOCCA} contains, on top of the above:
	\begin{itemize}
		\item winds by \citet{Giacobboetal2018},
		\item winds depending on surface gravity and effective temperature of a star by \citet{SchroederCuntz2005},
		\item (P)PISNe from \texttt{SEVN} simulations by \citet{SperaMapelli2017},
		\item an earlier treatment by \citet{Tanikawaetal2020} to model the evolution of extremely metal-poor and high mass POP III stars,
		\item and proper CV treatment and related dynamical mass transfer, magnetric braking and gravitational radiation critera by \citet{Bellonietal2018c}. 
	\end{itemize}
	The \texttt{SSE \& BSE} algorithms of \texttt{Nbody6++GPU} and \texttt{McLuster} contain, on top of the list of the commonalities between the three codes: 
	\begin{itemize}
		\item moderate and weak (P)PISNe by \citet{Leungetal2019c},
		\item and WD kicks from \citet{Fellhaueretal2003}.
	\end{itemize}
	We discuss future updates in section 5.2.
	
	\section{Initial models - \texttt{delayedSNe-Uniform \& rapidSNe-Sana}}
	\begin{table}
		\begin{tabular}{|l|l|}
			\hline \textbf{Parameter} & \texttt{Nbody6++GPU} \& \texttt{MOCCA}  \\
			\hline \hline \textbf{Particle number} & 110000 \\
			\hline \textbf{Binary fraction} $f_{\text{fb}}$ & $10.0\%$  \\
			\hline \textbf{Half mass radius} $r_{\text h}$ & $1.85$~pc  \\
			\hline \textbf{Tidal radius} $r_{\text{tid}}$ & $500$~pc   \\
			\hline \textbf{IMF} & Kroupa IMF, ($0.08-100$)~$\text M_{\odot}$ \\
			\hline \textbf{Metallicity} $Z$ & $0.00051$  \\
			\hline \textbf{Density model} & King model, $w_0 = 3.0$ \\
			\hline \textbf{Eccentricity distribution} & Thermal \\
			\hline \textbf{Semi-major axis distribution} & flat in log \\
			\hline \hline
		\end{tabular}
		\caption{Initial models for the (\texttt{MOCCA} \& \texttt{Nbody6++GPU}) simulations.}
		\label{Initial_conditions}
	\end{table}
	
	We choose two initial models, which we generate with \texttt{McLuster} \citep{Kuepperetal2011a}, that satisfy the following conditions. Firstly, we do not want these models to be too dense, as we prefer that the dynamics does not overly interfere with the stellar evolution in the star cluster pre-core collapse evolution and secondly, we want the models to have a large tidal radius in order to curtail initial mass loss from the cluster models. With this, we arrive at the structural parameters listed in Tab.\ref{Initial_conditions}. We have a total number of $1.1\times10^5$ particles (i.e. stars), of which $2.0\times 10^4$ are initially in primordial hard binaries. The number of binaries is thus $1.0\times 10^4$ and the binary fraction is $f_{\rm b} = 10$\% The initial half-mass radius $r_{\text{h,0}}$ is set to $1.85$~pc. The smaller particle number then introduces the problem of enhanced mass loss from the cluster. We therefore put the cluster on a circular orbit with a galacto-centric distance of $259.84$~kpc in a MW-like point mass potential of $2.92 \times 10^{12}$ $\mathrm{M}_{\odot}$. This gives an initial tidal radius $r_{\text{tid,0}}$ of $500$~pc in order to curtail this initial mass loss. The density model is a King model with a concentration parameter with $w_0 = 3.0$ \citep{Kingetal1966a} and since it is extremely tidally underfilling, it is very close to the corresponding isolated model. The metallicity of the cluster is set to a low, but realistic (metallicity of the GC NGC3201 \citep{Harris1996}) value of  $Z = 0.00051$, meaning that 0.051 per cent of the mass in the cluster stars is not hydrogen or helium. The IMF is set in a range from ($0.08$-$100.0$) $\mathrm{M}_{\odot}$, following \citet{Kroupa2001}. 
	\\
	The binaries are initially thermally distributed in their eccentricities as is the current standard in $N$-body simulations \citep{Kroupa2008}. This, in general, may overpredict the merger rates significantly \citep{Gelleretal2019}. 
	The binary semi-major axes follow flat distributions in the logarithm of the semi-major axis. The minimum and maximum of the semi-major axes distributions of the primordial binary population are set to the radius of the lowest mass star in the star cluster and $100$~AU, respectively. This distribution of binary semi-major axes for hard binaries is reproduced from an initial distribution that includes many more, wider binaries initially in \citet{Kroupa1995b}. 
	\\
	The difference between the two distinct initial models that we use in this work arises from the choice of binary mass-ratio distribution and SN mechanism. For one model we use the uniform binary mass-ratio distribution $q_{\mathrm{Uniform}}$ and the delayed SNe mechanism and for the other we use the Sana binary mass-ratio distribution $q_{\mathrm{Sana}}$ \citep{Kiminkietal2012,SanaEvans2011,Sanaetal2013a, Kobulnickyetal2014} along with activating the rapid SNe treatment \citep{Fryeretal2012} (both \texttt{Level B}: the parameters chosen are highlighted in \textcolor{orange}{orange} in Tab.\ref{Stellar_evolution_levels_parameters}, \ref{Nbody6++GPU_Stellar_evolution_levels_literature} and \ref{MOCCA_Stellar_evolution_levels_literature}). To clarify, in the $q_{\mathrm{Sana}}$ mass ratio distribution, all the stars that have a mass above 5.0 $\mathrm{M}_{\odot}$ get paired with a secondary, such that the mass ratios are uniformly distributed in the range of $0.1 \leq q \leq 1.0$. The rest of the stars are paired randomly in their mass ratios. In this way, $q_{\mathrm{Sana}}$ and $q_{\mathrm{Uniform}}$ are actually quite similar in theory and we will find out if this is case through the simulations over time. An important point is that through the pairing algorithm for $q_{\mathrm{Sana}}$ in \texttt{McLuster} (with \texttt{pairing}=3), we first select all stars and after that we pair them, so we strictly speaking do respect the IMF \citep{Ohetal2015}. 
	\\
	These two separate models will be referred to as \texttt{delayedSNe-Uniform} and \texttt{rapidSNe-Sana} henceforth. 
	In all other respects the stellar evolution settings of the two simulations are identical (\texttt{Level B}). 
	The stellar evolution levels and their definitions may be understood from Tab.\ref{Stellar_evolution_levels_parameters}, Tab.\ref{Nbody6++GPU_Stellar_evolution_levels_literature} (\texttt{Nbody6++GPU} settings) and Tab.\ref{MOCCA_Stellar_evolution_levels_literature} (\texttt{MOCCA} settings). 
	\\
	We do not enable any (P)PISNe schemes (parameters \texttt{psflag, piflag}) for the \texttt{Nbody} and \texttt{MOCCA} simulations due to the maximum of the IMF at $100\mathrm{M}_{\odot}$ and the low initial cluster density (because of models with very low central density are expected only a few expected stellar mergers that produce stellar masses large enough to be progenitors of (P)PISNe BHs, compare \citet{Kremeretal2020b}). Furthermore, the \texttt{Nbody6++GPU} models have the WD natal kicks switched on following \citet{Fellhaueretal2003} and the \texttt{MOCCA} simulations do not assign natal kicks to the WDs. Moreover, the winds in the \texttt{MOCCA} simulations with \texttt{edd\_factor}=0 ignore the so-called bi-stability jump (see Appendix A2), whereas the \texttt{Nbody6++GPU} simulations with \texttt{mdflag}=3 do not ignore it \citep{Belczynskietal2010}. 
	
	Following the original concept in \citet{Hurleyetal2002b}, we define time step parameters $p_1,p_2,p_3$, to determine how many steps are done during certain evolutionary phases of stars (Note that \citet{Banerjeeetal2020} use symbols \texttt{pts1, pts2 \& pts3} for these). Also \texttt{MOCCA} uses via BSE the same representation. $p_1$ describes the step used in the main sequence phase, $p_2$ in the sub-giant (BGB) and Helium main sequence phase, and $p_3$ in more evolved giant, supergiant, and AGB phases. For clarity we reproduce the equation in \citet{Hurleyetal2002b}, where $\delta t_k$ is the time step used to update the stellar evolution in the code, for stellar type $k$:
	\begin{equation}
	\delta t_k = p_k 
	\begin{cases}
	t_{\rm MS}                 &   k=0,1  \\
	(t_{\rm BGB} - t_{\rm MS})  & k=2      \\
	(t_{\rm inf,1}-t)          & k=3\,t\le t_x \\
	(t_{\rm inf,2}-t)          & k=3\,t>t_x \\
	t_{\rm He}                 & k=4           \\
	(t_{\rm inf,1}-t)          & k=5,6\,t\le t_x \\
	(t_{\rm inf,2}-t)          & k=5,6\,t>t_x \\
	t_{\rm HeMS}               & k=7          \\
	(t_{\rm inf,1}-t)          & k=8,9\,t\le t_x \\
	(t_{\rm inf,2}-t)          & k=8,9\,t>t_x \\
	\max(0.1,10.0 t)                 & k\ge 10 
	\end{cases}       
	\end{equation}
	The original choice in \citet{Hurleyetal2000} was $p_{0,1}=0.01$, $p_{2,7} = 0.05$, and $p_k = 0.02$ for all other $k$. During the following years, in widely used \texttt{Nbody6} codes and derivatives, and in standard BSE packages $p_{0,1}$ and $p_4$ have been increased to 0.05, probably to save some computing time. However, after comparison with \texttt{Startrack} \citep{Belczynskietal2008} models with high time resolution, \citet{Banerjeeetal2020} suggested  $p_{0,1} =0.001$, $p_2=0.01$ and $p_k = 0.02$ for all others. In Fig.4 in \citet{Banerjeeetal2020}, we can see the difference that these time-step choices produce, by producing spikes in the initial-final mass relation for large progenitor ZAMS masses (ignoring (P)PISNe). Currently such small $p_i$ does not pose any significant computational problem; but as seen in \citet{Banerjeeetal2020} such problems with too large $p_i$ only show up for very large stellar masses $M \gtrsim 100 \text M_{\odot}$.
	
	\section{Results}
	\subsection{Global dynamical evolution}
	
	\begin{figure}
		\includegraphics[width=\columnwidth]{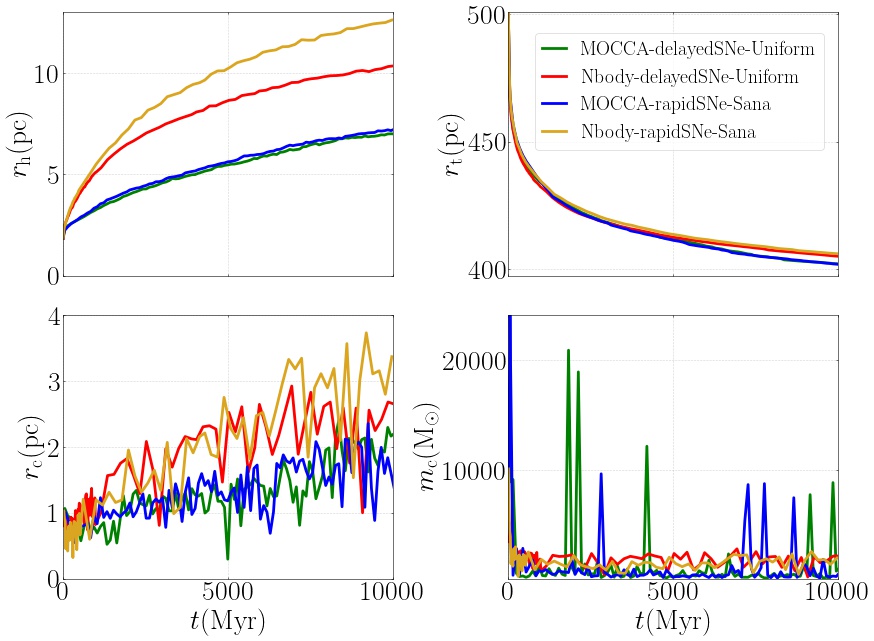}
		\caption{Time evolution of the half-mass radii $r_{\mathrm{h}}$(pc) (top-left), the tidal radii $r_{\mathrm{t}}$(pc) (top-right), core radii $r_{\mathrm{c}}$(pc) (bottom-left) and core masses $m_c$($\text M_{\odot}$) (bottom-right) for the four simulations. The \texttt{Nbody-delayedSNe-Uniform}, \texttt{Nbody-rapidSNe-Sana}, \texttt{MOCCA-delayedSNe-Uniform} and \texttt{MOCCA-rapidSNe-Sana} simulations are shown in \textcolor{red}{red}, \textcolor{yellow}{yellow}, \textcolor{green}{green} and \textcolor{blue}{blue}, respectively.}
		\label{Tidal_radius.png}
	\end{figure}
	
	\begin{figure}
		\includegraphics[width=\columnwidth]{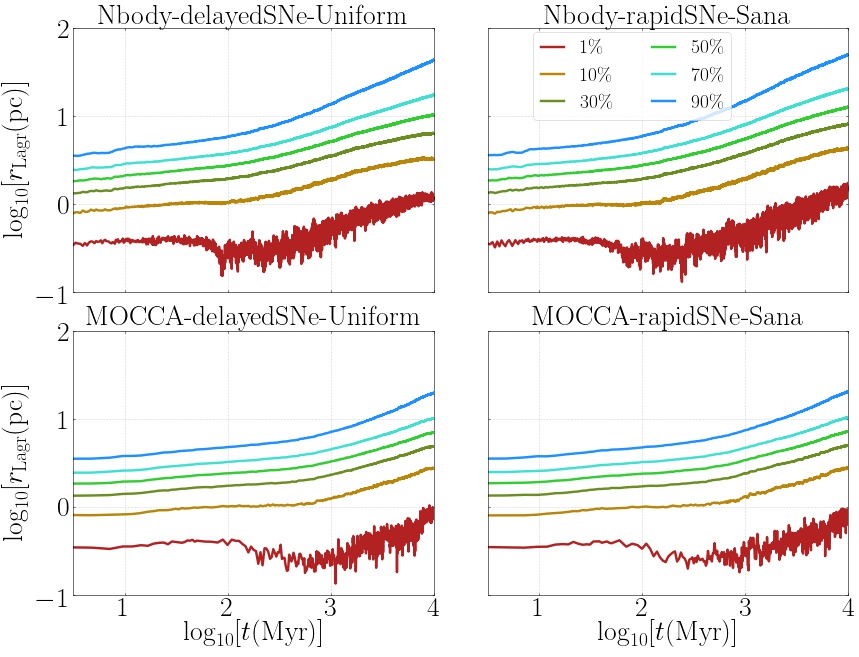}
		\caption{Time evolution of the logarithm of the Lagrangian radii $r_{\mathrm{Lagr}}$ $(1,10,30,50,70,90)$\% for the four simulations: \texttt{Nbody-delayedSNe-Uniform} (top-left), \texttt{Nbody-rapidSNe-Sana} (top-right), \texttt{MOCCA-delayedSNe-Uniform} (bottom-left) and \texttt{MOCCA-rapidSNe-Sana} (bottom-right).}
		\label{Global_lagr.jpg}
	\end{figure}
	
	\begin{figure}
		\includegraphics[width=\columnwidth]{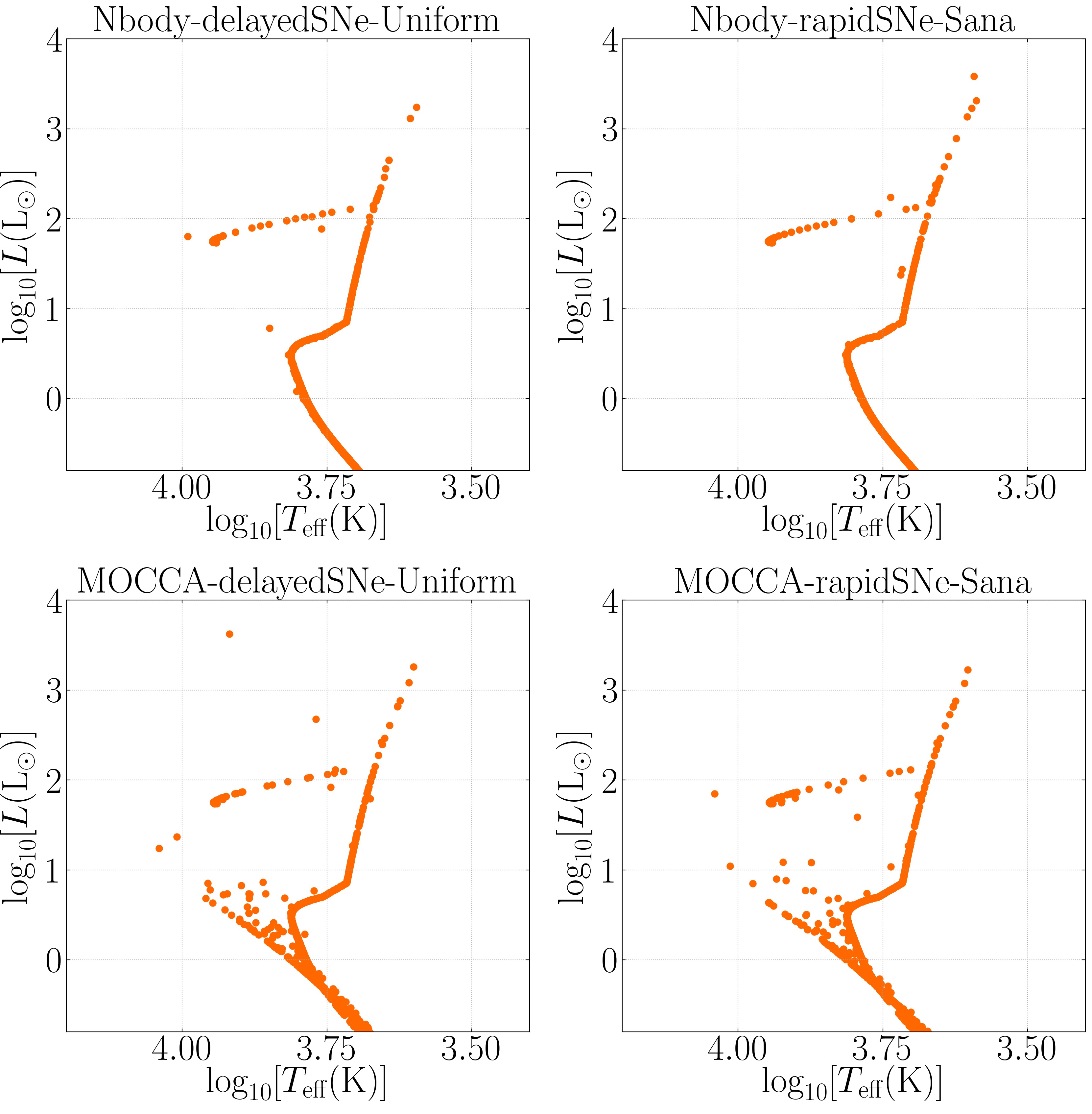}
		\caption{HRD for all four simulates at 10~Gyr. As can be seen from the HRDs of the \texttt{MOCCA} simulations, there are plenty of more blue stragglers in these than in the \texttt{Nbody6++GPU} simulations.}
		\label{HRD_complete.jpg}
	\end{figure}
	
	We run each of the two initial models with \texttt{Nbody6++GPU} and \texttt{MOCCA}. Hence we have four distinct simulations to compare and contrast. We discuss in the following Figs.~\ref{Tidal_radius.png} to \ref{HRD_complete.jpg}, to get an overview over the global evolution of the simulated star clusters.
	\\
	\\
	Fig.\ref{Global_lagr.jpg} shows us that the core collapse happens a bit later in the \texttt{MOCCA} simulations and this is connected with the problems with the timescale. According to H\^{e}non's principle, the rate of cluster evolution is governed by the heat flow through the half-mass radius. Therefore, for smaller $r_{\mathrm{h}}$ and half-mass relaxation time, $t_{\mathrm{h}}$, in \texttt{MOCCA} than in the \texttt{Nbody6++GPU} models, the MOCCA models have to evolve faster and provide more energy in the core than their \texttt{Nbody6++GPU} counterparts. This leads to more dynamical interactions in the core and a small delay in the core-collapse time. Primordial binaries become active earlier as an energy source than in the direct $N$-body simulations. This can also be seen from the core radii, $r_{\mathrm{c}}$, evolution of the cluster models and we see that the \texttt{MOCCA} simulations have a larger central density, which should lead to a larger number of dynamical interactions in \texttt{MOCCA} compared with the \texttt{Nbody6++GPU} runs. Likewise, this can be observed in the larger scatter in remnant masses in Fig.\ref{IFMR}. In combination with the smaller $r_{\mathrm{h}}$ in the \texttt{MOCCA} models, which have a similar total mass (similar $r_{\mathrm{t}}$ in all) to those of the \texttt{Nbody6++GPU} models, this means that the energy flow across $r_{\mathrm{h}}$ is much larger in \texttt{MOCCA} than in the \texttt{Nbody6++GPU} runs. The denser models in the \texttt{MOCCA} simulations are evidenced further in the number of binaries in the simulations. The time evolution of the logarithm of the binary fraction for the four simulations is shown in the top-row of Fig.\ref{Logarithmic_binary_fraction}.  
	Although the overall binary fractions are similar, the \texttt{Nbody6++GPU} simulations yield consistently larger fractions over 10~Gyr. This is due to more scattering events in \texttt{MOCCA} runs that disrupt binaries, which is mirrored by the denser cores and overall clusters in the \texttt{MOCCA} simulations, see Fig.\ref{Tidal_radius.png}. Moreover, looking at Fig.\ref{IFMR}, one can see from the larger scattering in the remnant masses of all compact objects in the \texttt{MOCCA} simulations that there must have been more interactions between the stars that led to mass gain or loss. This is further evidenced by the Hertzsprung-Russel diagram (HRD) in Fig.\ref{HRD_complete.jpg} from all four simulations. We see many more blue stragglers in the HRDs of \texttt{MOCCA} compared with the \texttt{Nbody6++GPU} simulations. This means that there must have been collisions or mass transfer to rejuvenate the stars in order to make them blue stragglers. The likelihood of these formation channels is generally larger in denser systems. 
	
	\subsection{Stellar evolution}
	\subsubsection{Compact binary fractions}
\begin{figure*}
		\includegraphics[width=\textwidth]{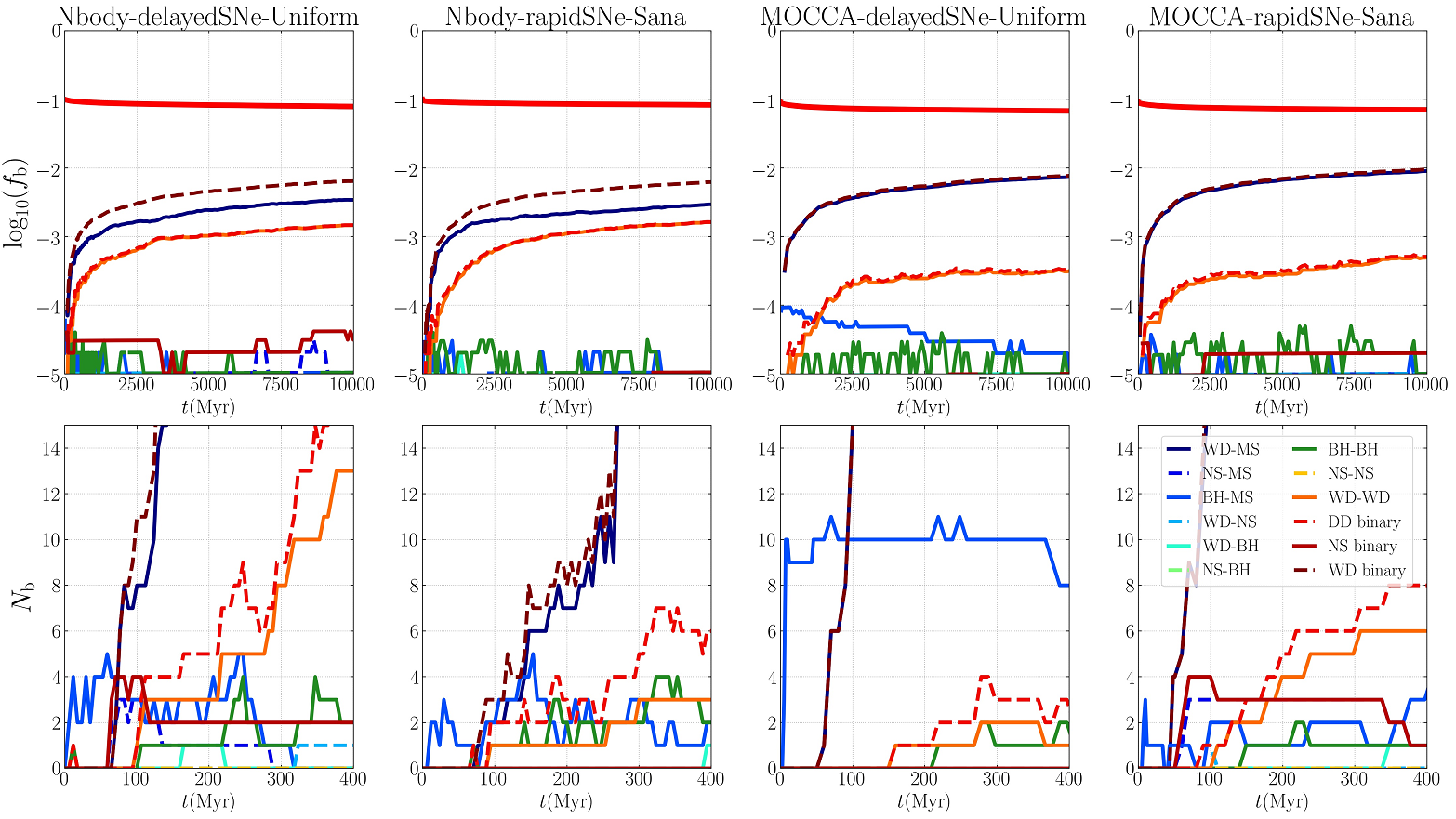}
		\caption{Time evolution over 10~Gyr of the logarithmic (compact) binary fractions (top row) for the \texttt{Nbody-delayedSNe-Uniform}, \texttt{Nbody-rapidSNe-Sana}, \texttt{MOCCA-delayedSNe-Uniform} and  \texttt{MOCCA-rapidSNe-Sana} simulations from left to right, respectively. Shown in the top row as a thick \textcolor{red}{red} line are the overall logarithmic binary fractions. On the bottom row for the first 400~Myr the absolute number of the double-degenerate (DD), NS, WD, WD-MS, NS-MS, BH-MS, WD-NS, WD-BH, NS-BH, BH-BH, NS-NS and WD-WD binaries are shown.}
		\label{Logarithmic_binary_fraction}
	\end{figure*}

	Fig.\ref{Logarithmic_binary_fraction} shows, in addition to the overall binary fraction, the binary fractions of several other compact binaries in which at least one member is a compact object. Both compact binary fractions are dominated by WD binaries, where in the \texttt{MOCCA} simulations the WD binaries are mostly found as WD-MS binaries. In the \texttt{Nbody6++GPU} simulations, there also many WD binaries consisting of secondaries other than MSs, many of them also being WDs. In all simulations the overall WD binary fraction, as well as the WD-MS binary fraction increases over the whole 10~Gyr in contrast to the total star cluster binary fraction.  The double-degenerate (DD) binary fraction for all simulations also increases continuously. This is dominated by WD-WD binaries, where the number of surviving WD-WD binaries in the \texttt{Nbody6++GPU} simulations is much larger than the number in the \texttt{MOCCA} simulations by a factor of about ten. This large discrepancy could be due to faster evolving and denser \texttt{MOCCA} star cluster simulations, which ionise or force to merge more binaries. This is also evidenced by the lower overall binary fractions in the \texttt{MOCCA} models: see also the discussion above. 
	\\
	Further differences in WD binary fractions, especially the WD-MS binaries in Fig.\ref{Logarithmic_binary_fraction}, might additionally arise from the WD kicks that are switched on in the \texttt{Nbody6++GPU} simulations but not in the \texttt{MOCCA} models. In general, these WD kicks are the same for WD types in \texttt{MOCCA} and are assigned an arbitrary kick speed of \texttt{vkickwd}, unlike in \texttt{Nbody6++GPU}, which draws kicks for HeWDs and COWDs from a Maxwellian of dispersion \texttt{wdksig1} and the kicks for the ONeWDs from a Maxwellian with dispersion \texttt{wdksig2}. Both Maxwellians are truncated at \texttt{wdkmax}=6.0~$\text{kms}^{-1}$, where typically \texttt{wdksig1}=\texttt{wdksig2}=2.0~$\text{kms}^{-1}$ following \citep{Fellhaueretal2003}. The presence of these kicks in the \texttt{Nbody6++GPU} models might lead to increased disruption of WD-MS binaries and thus lead to the observed lower abundances. However, since \texttt{MOCCA} and \texttt{Nbody6++GPU} lead to faster and slower global evolution of the star cluster models, respectively, it is difficult to disentangle what actually produces these differences. So far, no cluster simulations on the scale of our simulations presented here have been undertaken investigating the stability of WD binaries in the presence of kicks in detail using both \texttt{MOCCA} and \texttt{Nbody6++GPU} and these need to be performed in the future.
	\\
	From Fig.\ref{Logarithmic_binary_fraction} we can see that near the beginning of all simulations there are small numbers of BH-MS binaries produced for all four simulations, where the \texttt{delayedSNe-Uniform} simulations produce more BH-MS binaries overall. Over the 10~Gyr evolution of our cluster simulations, the \texttt{MOCCA-delayedSNe-Uniform} simulation produces the most surviving BH-MS binaries, but the logarithmic binary fraction is still continuously decreasing. All simulations produce BH-BH binaries in similar numbers where these start forming after about 100~Myr. This suggests that BH-BH binary systems formed in dynamic interactions, since the last BH formed in a SNe was about 80~Myr earlier. At the end of all simulations, we have a surviving BH-BH, whose orbital parameters and masses may be inspected in Tab.\ref{BH_BH_merger}. All of these binaries are located very close to the cluster density centre, with masses of the same order of magnitude, with the highest mass BH in a BH-BH (and all BH binaries) being found in the \texttt{MOCCA-rapidSNe-Sana} model with mass $M_{\mathrm{BH}}=31.032$~$\mathrm{M}_{\odot}$.  The semi-major axes $a$ of these BH-BH binaries are also all smaller than $100$~AU: the closest BH-BH binary found in the \texttt{Nbody-delayedSNe-Uniform} simulation having a semi-major axis value of $53.129$~AU. This is not small enough to have a merger within a Hubble time. The two BH-MS binaries in the \texttt{MOCCA-delayedSNe-Uniform} simulation both consist of an accreting BH with a low mass MS donor star of type \texttt{KW}=0. Therefore, these are not given in Tab.\ref{BH_BH_merger}. 
	\\
	The NS binaries are found further away from the density centre, the closest one coming from the \texttt{MOCCA-delayedSNe-Uniform} run with $r_{\mathrm{dens}}=2.018$~pc. The simulations do not produce any surviving NS-NS, NS-BH, or BH-WD binaries, the former of which are very elusive \citep{ArcaSedda2020,Chattopadhyayetal2020,Chattopadhyayetal2021,Drozdaetal2020}. The \texttt{MOCCA-rapidSNe-Sana} simulation produces one surviving BH-MS binary, whose parameters are given in Tab.\ref{BH_BH_merger}. All simulations produce NS binaries, where at 10~Gyr we have mostly only NS-MS binaries surviving, apart from the \texttt{Nbody6++GPU-delayedSNe-Uniform} simulation, which also produces one NS-COWD binary: see Tab.\ref{BH_BH_merger}. All NS masses in binaries are 1.26~$\mathrm{M}_{\odot}$ and thus these are either the result of a MIC, AIC or ECSNe.
	
	\begin{figure}
		\includegraphics[width=\columnwidth]{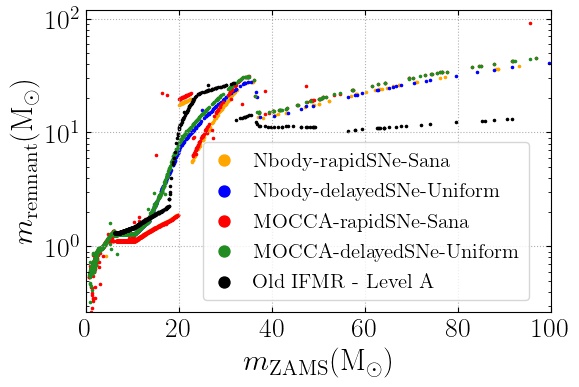}
		\caption{Initial-Final mass relation (IFMR) for the escaping compact objects. The \texttt{Nbody-delayedSNe-Uniform}, \texttt{Nbody-rapidSNe-Sana}, \texttt{MOCCA-delayedSNe-Uniform} and \texttt{MOCCA-rapidSNe-Sana} simulations are shown in \textcolor{red}{red}, \textcolor{green}{green}, \textcolor{blue}{blue} and \textcolor{yellow}{yellow}, respectively. The black points show BH masses from another $N$-body simulation with \texttt{Level A} parameters \citep{Belczynskietal2002}.}
		\label{IFMR_complete}
	\end{figure}
	
	\begin{figure*}
		\includegraphics[width=\textwidth]{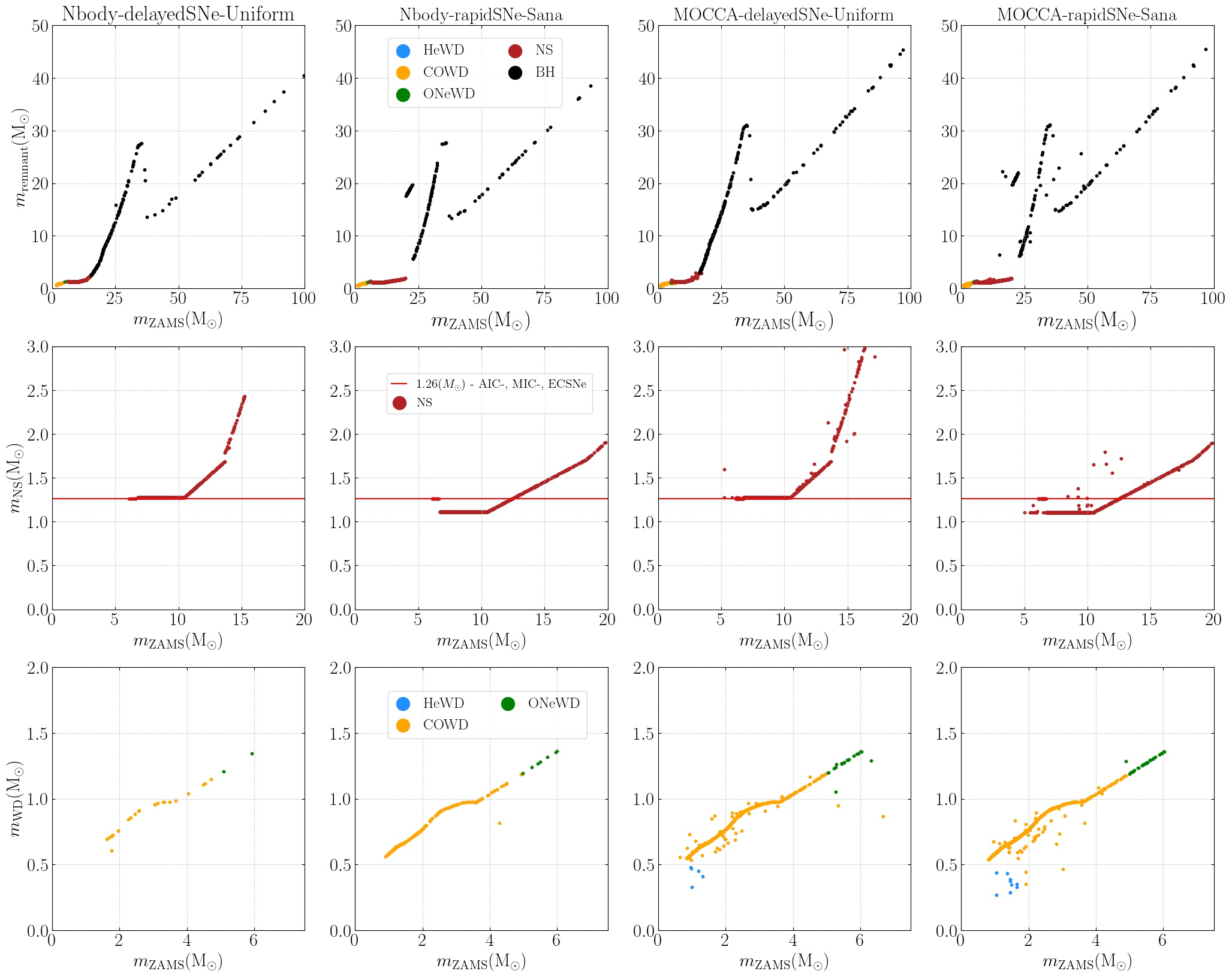}
		\caption{Initial-Final mass relations (IFMR) for the escaping compact objects. From left to right, there are shown the  \texttt{Nbody-delayedSNe-Uniform}, \texttt{Nbody-rapidSNe-Sana}, \texttt{MOCCA-delayedSNe-Uniform} and \texttt{MOCCA-rapidSNe-Sana} simulations, respectively. From top to bottom, there are plotted the IFMRs for the BHs, NS and WDs, the IFMRs of the NSs only and the IFMRs for the WDs, respectively. The top IFMRs show excellent agreement with \citet{Fryeretal2012,Banerjeeetal2020}. The bottom WD IFMR likewise compares well to \citet{Hanetal1995,Hurleyetal2000,HurleyShara2003}. Interestingly, the IFMR shows some NSs escaping at a mass of 1.26~$\text M_{\odot}$ (ECSNe, AIC or MIC) for all simulations even with small natal kicks following \citet{GessnerJanka2018}.}
		\label{IFMR}
	\end{figure*}
	
	\subsubsection{Remnant masses}
\begin{figure*}
		\includegraphics[width=\textwidth]{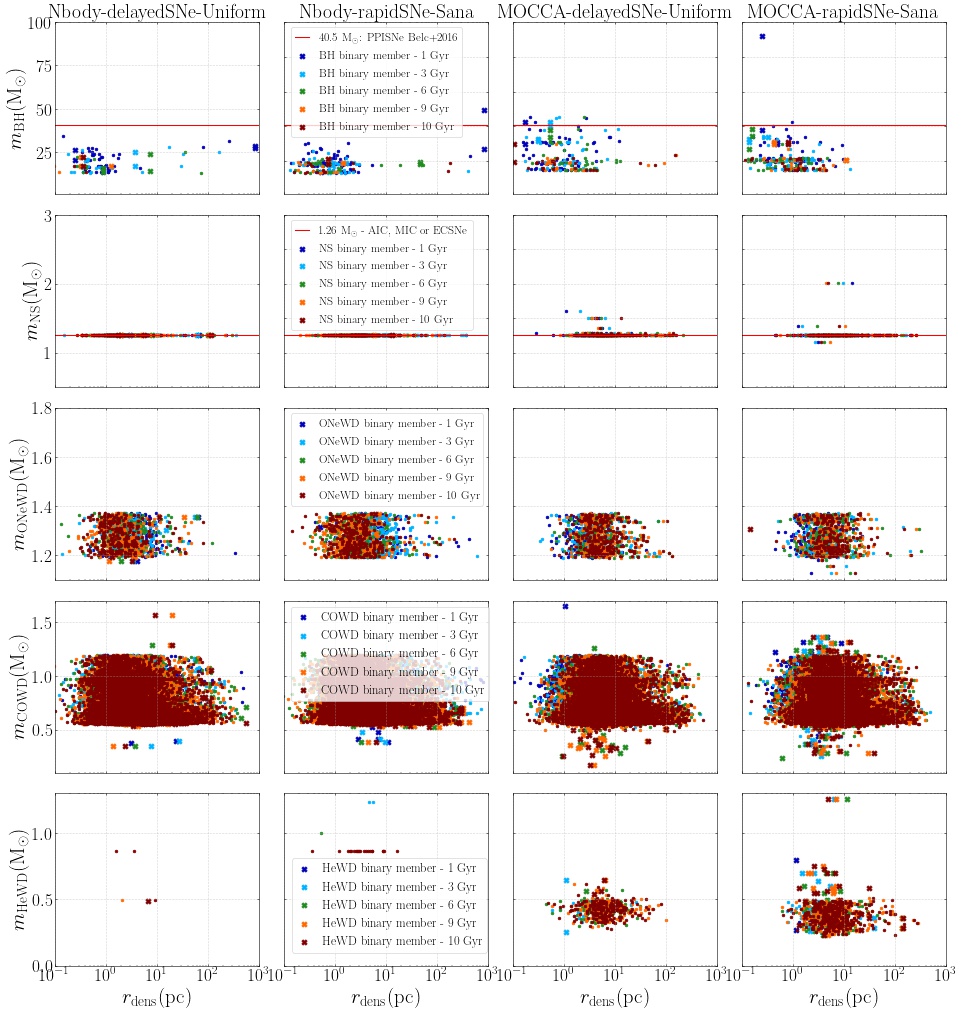}
		\caption{Plot showing the mass of the compact objects in relation to their distance to the density centre $r_{\mathrm{dens}}$(pc) for all four simulations at (1,3,6,9,10)~Gyr. From top to bottom the plots show the above information for the BHs, NSs, ONeWDs, COWDs and HeWDs, respectively. \textbf{BHs}: \texttt{MOCCA} retains more BHs than the \texttt{Nbody6++GPU} simulations and all four simulations retain a BH-BH binary at 10~Gyr. \textbf{NSs}: in the \texttt{Nbody6++GPU} runs only the ECSNe, AIC and MIC NSs are retained, whereas there is a larger spread in remnant masses in the \texttt{MOCCA} simulations (which might be due to a post-natal ECSNe, AIC, MIC NS accreting mass). \textbf{ONeWDs, COWDs}: the distributions across all four simulations are very similar. \texttt{HeWDs}: there are many more HeWDs retained in the \texttt{MOCCA} simulations than the \texttt{Nbody6++GPU} simulations at 10~Gyr.}
		\label{MOCCA_M_RI_cum_RI.png}
	\end{figure*}

\begin{figure*}
		\includegraphics[width=\textwidth]{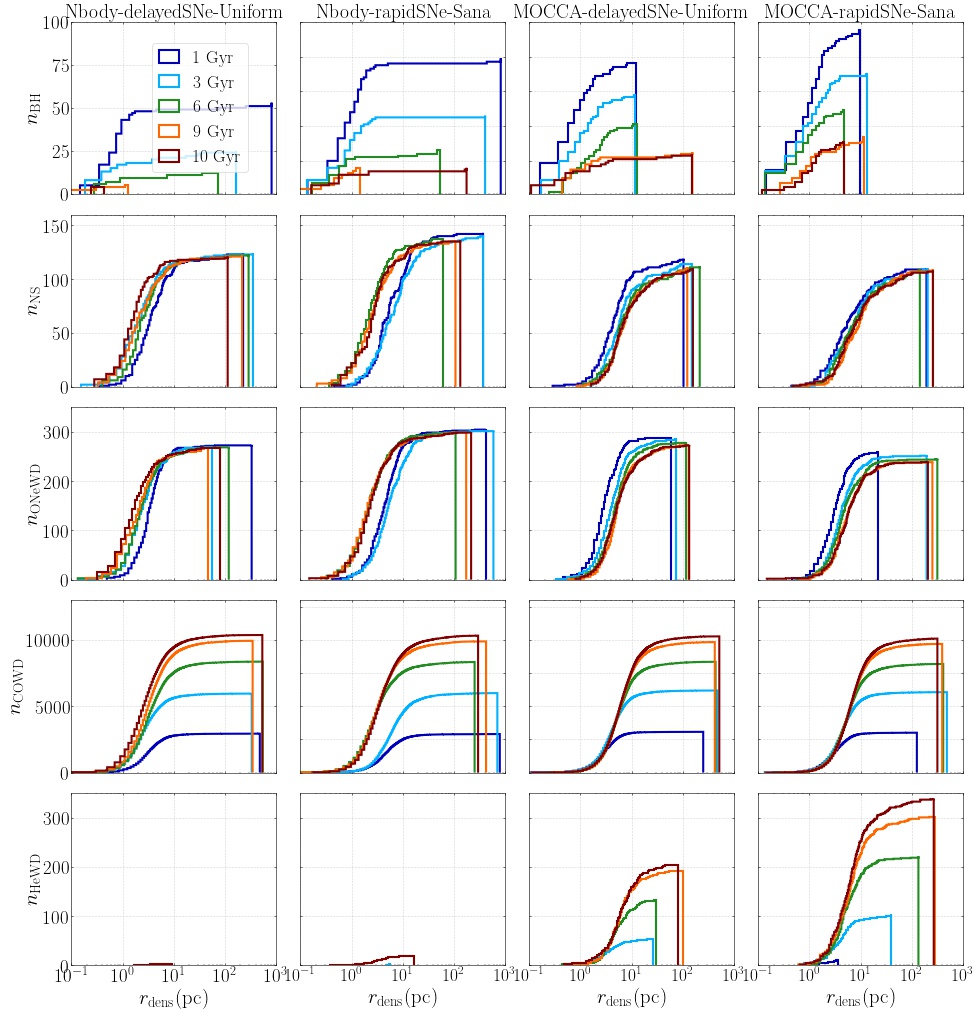}
		\caption{Cumulative distributions for compact object distances to the density centre $r_{\mathrm{dens}}$(pc) for all four simulations at (1,3,6,9,10)~Gyr. From top to bottom the plots show the above information for the BHs, NSs, ONeWDs, COWDs and HeWDs, respectively. \textbf{BHs}: \texttt{Nbody6++GPU} simulations retain consistently lower numbers of BHs with the \texttt{Nbody-delayedSNe-Uniform} having the lowest by far (4). \textbf{NSs}: the \texttt{Nbody6++GPU} simulations have consistently slightly larger numbers of NSs retained, but the distributions are very similar. \textbf{OneWDs} and \textbf{COWDs}: distributions and numbers for these objects are very similar. \textbf{HeWDs}: much lower numbers of HeWDs in the \texttt{Nbody6++GPU} simulations than in the \texttt{MOCCA} simulations. If the enabled WD kicks in \texttt{Nbody6++GPU} were the reason, then we would expect to have equally lower numbers of ONeWDs and COWDs as well, but we do not.}
		\label{Nbody_M_RI_cum_RI.png}
	\end{figure*}

	The remnant masses of the compact objects which have escaped the simulation are shown in the Initial-Final mass relation (IFMR) in  Fig.\ref{IFMR_complete}, where the initial mass is the ZAMS mass and the final mass denotes the compact remnant mass. These remnant masses are mainly determined by our choices of either the \texttt{delayed} SNe or the \texttt{rapid} SNe \citep{Fryeretal2012} and the lack of an enabled (P)PSINe mechanism. The masses of the compact objects in the \texttt{MOCCA} simulations appear to lie systematically above those of the \texttt{Nbody6++GPU} simulations. There exists one very high mass BH of mass $91.830$~$\rm M_{\odot}$ for the \texttt{MOCCA-rapidSNe-Sana} simulation, which escaped at $1.298$~Gyr.  This BH has a complex history and it was subject to an initial binary merger due to stellar evolution. The progenitor stellar mass was $95.618$~$\rm M_{\odot}$. If a (P)PISNe scheme was enabled, then we would never reach these high BH masses of $91.830$~$\mathrm{M}_{\odot}$. The resulting BH would have been capped at $40.5$~$\mathrm{M}_{\odot}$ if we had used \texttt{psflag}=1 \& \texttt{piflag=2} \citep{Belczynskietal2016}, for example. Also shown in this figure, is an old IFMR from \citet{Belczynskietal2002}. These black dots clearly lie below all the compact objects from the new delayed and rapid SNe prescriptions in the range of (30-100)$\mathrm{M}_{\odot}$. We also see that the difference in the delayed and rapid SNe prescription is mostly in the regime up to around $30.0$~$\mathrm{M}_{\odot}$ at our metallicity of $0.00051$. Therefore, the choice of \texttt{nsflag}/\texttt{compactmass} mostly affects the regime $<30.0$~$\mathrm{M}_{\odot}$. At larger ZAMS masses, all four simulations mostly coincide in their IFMRs. For the \texttt{rapidSNe} simulations, we see the double core-collapse hump, whereas for the \texttt{delayedSNe} simulations, we only see one hump \citep{Fryeretal2012}. 
	\\
	In Fig.\ref{IFMR}, we see a more detailed IFMR for each individual simulation, where we also zoom in on the NSs (middle row) and the WDs (bottom row) for all simulations. Apart from the already discussed larger spread in the remnant masses of the compact objects in the \texttt{MOCCA} simulations, the simulations show good consistency with each other, as well as the literature \citet{Fryeretal2012}. This is also true for the WD masses, which are unaffected by the delayed or rapid SNe mechanisms and which follow the original SSE algorithm \citep{Hanetal1995,Hurleyetal2000,HurleyShara2003}. To add more depth to the analysis, see Fig.\ref{MOCCA_M_RI_cum_RI.png} and Fig.\ref{Nbody_M_RI_cum_RI.png} for the masses of all the compact objects (BH, NS, ONeWD, COWD, HeWD) versus their distance to the density centre, $r_{\mathrm{dens}}$, as well as the cumulative histograms of the compact object distances for the \texttt{MOCCA} and the \texttt{Nbody6++GPU} simulations, respectively. There are objects in these plots that extend beyond the tidal radius. This is due to the fact that the escape criterion in \texttt{Nbody6++GPU} removes stars once they are further than two times the tidal radius from the density centre. Overall, there a lot more HeWDs both escaping and remaining inside the clusters of the \texttt{MOCCA} simulations over the full 10~Gyr. We know that HeWDs cannot be formed in the stellar evolution of single stars in a Hubble time. They can be formed only in binaries. In MOCCA models the central density is larger than in the $N$-body models, so it is expected that more frequent dynamical interactions force binaries to form HeWDs because of mass transfer. \\
	The COWD numbers and their distributions are similar for all simulations, but there are many more COWD-COWD binaries in the \texttt{Nbody6++GPU} simulations, mirroring findings in Fig.\ref{Logarithmic_binary_fraction}. The mass and $r_{\mathrm{dens}}$ distributions of the ONeWDs for the \texttt{MOCCA} and \texttt{Nbody6++GPU} simulations are likewise similar, but there are more outlying ONeWDs for the \texttt{MOCCA} simulations, indicating and underlying the point made early about the \texttt{MOCCA} simulations having more interactions across their full evolution: see Fig.\ref{Tidal_radius.png} and \ref{IFMR}. The \texttt{Nbody6++GPU} simulations retain slightly larger numbers of NSs inside the cluster than the \texttt{MOCCA} simulations. Additionally, the \texttt{Nbody6++GPU} simulations only retain NSs of masses 1.26~$\rm M_{\odot}$, which is the mass that is assigned for NSs produced by an ECSNe, AIC or MIC. The \texttt{MOCCA} simulations have a much larger spread in the NS masses again underpinning the point that the \texttt{MOCCA} simulations are denser and lead to more interactions between the stars. The BH masses are distributed very dissimilarly. Firstly, the \texttt{Nbody-delayedSNe-Uniform} simulation retains the least BHs up until 10~Gyr; two single BHs and the BH-BH binary (see Tab.\ref{BH_BH_merger}). This BH-BH binary is also the hardest of all BH-BH binaries remaining at 10~Gyr. The \texttt{MOCCA-rapidSNe-Sana} simulation retains the largest number of BHs up until 10~Gyr (around 20 of which two are in a BH-BH binary). This BH-BH binary is the most massive (combined mass of around 60~$\rm M_{\odot}$) and also the most distant to the density centre of this cluster. The \texttt{MOCCA-delayedSNe-Uniform} and the \texttt{Nbody-rapidSNe-Sana} retain similar numbers of BHs and they are also distributed similarly. 
	
	\subsubsection{Remnant natal kicks and escape speeds}

	\begin{table*}
		\centering
		\begin{tabular}{|l|l|l|l|l|l|l|l|l|}
			\hline \textbf{Simulation} & \textbf{type} $M_1$-$M_2$& $M_1$($\rm M_{\odot}$) &  $M_2$($\rm M_{\odot}$) & $e$ & $P(\mathrm{days})$ & $a(\mathrm{AU})$ & $r_{\mathrm{dens}}(\mathrm{pc})$ & $t_{\mathrm{GW}}(\mathrm{Gyr})$\\
			\hline \hline \texttt{Nbody-delayedSNe-Uniform}& \textbf{BH-BH} & 22.586 & 17.145 & 0.415 &  22452 & 53.129 & 0.355 & $2.268\times10^{10}$\\
			\hline \texttt{Nbody-delayedSNe-Uniform}& \textbf{MS-NS} & 0.871 & 1.260 & 0.479 & 5271 & 7.600 & 1.727 & /\\
			\hline \texttt{Nbody-delayedSNe-Uniform}& \textbf{NS-COWD} & 1.260 & 0.892 &  0.729 & 56863 & 37.361 & 5.535 & $2.712\times10^{12}$\\
			\hline \hline \texttt{Nbody-rapidSNe-Sana}& \textbf{BH-BH} & 18.275 & 20.969 & 0.953 & 24207 & 55.655 & 0.749 & $4.773\times10^{6}$\\
			\hline \texttt{Nbody-rapidSNe-Sana}& \textbf{NS-MS} & 1.260 & 0.553 & 0.766 & 133522 & 62.343 & 12.920 & / \\ 
			\hline \hline \texttt{MOCCA-delayedSNe-Uniform}& \textbf{BH-BH} & 29.910 & 19.747 & 0.940 & 31703 & 72.060 & 0.108 & $1.616\times10^{7}$ \\
			\hline \texttt{MOCCA-delayedSNe-Uniform}& \textbf{NS-MS} & 1.260 & 0.767 & 0.889 & 3517620 & 572.904 & 2.018 & /\\
			\hline \hline \texttt{MOCCA-rapidSNe-Sana}& \textbf{BH-BH} & 29.905 & 31.032 & 0.329 & 22269 & 60.963 & 0.811 & $1.598\times10^{10}$ \\
			\hline \texttt{MOCCA-rapidSNe-Sana}& \textbf{BH-MS} & 21.156 & 0.104 & 0.772 & 9223 & 23.845 & 3.3775 & /\\
			\hline \texttt{MOCCA-rapidSNe-Sana}& \textbf{NS-MS} & 1.260 & 0.343 & 0.801 & 153356 & 65.626 & 7.575 & /\\
			\hline \hline 
		\end{tabular}
		\caption{Table listing the orbital properties of some degenerate binaries surviving inside the cluster at time 10~Gyr with at least one member being a BH or a NS. Also shown is the expected merger timescale $t_{\mathrm{GW}}$ for the compact binaries computed from \citet{PetersMathews1963,Peters1964}. None of these compact binaries would be relevant for aLIGO or aVirgo detections.}
		\label{BH_BH_merger}
	\end{table*}
	
	\begin{figure}
		\includegraphics[width=\columnwidth]{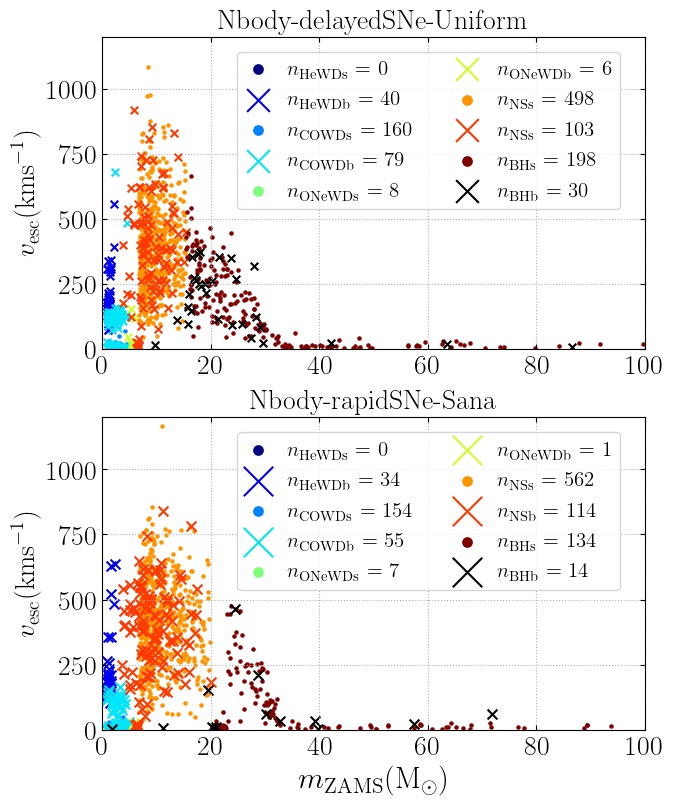}
		\caption{Plot showing the final escape speeds $v_{\mathrm{esc}}(\mathrm{kms}^{-1})$ of the compact objects (WDs, NSs, BHs) for the \texttt{Nbody-delayedSNe-Uniform} and the \texttt{Nbody-rapidSNe-Sana} simulations. Also shown in crosses are the compact objects that come from primordial ZAMS binary stars ($n_{\mathrm{HeWDb}}, n_{\mathrm{COWDb}}, n_{\mathrm{ONeWDb}}, n_{\mathrm{NSb}}, n_{\mathrm{BHb}}$), whereas the smaller dots display compact objects originating from ZAMS single stars ($n_{\mathrm{HeWDs}}, n_{\mathrm{COWDs}}, n_{\mathrm{ONeWDs}}, n_{\mathrm{NSs}}, n_{\mathrm{BHs}}$). The number counts $n_{\mathrm{HeWDb}}, n_{\mathrm{COWDb}}, n_{\mathrm{ONeWDb}}, n_{\mathrm{NSb}}, n_{\mathrm{BHb}}, n_{\mathrm{HeWDs}}, n_{\mathrm{COWDs}}, n_{\mathrm{ONeWDs}}, n_{\mathrm{NSs}}, n_{\mathrm{BHs}}$ are recorded in the plot legend.}
		\label{Nbody_V_Kick.jpg}
	\end{figure}
	
	\begin{figure}
		\includegraphics[width=\columnwidth]{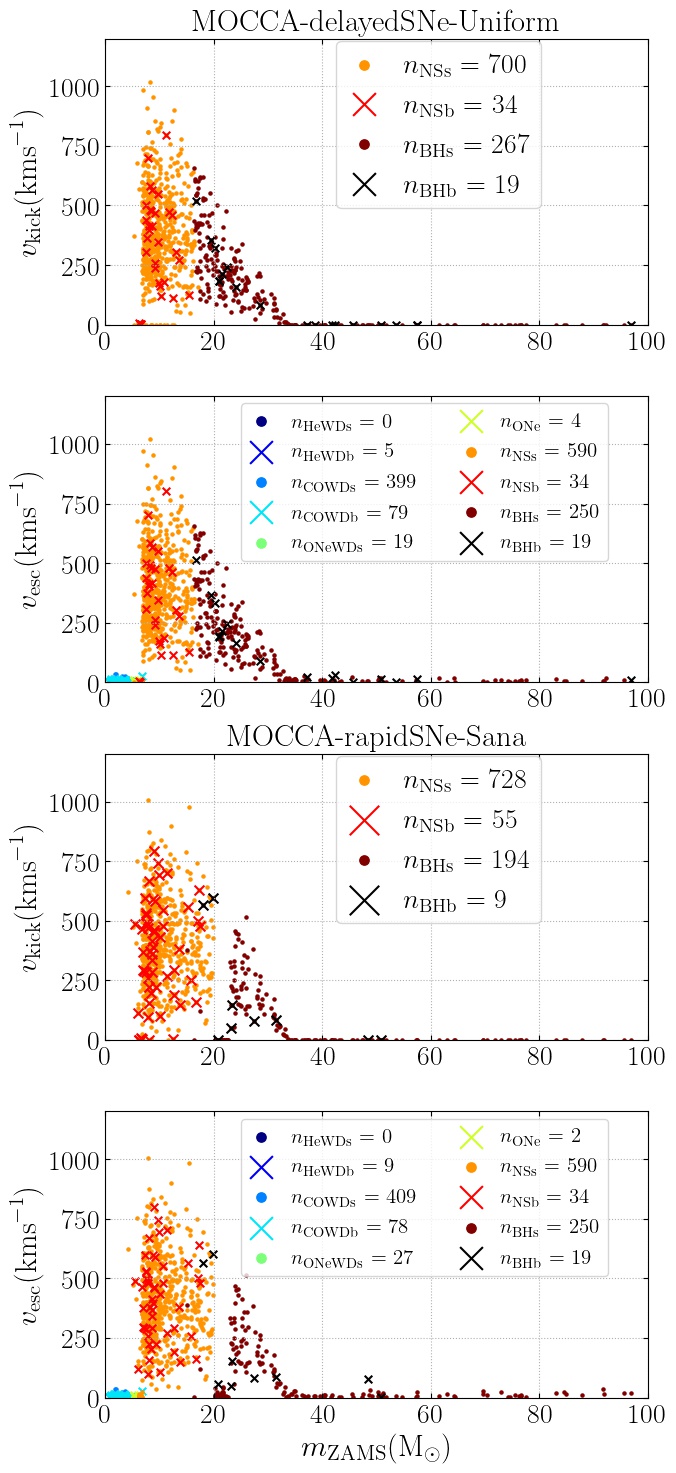}
		\caption{Plot showing the natal kick speeds $v_{\mathrm{kick}}(\mathrm{kms}^{-1})$ of the NSs and BHs (not recorded for WDs), as well as the final escape speeds $v_{\mathrm{esc}}(\mathrm{kms}^{-1})$ of all the compact objects (HeWDs, COWDs, ONeWDs, NSs, BHs) for the \texttt{MOCCA-delayedSNe-Uniform} (top two panels) and the \texttt{MOCCA-rapidSNe-Sana} (bottom two panels) simulations. Also shown in crosses are the compact objects that come from primordial ZAMS binary stars ($n_{\mathrm{HeWDb}}, n_{\mathrm{COWDb}}, n_{\mathrm{ONeWDb}}, n_{\mathrm{NSb}}, n_{\mathrm{BHb}}$), whereas the smaller dots display compact objects originating from ZAMS single stars ($n_{\mathrm{HeWDs}}, n_{\mathrm{COWDs}}, n_{\mathrm{ONeWDs}}, n_{\mathrm{NSs}}, n_{\mathrm{BHs}}$). The number counts $n_{\mathrm{HeWDb}}, n_{\mathrm{COWDb}}, n_{\mathrm{ONeWDb}}, n_{\mathrm{NSb}}, n_{\mathrm{BHb}}, n_{\mathrm{HeWDs}}, n_{\mathrm{COWDs}}, n_{\mathrm{ONeWDs}}, n_{\mathrm{NSs}}, n_{\mathrm{BHs}}$ are recorded in the plot legend. The compact objects with a zero kick velocity have a constant value of $0.0001$~$\mathrm{kms}^{-1}$ added to them to make them visible.}
		\label{MOCCA_V_Kick.jpg}
	\end{figure}

	In Fig.\ref{Nbody_V_Kick.jpg}, the escape speeds $v_{\mathrm{esc}}$ of the compact objects in relation to their ZAMS mass are shown for the \texttt{Nbody6++GPU} simulations. The absolute number of the objects per stellar type are shown and we distinguish between objects coming from either a ZAMS single star or ZAMS binary. This information, as well as the kick speeds $v_{\mathrm{kick}}$ for the NSs and BHs for the \texttt{MOCCA} simulations, is also shown in Fig.\ref{MOCCA_V_Kick.jpg}. For the \texttt{MOCCA} simulations, we computed the escape speeds $v_{\mathrm{esc}}$ from their escape energies infinitely far away from the cluster. 
	\\
	First, we discuss the WDs, for which we have the $v_{\mathrm{esc}}$ information readily available across all four simulations. All escaping HeWDs originate from ZAMS binaries in both simulations, which is expected from mass transfer in binaries and the production pathways of HeWDs in general. Their escape speeds reach a couple of hundred $\mathrm{kms}^{-1}$ in some instances for the \texttt{Nbody6++GPU} simulations. This is not the case for the \texttt{MOCCA} simulations. Comparing this with Fig.\ref{MOCCA_M_RI_cum_RI.png} and Fig.\ref{Nbody_M_RI_cum_RI.png}, there are still single HeWDs retained in both the \texttt{Nbody6++GPU} and \texttt{MOCCA} simulations, but a lot fewer for the \texttt{Nbody6++GPU} simulations than for the \texttt{MOCCA} simulations and on the other hand, many more HeWDs escape the \texttt{Nbody6++GPU} simulations than the \texttt{MOCCA} simulations. 
	\\
	Many more COWDs originating from ZAMS singles stars escape than those with a ZAMS binary origin in the \texttt{Nbody6++GPU} runs. The same is true for the \texttt{MOCCA} simulations, but here many more COWDs originating from ZAMS singles escape than from the \texttt{Nbody6++GPU} simulations. In the \texttt{Nbody6++GPU} simulations the escape speeds of the escaping COWDs from ZAMS binaries are much larger than those of the COWDS from ZAMS singles. This should be expected, because if the binary companion underwent a SNe event, the COWD or progrenitor might have adopted the binary's high orbital speed. In the \texttt{MOCCA} simulation, however, the COWDs (and all other WD types) from ZAMS singles and ZAMS binaries escape with highly uniform $v_{\mathrm{esc}}$. This needs to be investigated further in the future. In total, there are many more COWDs and ONeWDs retained for all simulations than those that escape (see Fig.\ref{MOCCA_M_RI_cum_RI.png} and \ref{Nbody_M_RI_cum_RI.png}). Consistently more ONeWDs escape the \texttt{MOCCA} simulations from singles and binary ZAMS stars. Future studies into the impact of WD natal kicks on binary stability, escape speeds and escaper number are needed going forward. 
	\\
	The BHs and NSs are affected by the delayed and rapid SNe as well as the fallback-scaled natal kicks, while the WDs are not. We see that compared with the \texttt{Nbody6++GPU} simulations, the distributions of the BH and NS escape speeds are very similar. The \texttt{KMECH}=1 in \texttt{Nbody6++GPU} and the \texttt{bhflag\_kick}=\texttt{nsflag\_kick}=3 settings in \texttt{MOCCA} for the fallback-scaled momentum conserving kicks, compare also Fig.\ref{McLuster_Remnant_masses_nsflag3_nsflag4.png} and \ref{McLuster_psflag.jpg}, lead to very similar distributions in escape speeds. It also shows that escape speeds and the natal kick speeds of the \texttt{MOCCA} simulations are very similar. To clarify again, $v_{\mathrm{kick}}$ and $v_{\mathrm{esc}}$ describe the actual natal kick velocity and the velocity at escape from the cluster, respectively. The speeds for the escaping NSs in all four simulations reach up to $10^3$~$\mathrm{kms}^{-1}$. 
	\\
	The NSs produced from AIC, ECSNe and MIC lead to very low escape speeds as a result of the very low natal kicks, which we assign by using  \texttt{ECSIG}=\texttt{sigmac}=3.0~$\mathrm{kms}^{-1}$ from \citet{GessnerJanka2018}. Even still, some of these NSs escape from all clusters without any significant acceleration. This may be due to evaporation, where a series of weak encounters finally leads to an escape of the NS, or by a strong dynamical ejection. Another reason might be their involvement in a binary, i.e., they were a member of a binary and the binary snaps due to the SN of its companion, causing the star to adopt the high orbital speed of the binary (similar to the proposed mechanism for the high $v_{\mathrm{esc}}$ for some HeWDs and COWDs in the \texttt{Nbody6++GPU} simulations). 
	\\
	The low mass BHs in the \texttt{delayedSNe} simulations also reach $10^3$~$\mathrm{kms}^{-1}$, whereas the low mass BHs just at the transition between the NSs and BHs in the \texttt{rapidSNe} simulations are very low, leading to a small gap in velocity distribution of the escaping BHs. This is due to the first of the two core-collapse humps in the remnant mass distribution of the rapid SNe scheme \citep{Fryeretal2012,Banerjeeetal2020}; the larger the fallback, the lower the natal kick of the NS or BH. Nevertheless, even some BHs in this gap escape all \texttt{rapidSNe} simulations, which is a result of the low masses of the clusters and thus the low escape speeds. In realistically sized GCs, these BHs would probably not escape, unless through some hard encounter. The larger the ZAMS mass, the lower the resulting escape speed and natal kicks are, due to increasing fallback. This is why at the high end of the BH mass spectrum, the velocities become very small (only a couple of $\mathrm{kms}^{-1}$) in all simulations. 
	
	\subsubsection{Binary parameters}
	
	\begin{figure}
		\includegraphics[width=\columnwidth]{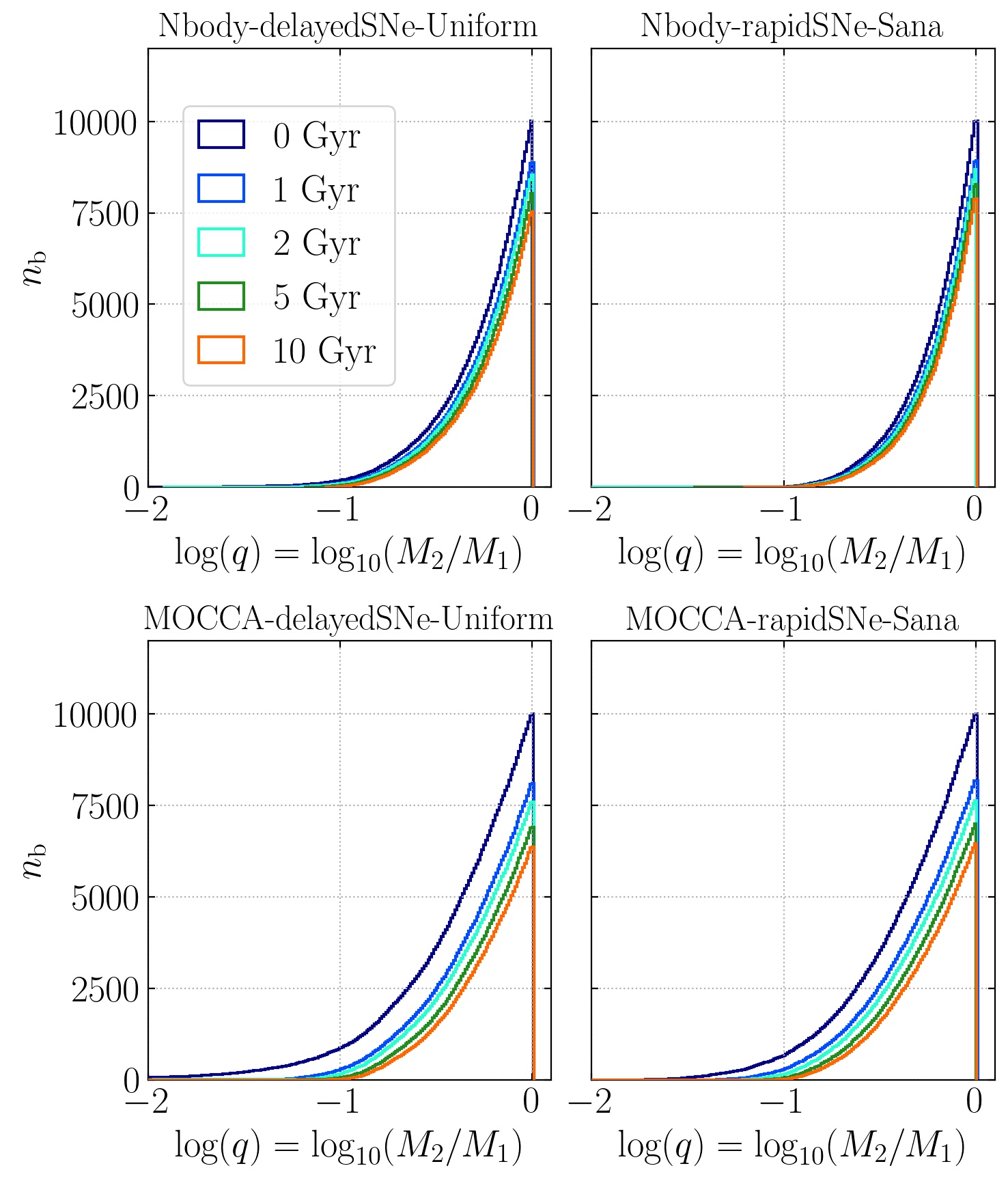}
		\caption{Cumulative histogram showing the mass ratio at times (1,2,5,9,10)~Gyr of the binaries for all four simulations. The mass ratio $q$ is calculated such that the lower mass $M_2$ is divided by the larger mass $M_1$.}
		\label{Mass_ratio}
	\end{figure}
	
\begin{figure}
		\includegraphics[width=\columnwidth]{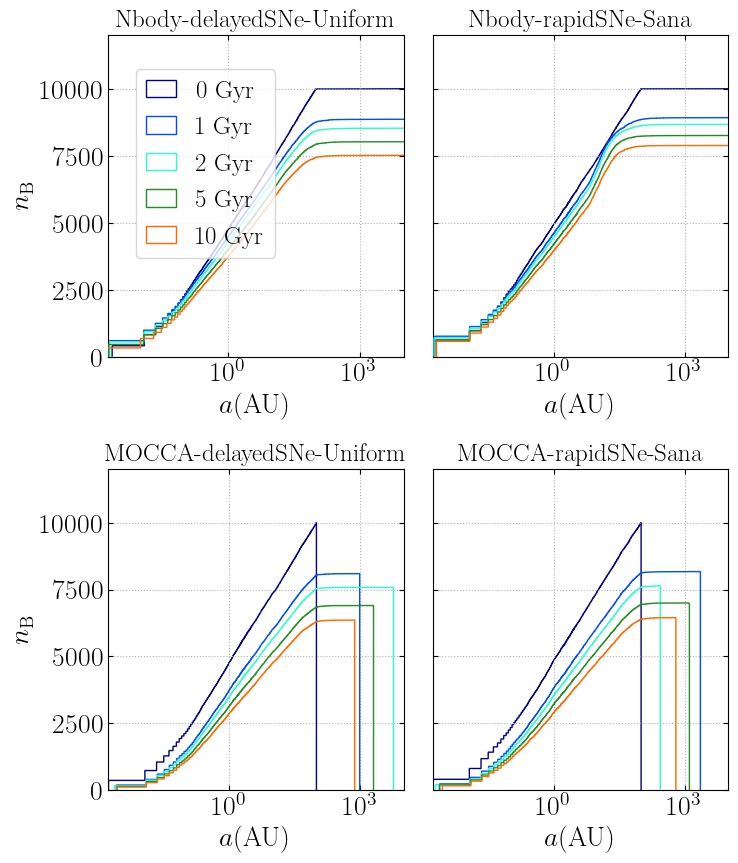}
		\caption{Cumulative histogram showing the semi-major axis $a$(AU) at times (1,2,5,9,10)~Gyr of the binaries for all four simulations.}
		\label{semi-major-axis}
	\end{figure}

\begin{figure}
		\includegraphics[width=\columnwidth]{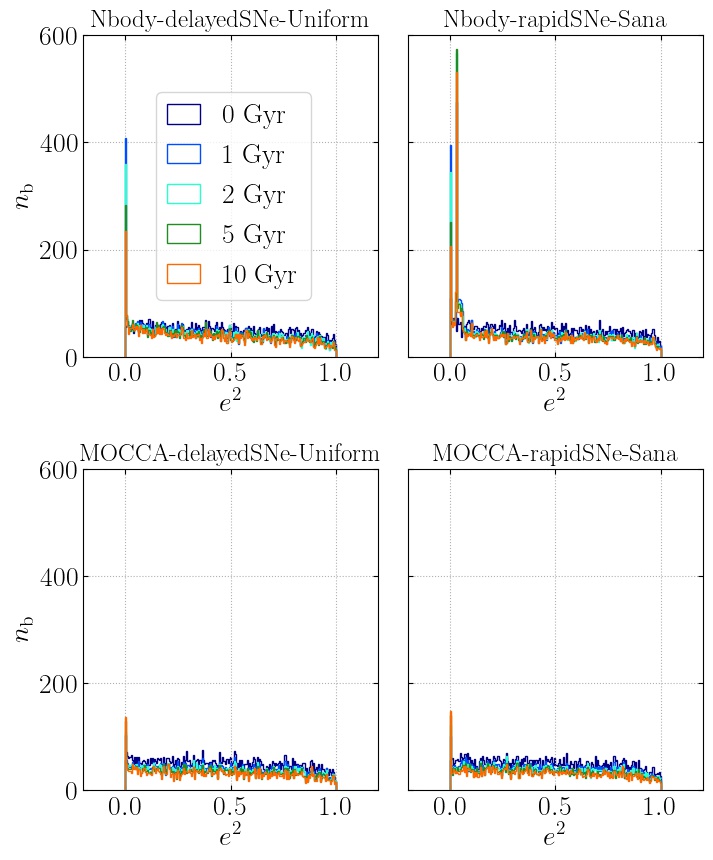}
		\caption{Histogram showing the eccentricity $e^2$ at times (1,2,5,9,10)~Gyr of the binaries for all four simulations ($N(<e)\propto e^2$) for a thermal distribution \citep{DuquennoyMayor1991}). The \texttt{Nbody-rapidSNe-Sana} simulation reveals a second peak, which might relate to regularisation or another complex origin.}
		\label{eccentricity}
	\end{figure}

	The only different initial binary parameters between the \texttt{delayedSNe-Uniform} and the \texttt{rapidSNe-Sana} simulations is the binary mass ratio distribution $q$, which is set to $q_{\text{Uniform}}$ and $q_{\text{Sana}}$, respectively. The binary mass ratios for all four simulations at times (1,2,5,9,10)~Gyr are presented in Fig.\ref{Mass_ratio}. The evolution across all simulations leads to very similar distributions at 10~Gyr with only a few very large binary mass ratios. We note that strictly speaking the $q_{\text{Uniform}}$ and $q_{\text{Sana}}$ initial distributions are very similar overall and thus it is not surprising, but rather reassuring, that this is indeed the case in the simulations. We also see similarities in the semi-major axes $a$ of the binaries as shown in Fig.\ref{semi-major-axis}. The shape of the curve is roughly what we would expect, since they are distributed flat in log($a$), however, for the \texttt{Nbody-rapidSNe-Sana} there is a small clustering at wide binaries in the cumulative distribution. This can more easily be seen as an unusual increase in the cumulative histogram of the binary eccentricities at low eccentricities in Fig.\ref{eccentricity}. This might be due to a change in regularisation, when the binaries move in and out of KS regularisation. Some testing has been done and we can confirm that this issue is definitely not related to stellar evolution and needs to be resolved in the future. Interestingly, this clustering does not seem to be present in the \texttt{Nbody-delayedSNe-Uniform} simulation. Therefore, it might be related to the hardware or technical parameters within the initialisation of the simulations. However, we did not change any of these between the two \texttt{Nbody6++GPU} simulations and therefore this seems unlikely. We need to explore this erratic issue further and resolve this. 
	
	\section{Summary, conclusion and perspective}
	\subsection{Summary: direct $N$-body (\texttt{NBODY6++GPU}) and Monte Carlo (\texttt{MOCCA}) simulations}
	We have compared direct $N$-body (\texttt{NBODY6++GPU}) and Monte Carlo (\texttt{MOCCA}) star cluster models for about 10~Gyr with our updated codes. We showcase the effect of parts of the updated stellar evolution, more specifically the delayed vs. rapid SNe as extremes for the convection-enhanced neutrino-driven SNe paradigm by \citet{Fryeretal2012} with standard momentum conserving fallback-scaled kicks in combination with metallicity dependent winds from \citet{Vinketal2001,VinkdeKoter2002,VinkdeKoter2005,Belczynskietal2010} and low-kick ECSNe, AIC and MIC \citep{Podsiadlowskietal2004b,Ivanovaetal2008,GessnerJanka2018,Leungetal2020a}. The BHs had no natal spins set (corresponding to the \texttt{Fuller} model in \citet{Banerjee2021a} from \citet{FullerMa2019,Fulleretal2019}). The initial model with the delayed SNe enabled had the binary mass ratios uniformly distributed ($q_{\text{Uniform}}$) and is dubbed \texttt{delayedSNe-Uniform}, whereas the initial model with the rapid SNe enabled, had the binary mass ratios distributed as inspired by observations following \citet{Kiminkietal2012,SanaEvans2011,Sanaetal2013a,Kobulnickyetal2014} ($q_{\text{Sana}}$) and is dubbed \texttt{rapidSNe-Sana}. The \texttt{MOCCA} models did not employ WD kicks, whereas the \texttt{Nbody6++GPU} models used WD natal kicks following \citet{Fellhaueretal2003}. The time-steps \texttt{pts1, pts2 \& pts3} of \texttt{MOCCA} represent fractions of stellar lifetimes in the main sequence, sub-giant, and more evolved phases that are taken as stellar-evolutionary time steps in the respective evolutionary stages and should, after calibrating them with \texttt{Startrack} \citep{Belczynskietal2008}, follow the suggestions by \citet{Banerjeeetal2020}: \texttt{pts1}=0.001, \texttt{pts2}=0.01 and \texttt{pts3}=0.02. In the \texttt{Nbody6++GPU} simulations, the time-steps \texttt{pts2 \& pts3} are all accounted for by \texttt{pts2}. Here, we chose \texttt{pts1}=0.05 and \texttt{pts1}=0.02. We make the following observations:
	\begin{itemize}
		\item Globally, the star cluster models evolve differently. The mass loss from \texttt{Nbody6++GPU} is slightly lower than that from the \texttt{MOCCA} simulations. The \texttt{Nbody6++GPU} simulations have consistently larger $r_{\mathrm{Lagr}}$ than the \texttt{MOCCA} simulations (see Fig.\ref{Tidal_radius.png}). In particular, the half-mass radii are significantly larger than those in the \texttt{MOCCA} simulations. Fig.\ref{Global_lagr.jpg} shows us that core collapse happens a bit later in the \texttt{MOCCA} simulations and this is connected with the time-scaling. In the Monte Carlo models the global cluster evolution rate is governed according to H\'{e}non's principle by the heat flow through the half-mass radius. So for smaller half-mass radius and half-mass relaxation time in \texttt{MOCCA} than in \texttt{Nbody6++GPU} models, the \texttt{MOCCA} models have to evolve faster and provide more energy in the core than for the \texttt{Nbody6++GPU} approach. This leads to more dynamical interactions in the core and a small delay in the core-collapse time. Primordial binaries become active earlier as an energy source than in $N$-body. The \texttt{MOCCA} simulations have smaller half-mass radius and mass and therefore the half-mass relaxation time is also smaller. This means that the MOCCA models are overall dynamically older and have evolved faster. Furthermore, from the core radii evolution of the cluster models, we see that \texttt{MOCCA} simulations have a larger central density, which should lead to a larger number of dynamical interactions in these models compared with the \texttt{Nbody6++GPU} runs. All of this is also connected to the treatment of unbound stars in \texttt{MOCCA}. In \texttt{MOCCA}, when a star acquires a high enough energy in relaxation/interaction to become unbound it is immediately removed from simulations. In \texttt{Nbody6++GPU} this is not the case as stars need time to travel across the star cluster system to be removed to a distance of twice the tidal radius from the density centre. Since $r_{\mathrm{t}}$ is very large in our simulations (see Tab.\ref{Initial_conditions}), this may take a very long timw (on the scale of Gyrs in some cases). During this time the star can undergo relaxation and become a bound star in the cluster yet again \citep{Baumgardt2001}. When this process is properly accounted for in \texttt{MOCCA} the evolution of Lagrangian radii in \texttt{MOCCA} and \texttt{Nbody6++GPU} are similar and a new version of the \texttt{MOCCA} code includes an upgrade to properly treat these escaped objects. \\
		
		\item From the core radii evolution of the cluster models, we see that \texttt{MOCCA} simulations have a larger central density over the whole simulation. This leads to a larger number of dynamical interactions in the \texttt{MOCCA} runs compared with the \texttt{Nbody6++GPU} runs, as can be inferred from the larger scatter in remnant masses in Fig.\ref{IFMR}.  
		Although the overall binary fractions are similar, the \texttt{Nbody6++GPU} simulations yield consistently larger fractions over 10~Gyr. Due to the denser \texttt{MOCCA} models, binaries will be disrupted and forced to merge at larger rates. Additionally, more blue straggler stars are show in the HRDs of the \texttt{MOCCA} simulations, as can be seen in Fig.\ref{HRD_complete.jpg}. This means that there must have been more interactions that lead to mass gain to produce these, i.e. this is a result of the denser \texttt{MOCCA} models. In Fig.\ref{IFMR_complete}, the masses of the escaping NSs for the \texttt{MOCCA-delayedSNe-Uniform} simulation are larger, simply because we found that the maximum NS mass was set to 3.0 $\rm M_{\odot}$, rather than 2.5 $\rm M_{\odot}$ in the other simulations. This maximum NS mass is taken as the upper limit of neutron star masses and follows from causality \citep{LattimerPrakash2004}. This is not a big a problem, however, since the IFMR for the delayed SNe is continuous in this regime. If we had instead set the maximum NS to 2.5 $\rm M_{\odot}$ then all the NSs in the mass range between 2.5 $\rm M_{\odot}$ and 3.0 $\rm M_{\odot}$ would be BHs with the same masses as the NSs. In the future gravitational million-body simulations, we will use 2.5 $\rm M_{\odot}$ in line with recent observations, such as \citet{Linares2018}. \\
		
		\item The differences in the time-step parameters (\texttt{pts1, pts2, pts3}) and the wind treatment  (\texttt{mdflag}=3$\neq$\texttt{edd\_factor}=0, where  \texttt{Nbody6++GPU} takes into account the bi-stability jump and the \texttt{MOCCA} simulations do not), in combination, might lead to the slight upward shift in values in the IFMR in Fig.\ref{IFMR_complete}, which otherwise shows excellent agreement in the BHs, NSs and WDs masses across all simulations for both \texttt{MOCCA \& Nbody6++GPU}. Further investigations should be done into systematic shifts of the remnant masses between the \texttt{MOCCA} and \texttt{Nbody6++GPU} code. Both of the IFMRs show excellent agreement with the theory from \citet{Fryeretal2012} and the \texttt{Nbody7} results from \citet{Banerjeeetal2020}. Comparisons with old (\texttt{Level A}) stellar evolution treatments reveal that these core-collapse neutrino-driven SNe schemes produce much larger BH masses for increasing ZAMS masses than what was previously available \citep{Belczynskietal2002} and provide a smooth transition to any of the available (P)PISNe treatments (see also Fig.\ref{McLuster_psflag.jpg}) if these are switched on. \\
		
		\item The fallback-scaled kick distributions for NSs and BHs likewise show excellent agreement across all masses as shown in  Figs.\ref{Nbody_V_Kick.jpg} and \ref{MOCCA_V_Kick.jpg}. All simulations retain NSs formed from an ECSNe, AIC or MIC of mass $1.26$~$\rm M_{\odot}$ \citep{Belczynskietal2008} as we see in  Figs.\ref{MOCCA_M_RI_cum_RI.png} and \ref{Nbody_M_RI_cum_RI.png}. But some of these also escape the cluster despite the low natal kick velocity that we set of \texttt{ecsig=sigmac}=$3.0$~$\mathrm{kms}^{-1}$ \citep{GessnerJanka2018} at similar escape speeds, which might be due to the low cluster densities, evaporation (a series of weak encounters), the kick itself or a combination of the above. Overall, the retention fractions and distributions, see Fig.\ref{MOCCA_M_RI_cum_RI.png}, \ref{Nbody_M_RI_cum_RI.png}, of the compact objects across all simulations are very similar. The HeWDs are the big exception which are mostly retained in the \texttt{MOCCA} simulations, in contrast to \texttt{Nbody6++GPU} where virtually all of them escape with large escape speeds. These escape speeds are, however, much larger than the largest permitted HeWD natal kick of $6.0$~$\mathrm{kms}^{-1}$ \citep{Fellhaueretal2003} that is set in the \texttt{Nbody6++GPU} simulations and they are also much larger than the escape speeds for the HeWDs from the \texttt{MOCCA} simulations (see Fig.\ref{MOCCA_V_Kick.jpg}). All of the escaped HeWDs originate from ZAMS binaries in both the \texttt{MOCCA} and the \texttt{Nbody6++GPU} simulations. Many more COWDs from single ZAMS stars escape the \texttt{MOCCA} simulations than the \texttt{Nbody6++GPU} simulations and the escape speeds are also much more similar and in many cases much lower than those of the \texttt{Nbody6++GPU} runs. COWDs from ZAMS binaries escape all the simulations in similar numbers. The same statements can be made about the ONeWDs. The reasons why the $v_{\mathrm{esc}}$ distributions are so dissimilar cannot be attributed only to the WD kicks in the \texttt{Nbody6++GPU} simulations, because the natal kicks are of very low velocity dispersion. Further studies with \texttt{MOCCA} and \texttt{Nbody6++GPU} on the effects that WD natal kicks have on binary stability and WD production and retention fraction in OCs, GCs and NSCs should be done going forward to shed more light on this particular aspect using the two modelling methods. 
	\end{itemize} 
	Overall, from the detailed comparison, we find very good agreement between the two modelling methods (\texttt{Nbody6++GPU} and \texttt{MOCCA}) when looking at, for example, the remnant mass distributions. This provides mutual support for both methods in star cluster simulations and the stellar evolution implementations in both codes. However, there are also some significant differences in the global evolution of the star cluster simulations with the two modelling methods. An example of these is the striking differences in blue straggler stars from Fig.\ref{HRD_complete.jpg}, the reasons for which are given above. The conclusion here relates to our initial models and the treatment of unbound stars in \texttt{MOCCA} vs. \texttt{Nbody6++GPU} simulations. In the future, we strongly suggest to not choose massively tidally underfilling initial cluster models with extremely large tidal radii, especially when using \texttt{MOCCA} simulations, to avoid problems with extremely large escape times for unbound objects. In any case, the results invite additional future comparative studies exploring the vast parameter space of star cluster simulations, also in the initial conditions, with direct $N$-body (\texttt{Nbody6++GPU}) and Monte Carlo (\texttt{MOCCA}) simulations using the updated stellar evolution. 
	
	\subsection{Perspective on future stellar evolution (\texttt{SSE \& BSE}) updates}
	We have identified the following pain points in our \texttt{SSE \& BSE} implementations in \texttt{Nbody6++GPU \& McLuster} and to a lesser extent \texttt{MOCCA}, where we still have some work to do. The version of \texttt{MOCCA} presented in this paper has the CV behaviour around the orbital period gap and the GR merger recoil and final post-merger spins, as well as some earlier implementation of modelling high mass and metal-poor Population III stars \citep{Tanikawaetal2020} available. An even more up-to-date version by \citet{Belloni2020b} also has an advanced treatment of the wind velocity factor $\beta_{\text W}$ as an option. Overall, we will include the stellar evolution routines listed below in the codes \texttt{MOCCA, Nbody6++GPU \& McLuster} in the next iteration of stellar evolution updates and refer to these necessary updates below as \texttt{Level D}, see also Appendix A. The (technical) details of these implementations are not shown in Tab.\ref{MOCCA_Stellar_evolution_levels_literature} and are reserved for a future publication in the interest of brevity. 
	\begin{enumerate}
		\item \textbf{CVs and the orbital period gap} The proper behaviour of the CVs around the so-called \textit{orbital period gap}, which is located at 2 hr $<P_{\mathrm{orb}}<$ 3 hr \citep{Knigge2006,Schreiberetal2010,Zorotovicetal2016}, cannot be reproduced by \texttt{Nbody6++GPU}, however, in \texttt{MOCCA} since the \texttt{BSE} modifications by \citet{Bellonietal2018c} and discussions by \citet{Bellonietal2017a} are accounted for, this behaviour can be modelled according to our best current understanding. The \texttt{BSE} algorithm of \texttt{Nbody6++GPU} is still in its original form to treat CVs and includes only a simple description of the evolution of accreting WD binary systems given that comprehensive testing of degenerate mass-transfer phases was beyond the original scope of \citet{Hurleyetal2002b}. The changes that need to be done and we are implementing at the moment in \texttt{Nbody6++GPU} require a lot of modifications. Firstly, the original mass transfer rate onto \textit{any} degenerate object (\texttt{KW} $\geq 10$) in \texttt{MOCCA} has been upgraded from \citet{WhyteEggleton1980,Hurleyetal2002b,Claeysetal2014} by including the formalism following \citet{Ritter1988}. The angular momentum loss in a close interacting CV that happens as a consequence of mass transfer is called the consequential angular momentum loss mechanism (CAML). Depending on the driving process behind the mass transfer it is either referred to as classical CAML (cCAML) \citep{KingKolb1995} or empirical CAML (eCAML) \citep{Schreiberetal2016}. The original \texttt{BSE} formalism can also be chosen \citep{Hurleyetal2002b}. The eCAML is more empirically motivated by including nova eruptions as the source of additional drag forces. Here the CAML is stronger for low mass WDs. Furthermore, \citet{Bellonietal2018c} introduced new, completely empirical normalisation factors for magnetic braking (MB) angular momentum loss and gravitational multipole radiation (GMR) angular momentum loss in the case of cCAML following \citet{Kniggeetal2011a} and in the case of eCAML, these normalisation factors for MB and GMR follow \citet{Zorotovicetal2016}. The merger between a MS star and its WD companion is now treated with the variable \texttt{qdynflag}, for which if set to 0 the merger assumes no CAML, if set to 1 the merger depends on classical cCAML and if set to 2 the merger depends on empirical CAML \citep{Schreiberetal2016}. Moreover, \citet{Bellonietal2018b} improved the stability criteria for thermally unstable mass transfer depending on a critical mass ratio between the primary and secondary star \citep{Schreiberetal2016} in the original \texttt{BSE} \citep{Hurleyetal2002b}, because the mass transfer rates for thermal timescale mass transfer are underestimated in the original \texttt{BSE}. All of these changes are further complemented by a large reduction in the time-steps for interacting binaries, depending on the factor that may be chosen freely. These upgrades in \texttt{MOCCA}, and soon to be included in \texttt{Nbody6++GPU}, will have the following impact. Firstly, the spins will be properly treated in response to the updated magnetic braking. Secondly, the inflation above and below the orbital period gap and the deflation in the orbital period gap of the donor primary star will be described correctly. Lastly, the processes of GR that lead to angular momentum loss and bloating \textit{below} the orbital period gap and of MB, which leads to angular momentum and bloating \textit{above} the orbital period period gap, will be accounted for. 
		\\
		\item \textbf{More on magnetic braking} As mentioned above, the MB mechanisms were updated in \citet{Bellonietal2018c}. The original version in \citet{Hurleyetal2002b} has been improved by \citet{Bellonietal2018c} to include the more rigorous treatment by \citet{Rappaportetal1983}, which may be switched on in \texttt{MOCCA}. Then, this new implementation was applied to CVs in GCs in the \texttt{MOCCA} study in \citet{Bellonietal2019}. This model was expanded further in \citet{Bellonietal2020a} by also adding the so-called reduced magnetic braking model, which extends the previous works to magnetic CVs.  An issue that remains in both \texttt{MOCCA} and \texttt{Nbody6++GPU} is the limit for applying MB, which arrives from the fact that MB is only expected to operate in MS stars with convective envelopes. This affects low-mass accreting compact object binaries, such as CVs and low-mass X-ray binaries. In \texttt{StarTrack} \citep{Belczynskietal2008}, there is such a mass limit imposed. At metallicities of $Z\geq 0.02$, the maximum mass is set to 1.25~$\mathrm{M}_{\odot}$ and for low metallicties at $Z\leq 0.001$, i.e. also at the metallicity used in the simulations of this paper, this limit should be 0.8~$\mathrm{M}_{\odot}$. Additionally, unlike \texttt{StarTrack}, the magnetic braking does not depend on the stellar type \texttt{KW} in \texttt{MOCCA} and in the \texttt{Nbody6++GPU} \texttt{BSE} algorithm, which should be the case, as the MB upper mass limit depends on it.
		\\
		\item \textbf{Extending \texttt{SSE} fitting formulae to extreme metal-poor (EMP) stars} - In $N$-body simulations that use  \texttt{SSE \& BSE} to model the stellar evolution, any extrapolation beyond $100$~$\mathrm{M}_{\odot}$ should be used with caution \citep{Hurleyetal2000}. However, this mass can be reached in the initial conditions when an IMF above $100$~$\mathrm{M}_{\odot}$ is used, e.g. \citet{Wangetal2021}, or can be reached through stellar collisions \cite{Kremeretal2020b}, especially in the beginning of the simulations \citep{Morawskietal2018,Morawskietal2019,DiCarloetal2019,DiCarloetal2021,Rizzutoetal2021a,Rizzutoetal2021b}. The fact the masses in these simulations sometimes reach masses largely in excess of the original upper mass limit to the fitting process employed in \citet{Hurleyetal2000} cannot simply be ignored. To this end, \citet{Tanikawaetal2020} devised fitting formulae for evolution tracks of massive stars from $8$~$\mathrm{M}_{\odot}$ up to $160$~$\mathrm{M}_{\odot}$ in extreme metal-poor environments ($10^{-8} \leq Z/Z_{\odot} \leq 10^{-2} $), which can be easily integrated into existing \texttt{SSE \& BSE} code variants. These formulae are based on reference stellar models that have been obtained from detailed time evolution of these stars using the \texttt{HOSHI} code \citep{Takahashietal2016,Takahashietal2019} and the 1-D simulation method described in \citet{Yoshidaetal2019}. In a further study with the same method \citet{Tanikawaetal2021a} provide fitting formulae of these stars that go up to even 1260~$\mathrm{M}_{\odot}$ and recently, these are now available up to 1500~$\mathrm{M}_{\odot}$ \citep{Hijikawaetal2021}. In general, \texttt{BSE\& SSE} variants need this implementation, which is already available in \texttt{MOCCA} (although not fully tested), to accurately model the evolution of these extremely-metal poor stars (e.g. Population III) star clusters, high mass stars in some extremely metal poor GCs and to use IMFs, which go beyond $100$~$\mathrm{M}_{\odot}$, e.g. \citet{Wangetal2021}, for these clusters. 
		Adding the \cite{Tanikawaetal2020} capability is especially interesting as  
		for the first time we might be able to model extremely massive stars (many hundreds and even thousands of $\mathrm{M}_{\odot}$) in massive GC environments. We note that there are likely some intrinsic differences between the standard \texttt{SSE} \citep{Hurleyetal2000} and the new fitting formulae by \citet{Tanikawaetal2020}, because the former were fitted to the \texttt{STARS} stellar evolution program \citep{Eggleton1971,Eggleton1972,Eggletonetal1973a,Eggletonetal1973b,Pols1995} results and latter to the afore-mentioned \texttt{HOSHI} code \citep{Takahashietal2016,Takahashietal2019}. This becomes particularly relevant when attempting to mix low mass stars ($\mathrm{M}_{\odot}\leq 8$) modelled with the traditional fitting formulae in the \texttt{SSE} code and high mass stars modelled by \citet{Tanikawaetal2020}. Moreover, the formulae by \cite{Tanikawaetal2020} are only valid for masses larger than 8~$\mathrm{M}_{\odot}$ and thus we need a sensible transition between \cite{Hurleyetal2000} and \cite{Tanikawaetal2020}. \\
		\\
		\item \textbf{Masses of merger products} In the most recent version of \texttt{StarTrack}, the merger products of certain stellar types were assigned new merger masses \citep{Olejaketal2020a}. The problem in the old \texttt{BSE} \citep{Hurleyetal2002b} arises from the fact that the mass of the product of a merger during dynamically unstable mass transfer, especially MS-MS merger, leads to $M\simeq M_{\mathrm{accretor}}$. There are many contact or over-contact MS-MS binaries that appear to be stable. On the other hand, there are also blue straggler stars and very massive stars ($>150$~$\mathrm{M}_{\odot}$) that are believed to be merger products, e.g. stars R136a, R136b and 136c in the large Magellanic cloud \citep{Bestenlehneretal2020} and the two stars WR 102ka in the Milky Way \citep{Hillieretal2001,Barniskeetal2008} are estimated to have masses exceeding 200~$\mathrm{M}_{\odot}$. To account for this, \citet{Olejaketal2020a} have introduced formalisms along the lines of $M = M_{\mathrm{accretor}} + f_{\mathrm{x}}\times M_{\mathrm{donor}}$, for a number of different merger scenarios involving different stellar types. Here $f_{\mathrm{x}}$ should be in the range of 0.5-1.0. This is still a very simple picture of stellar mergers and we need to elaborate on this approach. With the old \texttt{BSE} formalism, we may significantly reduce the cluster mass, which therefore also affects its evolution. This might be specially true when using the Sana orbital period distribution from \texttt{McLuster} initial conditions (\texttt{adis}=6) \citep{Kiminkietal2012,SanaEvans2011,Sanaetal2013a,Kobulnickyetal2014}, which has a lot of massive primordial MS-MS binaries with periods $P_{\mathrm{orb}}$ shorter than a few days.
		\\
		\item \textbf{GR merger recoil and final post-merger spins} The latest studies of IMBH growth with \texttt{Nbody6++GPU} \citep{DiCarloetal2019,DiCarloetal2020a,DiCarloetal2020b,DiCarloetal2021,Rizzutoetal2021a,Rizzutoetal2021b} do not include a general relativistic merger recoil treatment (in addition to missing PN terms). But \citet{ArcaSeddaetal2021} have included the recoil kicks by a posteriori analysis. The GR merger recoil is also missing from the \texttt{MOCCA} Survey Database I \citep{Askaretal2017a}. \texttt{Nbody7} and also the current development version of \texttt{Nbody6++GPU} contain a proper treatment of such velocity kicks. They depend on spins and mass ratio, and are caused due to asymmetric GW radiation during the final inspiral and merger process. Numerical relativity (NR) models \citep{Campanellietal2007,Rezollaetal2008,Hughes2009,vanMeteretal2010b} have been used to formulate semi-analytic descriptions for \texttt{MOCCA} and \texttt{Nbody} codes \citep{Morawskietal2018,Morawskietal2019,Banerjee2021a,BelczynskiBanerjee2020,ArcaSeddaetal2021,Banerjee2021c}. For (nearly) non-spinning BHs (\texttt{Fuller} model), the kick velocity is smaller than for high spins. In the case of large mass ratios the kick velocity is much smaller than for small mass ratios \citep{Morawskietal2018,Morawskietal2019} and therefore, in extreme cases these post-merger BHs might even be retained in open clusters \citep{Bakeretal2007b,Bakeretal2008,PortegiesZwartetal2010,Schoedel2014b, Baumgardtetal2018}. For non-aligned natal spins and small mass ratios on the other hand, the asymmetry in the GW may produce GR merger recoils that reach thousands of $\mathrm{kms}^{-1}$ \citep{Bakeretal2008,vanMeteretal2010b}. 
		\\
		Generally, the orbital angular momentum of the BH-BH dominates the angular momentum budget that contributes to the final spin vector of the post-merger BH and therefore, within limits, the final spin vector is mostly aligned with the orbital momentum vector \citep{Banerjee2021a}. In the case of physical collisions and mergers during binary-single interactions, the orbital angular momentum is not dominating the momentum budget and thus the BH spin can still be low. \citet{Banerjee2021a} also includes a treatment for random isotropic spin alignment of dynamically formed BHs. Additionally, \citet{Banerjee2021a} assumes that the GR merger recoil kick velocity of NS-NS and BH-NS mergers \citep{ArcaSedda2020,Chattopadhyayetal2021} to be zero but assigns merger recoil kick to BH-BH merger products from numerical-relativity fitting formulae of \citet{vanMeteretal2010b} (which is updated in \citealt{Banerjee2021c}). The final spin of the merger product is then evaluated in the same way as a BH-BH merger. 
		\\
		With the updates above, in addition to the BH natal spins discussed above, \texttt{Nbody6++GPU} will be able to fully model IMBH growth during the simulation (unlike in post-processing with \texttt{MOCCA} as in \citet{Morawskietal2018,Morawskietal2019}) in dense stellar clusters according to our best understanding. This is one of last remaining and important puzzle pieces in our \texttt{SSE \& BSE} implementations that helps us to simulate IMBH formation and retention in star clusters and the corresponding aLIGO/aVirgo GW signal \citep{Abbottetal2020b}.
		\\
		\item \textbf{Wind velocity factor} - the accretion of stellar winds in binaries depends on the wind velocity and a factor $\beta_{\text W}$. In the updated binary population synthesis (BPS) code \texttt{COSMIC} by \citet{Breiviketal2020a}, the value $\beta_{\text W}$ is allowed a broader range of values that actually do depend on stellar type following the \texttt{StarTrack} code by \citet{Belczynskietal2008}. In the \texttt{MOCCA \& Nbody6++GPU} versions presented in this paper $\beta_{\text W}$=0.125, where this represents the lower limit and should roughly correspond to the wind from the largest stars of $900$~$\text{R}_{\odot}$ \citep{Hurleyetal2002b}. In the future, $\beta_{\text W}$ will depend on the stellar type. 
		\\
		\item \textbf{Pulsars and magnetic spin field from NSs} The \texttt{COSMIC} BPS code \citep{Breiviketal2020a} includes new \texttt{BSE} additions that properly treat pulsars \citep{Kieletal2008a,Yeetal2019,Breiviketal2020a} in an attempt to mirror observations of spin periods and magnetic fields of young pulsars \citep{Manchesteretal2005}. Similarly, the \texttt{COMPAS} BPS code \citep{Stevensonetal2017a,Stevensonetal2017b} employs updated \texttt{BSE} and is used to study NS binaries, such as the elusive BH-NS \citep{Chattopadhyayetal2021} and NS-NS binaries \citep{Chattopadhyayetal2020} using updated pulsar prescriptions. These updates are also present in the earlier BPS code \texttt{BINPOP} by \citet{Kieletal2008b}, which is also based on the original \texttt{BSE} \citep{Hurleyetal2002b}. In detached binaries, a magnetic dipole radiation is assumed for the spin-period evolution whereas in non-detached binaries, a so-called magnetic field  burying as a response to mass transfer is implemented \citep{Kieletal2008a}, where the magnetic field decays exponentially depending on the accretion time and the mass that is transferred (equation (7) in \citet{Breiviketal2020a}). Mergers that include a NS produce a NS with a spin period and magnetic field that is drawn again from the same initial distribution, except for millisecond pulsars (MSPs) which stay MSPs after mergers. The magnetic field of a NS cannot be smaller than $5\times 10^7$~G \citep{Kieletal2008a}. In \texttt{Nbody6++GPU \& MOCCA}, we need these updates to properly account for the spin and the magnetic field evolution of all pulsars. 
		\\
		\item \textbf{Ultra-stripping in binary stars} After CE formation in a hard binary consisting of a NS or a BH and a giant star, the hydrogen-rich envelope of the giant star gets ejected, carrying large amounts of angular momentum with it \citep{Taurisetal2013b,Taurisetal2015b}. After the CE is ejected fully, the NS orbits a naked He star, after which further mass transfer via RLOF may happen \citep{Taurisetal2017} depending on the RLOF criteria mentioned above. This leads to stripping of the envelope of the He star until it reaches a naked core of mass $1.5$~$\rm M_{\odot}$ and explodes in a so-called ultra-stripped SNe (USSNe) \citep{Taurisetal2013b,Taurisetal2015b}. According to \citet{Taurisetal2017} most of these binaries survive the USSNe. \citet{Breiviketal2020a} have an implementation in \texttt{COSMIC}, which allows for this SNe pathway. In their models, the USSNe leads to an ejected mass of $0.1$~$\rm M_{\odot}$. The resulting kick velocity dispersion is much lower than the kick velocity dispersion following \citet{Hobbsetal2005}. In general, there should be a bi-modal kick distribution, where NSs with a mass above $1.33$~$\rm M_{\odot}$ receive large kicks and NSs with masses below that receive small kicks with a kick velocity dispersion of about 20.0~$\mathrm{kms}^{-1}$ \citep{Taurisetal2017}. Since the USSNe appears to be central to BH-BH, BH-NS and NS-NS merger rates \citep{Schneideretal2021}, we will work on implementations in \texttt{Nbody6++GPU \& MOCCA}.  Very recently, \citet{Schneideretal2021} found that through extreme stellar stripping in binary stars \citep{Taurisetal2013b,Taurisetal2015b, Taurisetal2017} in their \texttt{MESA} models \citep{Paxtonetal2011,Paxtonetal2015}, there is an overestimation by 90\% in the BH-BH mergers and 25-50\% in the BH-NS numbers if only any of the \citet{Fryeretal2012} prescriptions, rapid or delayed, are enabled. Overall, they predict a slight increase of 15-20\% more NS-NS mergers. This will definitely have to be explored in the future in N-body simulations.
	\end{enumerate}
	We are in the process of implementing the above into the \texttt{McLuster} version presented in this paper and results are reserved for a future publication.
	\\
	With the updates in the \texttt{SSE \& BSE} algorithms of \texttt{MOCCA \& Nbody6++GPU} presented in this paper, we are now able to fully model realistic GCs accurately across cosmic time with direct $N$-body simulation and also Monte-Carlo models according to our current understanding of stellar evolution of binary and single stars. Thus, the next step is to test these updates with new direct million-body \texttt{Dragon}-type GC simulations, following on from \citet{Wangetal2016}, and \texttt{Dragon}-like NSC simulations similar to \citet{Panamarevetal2019}, and compare these with \texttt{MOCCA} modelling. In addition to \texttt{Nbody6++GPU}, we will in the future also use the \texttt{PeTar} code by \citep{Wangetal2020b,Wangetal2020c,Wangetal2020d}. This code also uses up-to-date \texttt{SSE \& BSE} implementations in code structure similar to the original \texttt{SSE \& BSE} \citep{Hurleyetal2000,Hurleyetal2002b} and similar to \texttt{MOCCA}. These two direct $N$-body codes in combination with Monte-Carlo models from \texttt{MOCCA} all employing modern stellar evolution will yield unprecedented and exciting results into the dynamical and stellar evolution of star clusters of \textit{realistic} size. 
	\\
	Finally, we note that a successor to \texttt{SSE} called the Method of Interpolation for Single Star Evolution \texttt{METISSE} \citep{Agrawaletal2020} has recently been produced. This utilises advancements in astrophysical stellar evolution codes to provide rapid stellar evolution parameters by interpolation within modern grids of stellar models. Thus it offers the potential for an astrophysically more robust (and potentially faster) realistic alternative to the updated \texttt{SSE} implementation in \texttt{Nbody6++GPU} and \texttt{MOCCA}. However, a similar approach as presented by \citet{Agrawaletal2020} is not yet available for the \texttt{BSE} routines and thus we will have to wait for a binary stellar evolution version of \texttt{METISSE}. Similarly, the \texttt{SEVN} code \citep{SperaMapelli2017,Speraetal2019,Mapellietal2020a} and its binary version is still a work in progress and at this moment in time not ready to be fully implemented into our codes. Therefore, it is likely that the \texttt{SSE \& BSE} presented here and the large number of variants of these codes are destined to stay relevant in the modelling of stellar evolution of single and binary stars for quite some time.
	
	\section*{Acknowledgements}
	We thank the anonymous referee for constructive comments
	and useful suggestions that have helped to improve the manuscript. The authors gratefully acknowledge the Gauss Centre for Supercomputing e.V. for funding this project by providing computing time through the John von Neumann Institute for Computing (NIC) on the GCS Supercomputer JUWELS at Jülich Supercomputing Centre (JSC). As computing resources we also acknowledge the Silk Road Project GPU systems and support by the computing and network department of NAOC. This project has been initiated during meetings and cooperation visits at Silk Road Project of National Astronomical Observatories of China (NAOC); AWHK, AL, AA, SB, MG, JH are grateful for hospitality and partial support during these visits. AWHK is a fellow of the International Max Planck Research School for Astronomy and Cosmic Physics at the University of Heidelberg (IMPRS-HD). AWHK and RS acknowledges support by the DFG Priority Program ’Exploring the Diversity of Extrasolar Planets’ (SP 345/20-1 and 22-1). AWHK thanks Wolfram Kollatschny for his continuous support and mentorship throughout the research. AWHK furthermore extends his gratitude to Shu Qi, Xiaoying Pang, Taras Panamarev, Li Shuo, Katja Reichert, Bhusan Kayastha, Francesco Flamini Dotti and Francesco Rizzuto for productive discussions and accelerating the progress of this research significantly. This work was supported by the Volkswagen Foundation under the Trilateral Partnerships grants No. 90411 and 97778. MG, AL and AA were partially supported by the Polish National Science Center (NCN) through the grant UMO-2016/23/B/ST9/02732. The work of PB was also supported under the special program of the NRF of Ukraine "Leading and Young Scientists Research Support" - "Astrophysical Relativistic Galactic Objects (ARGO): life cycle of active nucleus", No.~2020.02/0346. PB acknowledges support by the National Academy of Sciences of Ukraine under the Main Astronomical Observatory GPU computing cluster project No.~13.2021.MM. PB also acknowledges the support from the Science Committee of the Ministry of Education and Science of the Republic of Kazakhstan (Grant No. AP08856149) and the support by Ministry of Education and Science of Ukraine under the collaborative grants M86-22.11.2021. PB acknowledges support by the Chinese Academy of Sciences (CAS) through the Silk Road Project at NAOC and the President’s International Fellowship (PIFI) for Visiting Scientists program of CAS. AA acknowledges support from the Swedish Research Council through the grant 2017-04217. AA also would like to thank the Royal Physiographic Society of Lund and the Walter Gyllenberg Foundation for the research grant: `Evolution of Binaries containing Massive Stars'. MAS acknowledges financial support by the Alexander von Humboldt Stiftung for the research project "Black Holes at all the scales". SB acknowledges the support from the Deutsche Forschungsgemeinschaft (DFG; German Research Foundation) through the individual research grant `The dynamics of stellar-mass black holes in dense stellar systems and their role in gravitational-wave generation' (BA 4281/6-1; PI: S. Banerjee). DB was supported by ESO/Gobierno de Chile and by the grant \#{2017/14289-3}, S\~ao Paulo Research Foundation (FAPESP). LW thanks the financial support from JSPS International Research Fellow (School of Science, The university of Tokyo). Parts of the research conducted by JH were supported within the Australian Research Council Centre of Excellence for Gravitational Wave Discovery (OzGrav), through project number CE170100004. JH would also like to acknowledge the generous support of the Kavli Visiting Scholars program at the Kavli Institute for Astronomy and Astrophysics at Peking University that made a visit to Beijing possible as part of this work.

	\section*{Data Availability}
	The data from the runs of these simulations will be made available upon reasonable request by the authors. The \texttt{Nbody6++GPU} and \texttt{McLuster} versions that are described in this paper will be made publicly available. The \texttt{MOCCA} version will be available upon reasonable request to MG. 
	
	\bibliographystyle{mnras}
	\bibliography{bibliography.bib}

	\appendix
	\section{Stellar and binary evolution levels \texttt{A, B, C}}
	The stellar evolution levels and the corresponding options are shown in Tables~\ref{Stellar_evolution_levels_parameters}, \ref{Nbody6++GPU_Stellar_evolution_levels_literature} and \ref{MOCCA_Stellar_evolution_levels_literature}. The foundation for evolving a single star in the \texttt{Nbody6++GPU} and \texttt{MOCCA} codes and all subsequent updates is provided by the state-of-art population synthesis code \texttt{SSE} \citep{Hurleyetal2000,Hurleyetal2013a}. In this code every evolutionary phase of the star receives an integer related to a certain stellar type \texttt{KW}. These stellar types are divided as such:
	\begin{itemize}
		\item \texttt{KW}$=0 \equiv \text{MS star} \quad M \leq 0.7\mathrm{M}_{\odot}$  
		\item \texttt{KW}$=1 \equiv \text{MS star} \quad M > 0.7\mathrm{M}_{\odot}$ 		
		\item \texttt{KW}$=2 \equiv \text{Hertzsprung Gap (HG)} $ 
		\item \texttt{KW}$=3 \equiv \text{First Giant Branch (GB)} $ 
		\item \texttt{KW}$=4 \equiv \text{Core Helium Burning (CHeB)} $ 
		\item \texttt{KW}$=5 \equiv \text{Early Asymptotic Giant Branch (EAGB)} $ 
		\item \texttt{KW}$=6 \equiv \text{Thermally Pulsating Asymptotic Giant Branch (TPAGB)} $ 
		\item \texttt{KW}$=7 \equiv \text{Naked Helium Star MS (HeMS)}$  
		\item \texttt{KW}$=8 \equiv \text{Naked Helium Star Hertzsprung Gap (HeHG)} $ 		
		\item \texttt{KW}$=9 \equiv \text{Naked Helium Star Giant Branch (HeGB)} $ 
		\item \texttt{KW}$=10 \equiv \text{Helium White Dwarf (HeWD)} $ 
		\item \texttt{KW}$=11 \equiv \text{Carbon-Oxygen White Dwarf (COWD)} $ 
		\item \texttt{KW}$=12 \equiv \text{Oxygen-Neon White Dwarf (ONeWD)} $ 
		\item \texttt{KW}$=13 \equiv \text{Neutron Star (NS)} $ 
		\item \texttt{KW}$=14 \equiv \text{Black Hole (BH)} $ 
		\item \texttt{KW}$=15 \equiv \text{massless remnant} $ 
	\end{itemize}
	We note, that \texttt{Nbody6++GPU} and \texttt{MOCCA} has another stellar type (for single stars), which is \texttt{KW}=$-1$, which assigns pre-MS stars \citep{Railtonetal2014}. This treatment is valid for stars in the range $0.1 - 8.0\, \text M_{\odot}$ at solar metallicity $Z_{\odot}$=0.02. 
	\\    
	The basis for this code are analytic fitting formulae, which are continuous over the entire stellar mass range and approximate the evolution of the stars in the $N$-body simulations depending on their mass $M$~($\text M_{\odot}$), metallicity $Z$ and age $t$~(Myr). The  delivered output is stellar luminosity $L$~($\text L_{\odot}$), 
	stellar radius $R$~($\text R_{\odot}$),
	stellar core radius $R_{\mathrm{c}}$~($\text R_{\odot}$) and core mass $M_{\mathrm{c}}$~($\text M_{\odot}$) and other parameters as a function of those parameters. The \texttt{SSE} code was fitted to detailed stellar models of up to $50\, \text M_{\odot}$ and originally tested to be valid for masses from $0.01$~$\text M_{\odot}$ up to $100$~$\text M_{\odot}$ (where anything above $50$~$\text M_{\odot}$ is an extrapolation and usage above $100$~$\text M_{\odot}$ was not recommended) and in metallicity ranges from $Z$=$0.0001$ up to $Z$=$0.03$ \citep{Hurleyetal2000}, where the solar metallicity is given by $Z_{\odot}$=$0.02$. They found that the fitted models were accurate to within 5\% of the detailed evolutionary tracks.
	\\ Stars rarely exist in isolation. In fact, it is expected that most stars are born as twins, so-called primordial binaries \citep{Kroupa1995a,SadavoyStahler2017,Belloni2018a}. Most of these primordial binaries are disrupted in a star cluster, leaving a star cluster with \textit{hard} binaries \citep{Milone2012b}, which has also been studied with, for example, \texttt{MOCCA} \citep{Leighetal2015}. The vicinity to another star or compact object radically changes the evolution of the star as many more processes, which may lead to mass gain or mass loss of the star come into play. The \texttt{BSE} code \citep{Hurleyetal2002b,Hurleyetal2013b} provides the foundation of binary stellar evolution on which all other recent updates stand.
	\\
	This appendix is devoted to summarise the extensive changes which have been made in the stellar evolution in \texttt{MOCCA \& Nbody6++GPU} since \citet{Hurleyetal2000,Hurleyetal2002b}. We categorise the existing stellar evolution routines in \texttt{levels}. This is because with the increasing number of recipes and complexity therein available, we found it difficult to document and communicate these quickly in our simulations. The stellar evolution options that are available in \texttt{Nbody6++GPU} and \texttt{MOCCA} as of the writing of this paper, are shown in Tables~\ref{Stellar_evolution_levels_parameters}, \ref{Nbody6++GPU_Stellar_evolution_levels_literature} and \ref{MOCCA_Stellar_evolution_levels_literature}, respectively. We divide the available stellar evolution recipes in \texttt{Nbody6++GPU \& MOCCA} as such:
	\begin{enumerate}
		\item \textbf{\texttt{Level A}} - Stellar evolution settings that mirror in part the settings in the \texttt{Dragon} simulations of GCs \citep{Wangetal2016} and NSCs \citep{Panamarevetal2019} and also the \texttt{MOCCA} Survey DataBase I \citep{Askaretal2017a}. Most of these are outdated and should be generally not be used anymore, see e.g \citet{Shutetal2021}. \\
		\item \textbf{\texttt{Level B}} - Stellar evolution settings that have been tested extensively and may be used without concern. A selection of these should be enabled in the next gravitational million-body simulations. \\
		\item \textbf{\texttt{Level C}} - Stellar evolution settings that are available in the codes, but those that are not present in \texttt{level B} have not yet undergone sufficient testing and are therefore deemed experimental as of the writing of this paper. \\
		\item \textbf{\texttt{Level D}} - Stellar evolution settings that will be added in the next iteration of stellar evolution updates, see also section 5.2 for details on these. 
	\end{enumerate}
	In the more distant future, we will sequentially add new levels (the next one would be \texttt{level E}), where we group further planned stellar evolution updates on top of the preceding level (in this case \texttt{level D}) in \texttt{Nbody6++GPU, MOCCA \& McLuster} together. We hope that this will greatly help in the documentation and aid the future user of the codes to properly choose \texttt{SSE \& BSE} settings in his or her simulations.
	
	\subsection{Dynamical mass transfer and other processes in binary stars}
	In \texttt{Nbody6++GPU}, the dynamical mass transfer and the stability thereof in Roche-lobe overflow (RLOF) between binary stars is computed by \texttt{roche.f}, which calls subroutines for magnetic braking \texttt{magbrk.f}, for gravitational radiation \texttt{grrad.f} and for coalescing of RLOF or common-evelope evolution (CEE) binaries \texttt{coal.f}. The tidal circularisation and tidal spin synchronisation and associated timescales are set in \texttt{bsetid.f}, which still follow the original treatment by \citet{Hurleyetal2002b} and sources therein. In \texttt{MOCCA}, all of the above is included in the original \texttt{evolv2b.f} \citep{Hurleyetal2002b} with lots of more recent updates regarding the proper evolution of cataclysmic variables (CVs) \citep{Bellonietal2018c}. These updates may be switched off, however, with the parameters \texttt{camlflagMZ}=\texttt{qdynflagMZ}=\texttt{qtherflagMZ}=0 \citep{Bellonietal2018c}. Therefore, we may still enable the same dynamical mass transfer and stability criteria in \texttt{Nbody6++GPU} and \texttt{MOCCA} based on \citep{Hurleyetal2002b}. Here, the stability of the mass transfer is determined by the original relations of radius-mass exponents $\zeta$ by \citep{Webbink1985b}, which give critical mass ratios of the donor and accretor star implemented in \citet{Hurleyetal2002b}. In semi-detached binaries, the primary loses some mass via winds and the secondary can accrete the material if passing through it. This Bondi-Hoyle accretion rate \citet{BondiHoyle1944} (\texttt{acc2} in both codes) is sensitive to the wind velocity factor $\beta_W$ \citep{Hurleyetal2002b}.  $\beta_{\text W}$ strongly depends on spectral type \texttt{KW}; the larger the star, the lower $\beta_{\text W}$ . In the \texttt{BSE} implementation of \texttt{Nbody6++GPU} (\textit{and} \texttt{PeTar \& Nbody7}) this is not the case, unlike in the latest versions of \texttt{MOCCA} \citep{Belloni2020b}, \texttt{StarTrack} \citep{Belczynskietal2008} and \texttt{COSMIC} \citep{Breiviketal2020a}. The latter is also implemented in the latest version of \texttt{CMC}   \citep{Kremeretal2020b}. 
	We set \texttt{beta}=0.125 in the simulations following \citet{Hurleyetal2002b}, where this represents the lower limit and should roughly correspond to the wind from the largest stars of $900 \, \text{R}_{\odot}$. The angular momentum factor for mass loss during RLOF in both codes is set by \texttt{gamm1} in \texttt{Nbody6++GPU} and \texttt{gamma} in \texttt{MOCCA} \citep{Hurleyetal2002b}. If positive \texttt{gamm1}=\texttt{gamma}$>0$, then the lost material carries with it a fraction gamma of orbital angular momentum. If set to \texttt{gamm1}=\texttt{gamma}=-1, then the material carries with it specific angular momentum of the primary and if set to \texttt{gamm1}=\texttt{gamma}=-2, then the material is lost from system as if it was a wind from the secondary. The factor to reduce the spin angular momentum change owing to wind accretion is \texttt{xi} and the fraction of accreted matter retained in nova eruption is \texttt{epsnov} in both codes \citep{Hurleyetal2002b}.
	\\
	Accretion rates onto a NS or BH (Eddington and Super-Eddington) are controlled by the parameter \texttt{eddfac} in both codes. Super-Eddington accretion rates are set by  (\texttt{eddfac}=100.0) \citep{CameronMock1967}. The Chandrasekhar mass of a WD is set to \texttt{MCH} = 1.44~$\text{M}_{\odot}$ \citep{Mazzalietal2007,Boshkayevetal2013}. The maximum NS mass is set to \texttt{mxns}$\leq 2.5 \, \text{M}_{\odot}$ \citep{LattimerPrakash2004,Linares2018,Baymetal2018,Baymetal2019}. In the \texttt{mix.f} and \texttt{coal.f} subroutines of \texttt{Nbody6++GPU}, \citep{Rizzutoetal2021a} implemented a variable \texttt{FctorCl}, that controls the mass accretion if a big star (\texttt{KW}$\, \leq \,$9) merges with a BH or NS. If \texttt{FctorCl}=1, then the whole star is accreted onto the BH or NS. Likewise, if \texttt{FctorCl}=0, then no mass is accreted. \texttt{MOCCA} has a similar variable available called \texttt{tzo}. We include a post-Newtonian (PN) orbit averaged dynamics treatment according to \citet{PetersMathews1963,Peters1964} for binaries containing a NS or BH in \texttt{grrad.f} in \texttt{Nbody6++GPU} and \texttt{evolv2b.f} in \texttt{MOCCA}.
	\\
	The routine \texttt{comenv.f} and the respective parameters (second row in Tab.~\ref{Nbody6++GPU_Stellar_evolution_levels_literature} and Tab.~\ref{MOCCA_Stellar_evolution_levels_literature} for \texttt{Nbody6++GPU} and \texttt{MOCCA}, respectively) deal with the common envelope evolution following \citet{Hurleyetal2002b}, which in turn follows \citet{DewiTauris2000,Taurisetal2000}. CEE is one of the possible outcomes of RLOF between close binary stars \citep{Paczynski1976,Ivanovaetal2013,Ivanova2016,Ivanovaetal2020}. At the end of CEE the envelope of the primary (in some cases also of the secondary) is stripped away and CEE terminates. It is described by two parameters $\alpha_{\mathrm{CE}}$ and $\lambda_{\mathrm{CE}}$; the first one parameterizes what fraction of the orbital energy is used to liberate the envelope; the second one is a factor scaling the binding energy of the envelope. Both codes also allow the addition of some fraction of recombination energy to the binding energy in order to lower the threshold for loss of the envelope, depending on the stellar type. The procedure used is similar, but not identical to \citet{Claeysetal2014}.
	\\
	Still today, both $\lambda_{\rm CE}$ and $\alpha_{\rm CE}$ remain highly uncertain \citep{Morawskietal2018,Morawskietal2019,GiacobboMapelli2018,GiacobboMapelli2019b,Deetal2020,Santoliquidoetal2020,Eversonetal2020,Langeretal2020}. However, for low-mass stars, given their relatively large numbers in observed samples, such as the post-CE binaries identified by the Sloan Digital Sky Survey \citep{Rebassa-Mansergasetal2012}, reconstruction techniques and binary population synthesis have allowed us to infer, to some extent, a low value for $\alpha_{CE}$, which is $\sim 0.2-0.3$ \citep{Zorotovicetal2010b,ToonenNelemans2013,Camachoetal2014,Cojocaruetal2017}.
	
	\subsection{Stellar winds}
	The routine \texttt{mlwind.f} and the respective parameters (second row in Tab.~\ref{Nbody6++GPU_Stellar_evolution_levels_literature} and Tab.~\ref{MOCCA_Stellar_evolution_levels_literature} for \texttt{Nbody6++GPU} and \texttt{MOCCA}, respectively) deal with the mass loss from stars via winds and outflows. In \texttt{Nbody6++GPU} and \texttt{MOCCA} the choices of wind prescriptions are determined by \texttt{mdflag} and \texttt{edd\_factor}, respectively. Stellar winds and their correct descriptions for our purposes are very important, because they are critical in determining the mass of the compact object progenitors and thus they have a large influence on the compact object mass distributions in the cluster themselves \citep{Belczynskietal2010,Giacobboetal2018,Kremeretal2020b}. In \texttt{Nbody6++GPU} and \texttt{MOCCA}, the options of winds are very different in many places and therefore, these are listed independently below.
	\\
	First of all, for \texttt{Nbody6++GPU} and \texttt{mdflag}$\leq$2 we apply the mass loss of \citet{NieuwenhuijzendeJager1990} for massive stars over the entire HRD with a metallicity factor from \citet{Kudritzkietal1989}. In the case of giant stars, \texttt{Nbody6++GPU} calculates the mass loss from \citet{KudritzkiReimers1978} (with \texttt{neta}=0.477 suggested from \citet{McDonaldZijlstra2015}). Similarly, for the AGB stars and \texttt{mdflag}$\, \leq \,$2 \texttt{BSE} follows \citet{VassiliadisWood1993} and we apply the reduced Wolf-Rayet (WR)-like mass loss for small H-envelope masses from \citet{Reimers1975b,HamannKoesterke1998,Hurleyetal2000}. If \texttt{mdflag}$=$2, then the treatment of luminous blue variable (LBV) winds are added, which follow \citet{HumphreysDavidson1979,HumphreysDavidson1994}. For \texttt{mdflag}$>$2, these winds follow the LBV winds of \citet{Belczynskietal2020}. If \texttt{mdflag}=3, then for massive and hot O and B-type stars, the code switches on the metallicity dependent winds by \citet{Vinketal2001,VinkdeKoter2002,VinkdeKoter2005, Belczynskietal2010}, who established their mass-loss rates for O and B-type from a grid of wind models across a wide range of metallicities ($10^{-5} < Z/Z_{\odot} < 10$). Caution is advised against the so-called bi-stability jump, which is the drastic change of the character of the driving (ionisation) line, because of a sudden change in the wind ionisation. There is the option available to have these winds without the bi-stability jump \citet{Belczynskietal2010} (temperature shifted to the edge of the jump) in \texttt{Nbody6++GPU} (\texttt{mdflag}=4). For more evolved stars starting from naked He stars with \texttt{KW}$\geq$7, with \texttt{mdflag}$\geq$3 the metallicity dependent WR wind factor from \citet{VinkdeKoter2005} is used. For H-rich low mass stars, the mass loss rates remain unchanged \citet{Hurleyetal2000}. 
	\\
	In the \texttt{MOCCA} version of \texttt{BSE}, with \texttt{edd\_factor}=0, we use fixed $\alpha$ from \citet{Giacobboetal2018} in the prescriptions by \citet{Belczynskietal2010}. If \texttt{edd\_factor}=1, then the electron-scattered Eddington factor is taken from \citet{GraefenerHamann2008} and the exponent of the dependence on metallicity is then calculated from \citet{Chenetal2015} instead.
	The rest of the \texttt{mlwind.f} routine uses the same prescriptions for the stars for both \texttt{edd\_factor}=0 and \texttt{edd\_factor}=1. The LBV-like mass loss beyond the Humphreys-Davidson limit follows \citet{HumphreysDavidson1994,Belczynskietal2010}. We apply the mass loss of \citet{NieuwenhuijzendeJager1990} for massive stars over the entire HRD with a metallicity factor from \citet{KudritzkiReimers1978}. In the case of giant stars, \texttt{MOCCA} calculates the mass loss from \citet{KudritzkiReimers1978}. If \texttt{neta}$>$0 (\texttt{neta}=0.477 from \citet{McDonaldZijlstra2015} is suggested), then this mass loss is based on \citet{Reimers1975b} and if \texttt{neta}$<$0 it follows a more realistic setting by \citet{SchroederCuntz2005}, which takes into account the effective temperature and surface gravity of the star (here \texttt{neta}=0.172 is suggested). The winds of the AGB stars follow \citet{VassiliadisWood1993} and we apply the reduced Wolf-Rayet (WR) like mass loss for small H-envelope masses from \citet{Reimers1975b,HamannKoesterke1998,Hurleyetal2000}. For massive and hot O and B-type stars, the code switches on the metallicity dependent winds by \citet{Vinketal2001,VinkdeKoter2002,VinkdeKoter2005,Belczynskietal2010}. For more evolved stars starting from naked He stars with \texttt{KW}$\geq$7, the \texttt{MOCCA BSE} uses the metallicity dependent WR wind factor from \citet{VinkdeKoter2005}. We note that the \texttt{MOCCA} \texttt{BSE} does not account for the aforementioned bi-stability jump, so overall the treatment of the winds from \texttt{MOCCA} and \texttt{Nbody6++GPU} are most similar for \texttt{mdflag}=4 $\simeq$ \texttt{edd\_factor}=0. 
	\\ 
	We note, that today the wind mass loss from very large mass stars in the regime of WR stars still remains very uncertain and is difficult to model \citep{SanderVink2020,Sanderetal2020,HigginsVink2019,Higginsetal2021,Vink2021}. The same can also be said in general about stars on the lower mass end \citep{Decin2020}. It is likely that we will need to revise our stellar wind mass loss and terminal velocity models many times in the future with this in mind, especially, when we aim to properly model aLIGO/aVirgo GW source progenitor stars. 
	
	\subsection{Remnant masses of compact objects}
	
	The routine \texttt{hrdiag.f} and the respective parameters (first row in Tab.~\ref{Nbody6++GPU_Stellar_evolution_levels_literature} and Tab.~\ref{MOCCA_Stellar_evolution_levels_literature} for \texttt{Nbody6++GPU} and \texttt{MOCCA}, respectively) deal with the post-SNe remnant masses of the NSs and BHs. In \texttt{Nbody6++GPU} and \texttt{MOCCA} the choices of the NS and BH remnant masses are determined by \texttt{nsflag} and \texttt{compactmass}, respectively. The updated stellar evolution now incorporates a selection of possible SNe pathways, which lead to a variety of remnant masses. In the present versions of the \texttt{hrdiag.f} routine, any of the five remnant-mass schemes following \citet{EldridgeTout2004b,Belczynskietal2002,Belczynskietal2008,Fryeretal2012} may be chosen. In this paper the  \textit{rapid} (\texttt{nsflag}=\texttt{compactmass}=3)  and \textit{delayed} (\texttt{nsflag}=\texttt{compactmass}=4) SNe mechanisms are used as extremes for the convection-enhanced neutrino-driven SNe paradigm \citep{Fryeretal2012}. 
	\\
	In \texttt{hrdiag.f}, we can also set the pulsating pair instability SNe (PPISNe) resulting from electron-positron pair production and subsequent decreasing pressure support in massive He cores. These electron-positron pairs effectively remove pressure from outward photons, until the oxygen in the stellar core ignites in a flash, which creates a pulse and a thermonuclear reaction in the outward direction, after which the core stabilises. In even more massive He cores, the core does not stabilise and creates many of the above pulses, which leads to a failed or disrupted SNe, as the star is completely destroyed in the process. This is known as pair instability SNe (PISNe). Both of these processes are theoretically well understood  \citep{Belczynskietal2016,Woosley2017,Kremeretal2020b,Breiviketal2020a,Leungetal2019c,Leungetal2020b}. In \texttt{Nbody6++GPU} and \texttt{MOCCA} \texttt{psflag} and \texttt{piflag} determine the BH remnant masses that are produced by a (P)PISNe. By setting \texttt{psflag}=\texttt{piflag}=0, the progenitor star in the He core mass range of $65.0 \leq m_{\text He}/\mathrm{M}_{\odot} \leq 135.0$ is destroyed in the SN explosion (\texttt{KW}=15). With \texttt{psflag}=1 or \texttt{piflag}=2 the maximum He core mass is set to $45.0$~$\mathrm{M}_{\odot}$, below which the PISNe is not activated \citep{Belczynskietal2016}. In their scheme, the BH mass from a PPISNe is set to 40.5~$\mathrm{M}_{\odot}$ from $45.0$~$\mathrm{M}_\odot$ minus a 10$\%$ neutrino mass loss \citep{Timmes1996}. In the range of $45.0 \leq m_{\text He}/\mathrm{M}_{\odot} \leq 135.0$ the star is destroyed by PISNe. Additionally, for \texttt{Nbody6++GPU} \texttt{psflag}=2,3 the so-called \textit{moderate} (P)PISNe and \textit{weak} (P)PISNe following \citep{Leungetal2019c} may be set. These models again assume a 10\% neutrino loss in the PPISNe and set for He core mass range of $40.0 \leq m_{\text He}/\mathrm{M}_{\odot} \leq 65.0$ linearly increasing BH remnant masses dependent on the initial stellar mass. In the mass range of $60.0 \leq m_{\text He}/\mathrm{M}_{\odot} \leq 62.5$, the BH remnant masses (including 10\% neutrino loss) are $50.04$~$\mathrm{M}_\odot$ for the weak and $46.08$~$\mathrm{M}_\odot$ for the moderate PPISNe, respectively. These two (P)PISNe presciptions are not yet available in \texttt{MOCCA}. With \texttt{piflag}=1 we activate the remnant mass scheme by \citet{SperaMapelli2017} in \texttt{MOCCA}, who fit the compact remnants as a function of the final He mass fraction and final He core mass \citep{Woosley2017}. However, they fitted the data using the \texttt{SEVN} code \citep{Speraetal2015} and not any variant of the \texttt{BSE} and so this should be used with caution in \texttt{Nbody6++GPU \& MOCCA}.
	\\
	At the lower end of the progenitor mass spectrum, \texttt{Nbody6++GPU} and \texttt{MOCCA} have implementations of electron-capture SNe (ECSNe) \citep{Nomoto1984,Nomoto1987,Podsiadlowskietal2004b,Kieletal2008a,Ivanovaetal2008,GessnerJanka2018,Leungetal2020a}, which are activated using \texttt{ecflag}=1 in both codes for progenitor stars in the range of $8 \leq m/\mathrm{M}_{\odot} \leq 11$. Detailed studies of the behaviour of these stars in direct $N$-body simulations may be found in \citet{Banerjee2018,FragioneBanerjee2020} and in \texttt{CMC} models in \citet{Yeetal2019}. The progenitor stars build up He cores in a theoretical uncertain range of $1.4 \leq m_{\rm He}/\mathrm{M}_{\odot} \leq 2.5$ \citep{Hurleyetal2002b,Podsiadlowskietal2004b,Belczynskietal2008}, where in \texttt{Nbody6++GPU} and \texttt{MOCCA} we take $1.6 \leq m_{\rm He}/\mathrm{M}_{\odot} \leq 2.25$ from \citet{Hurleyetal2002b}. In these cores, Ne and Mg capture electrons, thus effectively removing electron pressure from the cores, and if the stellar core mass (\texttt{mcx}) surpasses the ECSNe critical mass of $1.372$~$\mathrm{M}_{\odot}$ \citep{Ivanovaetal2008}, the star collapses almost instantaneously, unlike the neutrino-driven core-collapse explosions. This instantaneous explosion also means that the ECSNe NS has no fallback mass leaving behind NSs with a characteristic mass of $m = 1.26$~$\mathrm{M}_{\odot}$ \citep{Belczynskietal2008}. In binaries, accretion may lead to a accretion-induced collapse (AIC) \citep{NomotoKondo1991,SaioNomoto2004}, when an ONeWD accretes material from a COWD or ONeWD and the resulting ONeWD exceeds the ECSNe critical mass \citep{NomotoKondo1991,Hurleyetal2002b}. Similarly, if this mass is surpassed by a COWD-COWD or ONeWD-ONeWD merger, then the result is a merger-induced collapse (MIC) \citep{SaioNomoto1985}, which is treated the same as an AIC if the ECSNe critical mass is surpassed. The kicks for the ECSNe, AIC and MIC are all drawn from the same Maxwellian, see below. All the above paths generally produce NSs in binaries, which can often lead to subsequent RLOF and the production of low-mass X-ray binaries (LMXBs; in GCs see \citet{Clark1975}) and millisecond pulsars (MSPs; in GCs see \citet{Manchesteretal2005}).

	\subsection{Compact object natal kick distributions}
	
	The routines \texttt{kick.f} in \texttt{Nbody6++GPU} and \texttt{kickv.f} in  \texttt{MOCCA} and the respective parameters (fourth row in Tab.~\ref{Nbody6++GPU_Stellar_evolution_levels_literature} and Tab.~\ref{MOCCA_Stellar_evolution_levels_literature} for \texttt{Nbody6++GPU} and \texttt{MOCCA}, respectively) deal with the (fallback-scaled) kick distributions of the compact objects. The purpose of updating this routine is to retain some of the compact objects in dense clusters of all sizes (OCs, GCs, NSCs) in order of increasing escape velocity $v_{\mathrm{esc}}$ \citep{PortegiesZwartetal2010,Schoedel2014b,Baumgardtetal2018} based on physically motivated SNe mechanisms. This is crucial since the simulations need to properly treat the formation of NSs and BHs in these environments \citep{KuranovPostnov2006,PortegiesZwartetal2010,Giesersetal2018,Giesersetal2019} and it makes the formation and survival of complex compact binaries such as NS-NS, and BH-BH possible \citep{FryerKalogera1997,Banerjeeetal2020}.
	\\
	How these kicks are constrained remains uncertain and is highly theoretical. The origin of these kicks come from asymmetries either due to further in-falling material or accretion onto the proto-NS core and/or strong neutrino-driven convection during the long phase after the stalling of the first shockwave, which has bounced off of the proto-NS core. Traditionally, the kicks for the NSs are given by \citet{Hobbsetal2005}, i.e. following a Maxwellian with a velocity dispersion of 265.0~$\text{kms}^{-1}$. However, before this work, a dispersion of 190.0~$\text{kms}^{-1}$ by \citet{HansenPhinney1997} was also frequently used. Drawing natal kicks from these Maxwellians with these velocity dispersions would effectively kick all NSs out of the cluster, which can be observed in the output of the \texttt{Dragon} simulations by \citet{Wangetal2016}: they use a high and a low velocity dispersion, 265.0~$\text{kms}^{-1}$ from \citet{Hobbsetal2005} and 30.0~$\text{kms}^{-1}$ inspired by \citet{Manchesteretal2005}, respectively.
	\\
	The LIGO/Virgo detections of the gravitational wave sources coming from a NS-NS binary \citep{Abbottetal2017a,Abbottetal2017b,Abbottetal2020a} or other NS binaries observed in star clusters \citep{Benacquistaetal2013} inspired the update of the natal kicks for these NSs. To this end, for the ECSNe, AIC and MIC, the kick distribution is now a Maxwellian with a velocity dispersion of $3.0$~$\text{kms}^{-1}$ (\texttt{ECSIG} in \texttt{Nbody6++GPU} and \texttt{sigmac} in \texttt{MOCCA}) following \citet{GessnerJanka2018}, who used detailed 2-D and 3-D simulations to model these processes. We note that other groups, for example, the \texttt{COSMIC} developers \citep{Breiviketal2020a} use $20.0$~$\text{kms}^{-1}$ and the \texttt{MOBSE} team \citep{Giacobboetal2018} use $15.0$~$\mathrm{kms}^{-1}$ in previous simulations. The justification for the low velocity dispersions are that the ECSNe, AIC are MIC are modelled as instantaneous events \citep{Hurleyetal2002b,Podsiadlowskietal2004b,Ivanovaetal2008}. 
	\\
	All other NSs and BHs that do not undergo ECSNe, AIC or MIC have their kicks traditionally scaled by the before-mentioned fallback onto the proto-remnant core \citep{Belczynskietal2008,Fryeretal2012}, which most importantly implies that the larger the fallback, the lower the natal kick is and if $f_b$=1, then the natal kick is zero. This would be called a direct collapse or a \textit{failed} SN. The variables to set the kicks are \texttt{KMECH} in \texttt{Nbody6++GPU} (which also necessitates setting \texttt{bhflag}$\geq$2 for all \texttt{KMECH}) and \texttt{bhflag\_kick} for the BHs and \texttt{nsflag\_kick} for the NSs in \texttt{MOCCA}. Therefore, in \texttt{MOCCA} we may enable separate kick mechanisms with different kick velocity dispersions (\texttt{sigmans, sigmabh}), whereas all the kicks in \texttt{Nbody6++GPU} excluding the ECSNe, AIC and MIC are drawn from the same Maxwellian with dispersion \texttt{disp}. 
	\\
	On top of the standard momentum-conserving kick mechanism (\texttt{KMECH}=1, \texttt{bhflag\_kick}=\texttt{nsflag\_kick}=3), there are the convection-asymmetry-driven (\texttt{KMECH}=2, \texttt{bhflag\_kick}=\texttt{nsflag\_kick}=4) \citep{Schecketal2004,FryerYoung2007, Schecketal2008}, collapse-asymmetry-driven (\texttt{KMECH}=3, \texttt{bhflag\_kick}=\texttt{nsflag\_kick}=5) \citep{BurrowsHayes1996, Fryer2004,MeakinArnett2006,MeakinArnett2007} and neutrino-driven natal kicks (\texttt{KMECH}=4, \texttt{bhflag\_kick}=\texttt{nsflag\_kick}=6) \citep{Fulleretal2003,FryerKusenko2006,Banerjeeetal2020} options, where the authors assume \textit{one} dominant kick mechanism in the SNe. In \texttt{MOCCA} and \texttt{Nbody6++GPU}, we also make this assumption. The equations for the kick velocity of the compact object in \texttt{Nbody6++GPU} and \texttt{MOCCA} mirror those in \texttt{Nbody7} \citep{Banerjeeetal2020}. We note that both \texttt{MOCCA} and \texttt{Nbody6++GPU} both have implementations for WD natal kicks \citep{Fellhaueretal2003,Jordanetal2012b,Vennesetal2017}, but they are not the same. In \texttt{MOCCA}, these WD kicks are the same for WD types and are assigned an arbitrary kick speed of \texttt{vkickwd}, unlike in \texttt{Nbody6++GPU}, which draws kicks for HeWDs and COWDs from a Maxwellian of dispersion \texttt{wdksig1} and the kicks for the ONeWDs from a Maxwellian with dispersion \texttt{wdksig2}. Both Maxwellians are truncated at \texttt{wdkmax}=6.0~$\text{kms}^{-1}$, where typically \texttt{wdksig1}=\texttt{wdksig2}=2.0~$\text{kms}^{-1}$ following \citet{Fellhaueretal2003}.
	
	\subsection{Compact objects natal spins}
	The aforementioned routines \texttt{kick.f} in \texttt{Nbody6++GPU} and \texttt{kickv.f} in \texttt{MOCCA} and the respective parameters (fourth row in Tab.~\ref{Nbody6++GPU_Stellar_evolution_levels_literature} and Tab.~\ref{MOCCA_Stellar_evolution_levels_literature} for \texttt{Nbody6++GPU} and \texttt{MOCCA}, respectively) also deal with the natal spins distributions of the BHs. In \texttt{Nbody6++GPU} these spins are controlled by the variable \texttt{bhflag}. The latest version of \texttt{Nbody6++GPU} includes updated \textit{metallicity-dependent} treatments of BH natal spin (the natal NS spins are not changed from the original \texttt{BSE}), which follow those of \citet{Belczynskietal2020,Banerjee2021a}. This is needed, because the spin angular momentum of the parent star does not necessarily translate directly into the natal spin angular momentum of the BH. We define a dimensionless parameter that accounts for the natal spin angular momentum following \citet{Kerr1963}. Like \citet{Banerjee2021a}, we assume the magnitude of this parameter for the BHs directly at their birth without any mass accretion of GR coalescence processes. The simplest model of BH natal spins, the \texttt{Fuller} model, produces zero natal spins \citep{Banerjee2021a} (\texttt{bhflag}=2), as here the Tayler-Spruit magnetic dynamo can essentially extract all of the angular momentum of the proto-remnant
	core, leading to nearly non-spinning BHs \citep{Spruit2002,FullerMa2019,Fulleretal2019}. The second spin model is the \texttt{Geneva} model \citep{Eggenbergeretal2008,Ekstroem2012,Banerjee2021a} (\texttt{bhflag}=3). The basis for this model is the transport of the angular momentum from the core to the envelope. This is only driven by convection, because the \texttt{Geneva} code does not have magnetic fields in the form of the Taylor-Spruit magnetic dynamo. This angular momentum transport is comparatively inefficient and leads to \textit{high} natal spins for low to medium mass parent O-type stars, whereas for high mass parent O-type stars, the angular momentum of the parent star may already haven been transported away in stellar winds and outflows and thus the natal BH spins may be low. The third and last spin model is the \texttt{MESA} model (\texttt{bhflag}=4), which also accounts for magnetically driven outflows and thus angular momentum transport \citep{Spruit2002,Paxtonetal2011,Paxtonetal2015,Fulleretal2019,Banerjee2021a}. This generally produces BHs with much \textit{smaller} natal spins than the \texttt{Geneva} model described above. 
	
	\begingroup
	
	\setlength{\tabcolsep}{8.0pt} 
	\renewcommand{\arraystretch}{1.31} 

	\definecolor{cadmiumgreen}{rgb}{0.0,0.42,0.24}
	\begin{landscape}
		\begin{table}
			\centering
			\begin{tabular}{|l|l|l|l|l||l|l|l|l|}
				\hline
				\multicolumn{2}{|c|}{\texttt{Code}} & \multicolumn{3}{c||}{\large{\texttt{Nbody6++GPU}}} & \multicolumn{4}{c|}{\large{\texttt{MOCCA}}} \\
				\hline
				\hline
				\multicolumn{2}{|c|}{\texttt{BSE} \& \texttt{SSE} ?} & \multicolumn{3}{c||}{\textbf{Stellar evolution level}} & \multicolumn{1}{c|}{\textbf{}} & \multicolumn{3}{c|}{\textbf{Stellar evolution level}} \\
				\hline
				\hline \textbf{Relevance} & \textbf{Routine} & \makecell{\texttt{Level A}} &  \makecell{\texttt{Level B}} & \makecell{\texttt{Level C}} & \textbf{Routine} & \makecell{\texttt{Level A}} &  \makecell{\texttt{Level B}} & \makecell{\texttt{Level C}} \\
				\hline \hline \textbf{SSE} & \texttt{hrdiag.f} & \makecell{\texttt{ecflag} = 0 \\ \texttt{wdflag}=1 \\ \texttt{nsflag} = 1 \\ \texttt{psflag} = 0} &  \makecell{ \textcolor{orange}{\texttt{ecflag} = 1} \\ \textcolor{orange}{\texttt{wdflag}=1} \\  \texttt{nsflag} = 2,\textcolor{orange}{3,4}  \\ \textcolor{orange}{\texttt{psflag} = 0},1}  &  \makecell{\texttt{psflag} = 2,3} & \texttt{hrdiag.f} & \makecell{\texttt{ecflag} = 0 \\ \texttt{wdflag}=1 \\ \texttt{compactmass} = 1 \\ \texttt{piflag} = 0} &  \makecell{\textcolor{orange}{ \texttt{ecflag} = 1} \\ \textcolor{orange}{\texttt{wdflag}=1} \\  \texttt{compactmass} = 2,\textcolor{orange}{3,4}  \\ \textcolor{orange}{\texttt{piflag} = 0},2}  &  \makecell{\texttt{piflag} = 1}\\ 
				\hline \textbf{SSE} & \texttt{mlwind.f} & \makecell{\texttt{mdflag} = 1 \\ \texttt{neta} = 0.5 \\ \texttt{bwind} = 0.0 \\ \texttt{flbv} = 1.5} & \makecell{\textcolor{orange}{\texttt{mdflag}} = 2,\textcolor{orange}{3} \\ \textcolor{orange}{\texttt{neta} = 0.5} \\ \textcolor{orange}{\texttt{bwind} = 0.0} \\ \textcolor{orange}{\texttt{flbv} = 1.5} } & \makecell{\texttt{mdflag} = 4 \\ \texttt{neta} = 0.477 \\ \texttt{bwind} = $ 0.0$ \\ \texttt{flbv} = 1.5 } & \texttt{mlwind.f} &  \makecell{\texttt{edd\_factor} = 0 \\ \texttt{neta} = 0.5 \\  \texttt{bwind} = 0.0 \\ \texttt{flbv} = 1.5} & \makecell{\textcolor{orange}{\texttt{edd\_factor} = 0} \\ \textcolor{orange}{\texttt{neta} = 0.5} \\ \textcolor{orange}{\texttt{bwind} = 0.0} \\ \textcolor{orange}{\texttt{flbv} = 1.5}} & \makecell{\texttt{edd\_factor} = 1 \\ \texttt{neta} = 0.477, -0.172 \\ \texttt{bwind} = 0.0 \\ \texttt{flbv} = 1.5 } \\
				\hline \hline \textbf{BSE} & \texttt{comenv.f} & \makecell{\texttt{LAMBDA} = 0.5 \\ \texttt{ALPHA1} = 3.0} & \makecell{\textcolor{orange}{\texttt{LAMBDA} = 0.5} \\ \textcolor{orange}{\texttt{ALPHA1} = 3.0}}  & \makecell{\texttt{LAMBDA} = 0.0 \\ \texttt{ALPHA1} = 1.0} & \texttt{comenv.f} & \makecell{\texttt{lambda} = 0.5 \\ \texttt{alpha} = 3.0} & \makecell{\textcolor{orange}{\texttt{lambda} = 0.5} \\ \textcolor{orange}{\texttt{alpha} = 3.0}}  & \makecell{\texttt{lambda} = 0.0 \\ \texttt{alpha} = 1.0} \\
				\hline \textbf{BSE} & \texttt{kick.f} & \makecell{\texttt{bhflag} = 0 \\ \texttt{KMECH} = 1\\ \texttt{disp} = 30.0, 190.0, 265.0 \\ \texttt{ecsig} = 20.0 \\ \texttt{wdsig1} = 2.0 \\ \texttt{wdsig2} = 2.0 \\ \texttt{wdkmax} = 6.0 \\ \texttt{vfac} = 0.0} & \makecell{\textcolor{orange}{\texttt{bhflag} = 2} \\ \textcolor{orange}{\texttt{KMECH} = 1},2,3\\ \textcolor{orange}{\texttt{disp} = 265.0} \\ \textcolor{orange}{\texttt{ecsig} = 3.0} \\ \textcolor{orange}{\texttt{wdsig1} = 2.0} \\ \textcolor{orange}{\texttt{wdsig2} = 2.0} \\ \textcolor{orange}{\texttt{wdkmax} = 6.0} \\ \textcolor{orange}{\texttt{vfac} = 0.0}} & \makecell{\texttt{bhflag} = 3,4 \\ \texttt{KMECH} = 4\\ \texttt{disp} = 265.0 \\ \texttt{ecsig} = 3.0 \\ \texttt{wdsig1} = 2.0 \\ \texttt{wdsig2} = 2.0 \\ \texttt{wdkmax} = 6.0 \\ \texttt{vfac} = 0.0} & \texttt{kick.f} & \makecell{\texttt{sigmans} = 30.0, 190.0, 265.0 \\ \texttt{sigmabh} = 30.0, 190.0, 265.0  \\ \texttt{bhflag\_kick} = 0,1,2 \\ \texttt{nsflag\_kick} = 0,1,2 \\ \texttt{sigmac} = 20.0} & \makecell{\textcolor{orange}{\texttt{sigmans} = 265.0} \\ \textcolor{orange}{\texttt{sigmabh} = 265.0}  \\ \textcolor{orange}{\texttt{bhflag\_kick} = 3},4,5 \\ \textcolor{orange}{\texttt{nsflag\_kick} = 3},4,5 \\ \textcolor{orange}{\texttt{sigmac} = 3.0}} & \makecell{\texttt{sigmans} = 265.0 \\ \texttt{sigmabh} = 265.0 \\ \texttt{bhflag\_kick} = 6 \\ \texttt{nsflag\_kick} = 6 \\ \texttt{sigmac} = 3.0}\\
				\hline \textbf{BSE} & \texttt{roche.f} &   \makecell{\texttt{acc2} = 1.5 \\ \texttt{beta} = 0.125 \\ \texttt{epsnov} = 0.001 \\ \texttt{eddfac} = 100.0 \\ \texttt{gamm1} = -1.0 \\ \texttt{xi} = 1.0} &   \makecell{\textcolor{orange}{\texttt{acc2} = 1.5} \\ \textcolor{orange}{\texttt{beta} = 0.125} \\ \textcolor{orange}{\texttt{epsnov} = 0.001} \\ \textcolor{orange}{\texttt{eddfac} = 100.0} \\ \textcolor{orange}{\texttt{gamm1} = -1.0} \\ \textcolor{orange}{\texttt{xi} = 1.0}}  &   \makecell{\texttt{acc2} = 1.5 \\ \texttt{beta} = 0.125 \\ \texttt{epsnov} = 0.001 \\ \texttt{eddfac} = 100.0 \\ \texttt{gamm1} = 0.0,-2.0 \\ \texttt{xi} = 1.0} & \texttt{evolv2b.f} & \makecell{\texttt{acc2} = 1.5 \\ \texttt{beta} = 0.125 \\ \texttt{epsnov} = 0.001 \\ \texttt{eddfac} = 100.0 \\ \texttt{gamma} = -1.0 \\ \texttt{xi} = 1.0} &   \makecell{\textcolor{orange}{\texttt{acc2} = 1.5} \\ \textcolor{orange}{\texttt{beta} = 0.125} \\ \textcolor{orange}{\texttt{epsnov} = 0.001} \\ \textcolor{orange}{\texttt{eddfac} = 100.0} \\ \textcolor{orange}{\texttt{gamma} = -1.0} \\ \textcolor{orange}{\texttt{xi} = 1.0}}  &   \makecell{\texttt{acc2} = 1.5 \\ \texttt{beta} = 0.125 \\ \texttt{epsnov} = 0.001 \\ \texttt{eddfac} = 100.0 \\ \texttt{gamma} = 0.0,-2.0,3.0 \\ \texttt{xi} = 1.0}\\
				\hline \hline
			\end{tabular}
			\caption{Table showing our stellar evolution \texttt{levels A, B} and \texttt{C} with respect to changes in the stellar evolution routines and the parameters in the codes \texttt{Nbody6++GPU} (left) and \texttt{MOCCA} (right). The parameters used in the simulations of this study (\texttt{delayedSNe-Uniform \& rapidSNe-Sana}) are shown in \textcolor{orange}{orange}. All excluded stellar evolution routines not shown in the table are largely identical to the original \texttt{SSE} and \texttt{BSE} \citep{Hurleyetal2000,Hurleyetal2002b}. The exact meaning of the parameters and the literature basis for the choice of these are given in Tab.\ref{Nbody6++GPU_Stellar_evolution_levels_literature} for \texttt{Nbody6++GPU} and in Tab.\ref{MOCCA_Stellar_evolution_levels_literature} for \texttt{MOCCA}. \texttt{Level C} includes stellar evolution settings that are available in the codes, but those that are not present in \texttt{level B} have not yet undergone sufficient testing and are therefore deemed experimental as of the writing of this paper.
				The \texttt{Level A} stellar evolution should \textbf{not} be used anymore if there are other options in \texttt{Level B} available for that parameter and is included to give a reference guide, especially when analysing data from older simulations such as those from \citet{Wangetal2016} and \citet{Panamarevetal2019}.}
			\label{Stellar_evolution_levels_parameters}
		\end{table}
	\end{landscape}
	\endgroup
	
	\begingroup
	\begin{landscape}
		\begin{table}
			\centering
			\begin{tabular}{|l|l|l|l|}
				\hline \textbf{Relevance} & \textbf{Routine} & \textbf{Parameter} & \textbf{Parameter option and source} \\
				\hline \hline \textbf{SSE} & \texttt{hrdiag.f} & \makecell[l]{\texttt{ecflag} - Enables or disables ECSNe \\ \texttt{nsflag} - Choices for how NS/BH masses are calculated \\ \texttt{psflag} - No, strong, weak or moderate PPISNe \\ \texttt{wdflag} - Choices for how the white dwarfs are cooled} & \makecell[l]{\texttt{ecflag} = 0 $\rightarrow$ No ECSNe, $\quad$ \textcolor{orange}{\texttt{ecflag} = 1 $\rightarrow$ ECSNe from} \citep{Belczynskietal2008} \\ \texttt{nsflag} =  0 $\rightarrow$ Use original SSE NS/BH mass \citep{Hurleyetal2000} \\ \texttt{nsflag} = 1 $\rightarrow$ Use FeNi core mass from \citep{Belczynskietal2002} \\ \texttt{nsflag} = 2 $\rightarrow$ Use FeNi core mass from \citep{Belczynskietal2008} \\  \textcolor{orange}{\texttt{nsflag} = 3 $\rightarrow$ Remnant masses with \textbf{rapid} SNe} \citep{Fryeretal2012} \\   \textcolor{orange}{\texttt{nsflag} = 4 $\rightarrow$ Remnant masses with \textbf{delayed} SNe} \citep{Fryeretal2012} \\  \texttt{nsflag} = 5 $\rightarrow$ Remnant masses from \citep{EldridgeTout2004b}\\ \textcolor{orange}{\texttt{psflag} = 0 $\rightarrow$ No PPISNe/PISNe} \\ \texttt{psflag} = 1 $\rightarrow$ \textbf{Strong} PPISNe/PISNe \citep{Belczynskietal2016} \\ \texttt{psflag} = 2 $\rightarrow$ \textcolor{red}{\textbf{Weak} PPISNe/PISNe} \citep{Leungetal2019b,Leungetal2020c} \\ \texttt{psflag} = 3 $\rightarrow$ \textcolor{red}{\textbf{Moderate} PPISNe/PISNe} \citep{Leungetal2019b,Leungetal2020c} \\ \texttt{wdflag} = 0 $\rightarrow$ Mestel cooling \citep{Mestel1952a} \\ \textcolor{orange}{\texttt{wdflag} = 1 $\rightarrow$ Modified Mestel cooling} \citep{Toutetal1997}} \\
				\hline \textbf{SSE} & \texttt{mlwind.f} & \makecell[l]{\texttt{mdflag} - Sets the wind mass-loss prescription \\ \texttt{neta} - Reimers mass-loss coefficient for giant winds \\  \texttt{bwind} - Companion-enhanced mass-loss factor \\ \texttt{flbv} - Coefficient for LBV wind rate if \texttt{mdflag}$>2$} & \makecell[l]{\texttt{mdflag} = 1 $\rightarrow$ Mass loss from \citep{Hurleyetal2000} \\ \texttt{mdflag} = 2 $\rightarrow$ Mass loss added for LBVs \citep{HumphreysDavidson1979} \\ $\qquad\qquad$ $\&$ \citep{Belczynskietal2010} \\  \textcolor{orange}{\texttt{mdflag} = 3 $\rightarrow$ Mass loss from} \citep{Belczynskietal2010} \textcolor{orange}{+ Metallicity factor from} \citep{Vinketal2001} \\ $\qquad\quad$ \textcolor{orange}{+ Mass loss for hot, massive H-rich O/B stars from} \citep{Vinketal2001} \\ \texttt{mdflag} = 4 $\rightarrow$ Mass loss without bi-stability jump \citep{Belczynskietal2010} \\ $\qquad\quad$+ Metallicity factor from \citep{Belczynskietal2010} \\ \textcolor{orange}{\texttt{neta} = 0.5} \citep{KudritzkiReimers1978}, \texttt{neta} = 0.477 \citep{McDonaldZijlstra2015} \\ \textcolor{orange}{\texttt{bwind} = 0.0}, \citep{Hurleyetal2002b} \quad \textcolor{orange}{\texttt{flbv} = 1.5} \citep{Belczynskietal2010}} \\
				\hline \hline \textbf{BSE} & \texttt{comenv.f} & \makecell[l]{\texttt{LAMBDA} - Structural parameter for giant envelope \\ and control of recombination energy used \\ \texttt{ALPHA1} - factor scaling the amount of orbital energy \\ used for envelope liberation in CEE} & \makecell[l]{ \textcolor{orange}{\texttt{lambda} = 0.5, \texttt{alpha} = 3.0} \citep{GiacobboMapelli2018} ($\lambda_{\mathrm{CE}}$: depends on stellar type \\ and recombination energy; $\alpha_{\mathrm{CE}}$ = \texttt{ALPHA1}) \\ \texttt{lambda} = 0.0, \texttt{alpha} = 1.0 ($\lambda_{\mathrm{CE}}$: depends on stellar type, \\ NO recombination energy used; $\alpha_{\mathrm{CE}}$ = \texttt{ALPHA1}, see Appendix A1)}\\
				\hline \textbf{BSE} & \texttt{kick.f}  & \makecell[l]{\texttt{bhflag} - BH kicks, same as for CC NSs \\ $\qquad\qquad$ but reduced for momentum cons. if fallback \\ \texttt{KMECH} - NS, BH kick mechanism: standard momentum cons. \\ $\qquad\qquad$ convection-asymm., collapse-asymm. \& neutrino driven \\ \texttt{disp} - Dispersion in a Maxwellian velocity distribution \\ $\qquad\qquad$ or the maximum value for a flat distribution  \\ \texttt{ecsig} - Dispersion for ECSNe \\ \texttt{wdsig1} - Dispersion for He and CO WDs \\ \texttt{wdsig2} - Dispersion for ONe WDs \\ \texttt{wdkmax} - Maximum WD kick velocity,  $\quad$ \texttt{vfac} - Option to scale by \texttt{VSTAR} } &  \makecell[l]{\texttt{bhflag} = 0 $\rightarrow$ No BH kicks $\quad$ \textcolor{black}{\texttt{bhflag} = 1 $\rightarrow$ unscaled kicks: BHs \& NSs} \citep{Belczynskietal2002}
					\\ \textcolor{orange}{\texttt{bhflag} = 2 $\rightarrow$ fallback-scaled kicks for BHs and NSs (\texttt{KMECH})} \\ \texttt{bhflag} = 3 $\rightarrow$ \texttt{KMECH} kicks + BH natal spins \texttt{Geneva} models \citep{Banerjeeetal2020,Banerjee2021a} \\ \texttt{bhflag} = 4 $\rightarrow$  \texttt{KMECH} kicks + BH natal spins \texttt{MESA} models \citep{Banerjeeetal2020,Banerjee2021a} \\ \textcolor{orange}{\texttt{bhflag} $>4$ $\rightarrow$ \texttt{KMECH} kicks + BH natal spins \texttt{Fuller} model} \citep{Banerjeeetal2020,Banerjee2021a} \\ \textcolor{orange}{\texttt{KMECH} = 1 $\rightarrow$ standard mom. conserving kick} \citep{Belczynskietal2008} \\ \texttt{KMECH} = 2 $\rightarrow$ Convection-asymm.-driven kick \citep{Banerjeeetal2020}\\ \texttt{KMECH} = 3 $\rightarrow$ Collapse-asymm.-driven kick \citep{Banerjeeetal2020}\\ \texttt{KMECH} = 4 $\rightarrow$ Neutrino-driven kick\citep{Banerjeeetal2020}\\ \texttt{disp} = 30.0 \citep{Wangetal2016}, \texttt{disp} = 190.0 \citep{HansenPhinney1997}, \\ \textcolor{orange}{\texttt{disp} = 265.0} \citep{Hobbsetal2005} \\ \textcolor{orange}{\texttt{ecsig} = 3.0} \citep{GessnerJanka2018}, \textcolor{orange}{\texttt{wdsig1}=\texttt{wdsig2} = 2.0, \quad \texttt{wdkmax} = 6.0} \citep{Fellhaueretal2003}} \\
				\hline \textbf{BSE} & \texttt{roche.f} & \makecell[l]{\texttt{acc2} - Bondi-Hoyle wind accretion efficiency factor \\ \texttt{beta} - Wind velocity factor \\ \texttt{epsnow} - Fraction of material accreted onto a WD \\ $\qquad\qquad$ ejected in a nova \\ \texttt{eddfac} - Eddington limit for accretion onto \\ $\qquad\qquad$ a degenerate object \\ \texttt{gamm1} - Choices for angular momentum changes \\ $\qquad\qquad$ owing to RLOF mass-loss \\ \texttt{xi} - Factor to reduce spin angular momentum \\ $\qquad\qquad$ change owing to wind accretion} & \makecell[l]{\textcolor{orange}{\texttt{acc2} = 1.5} \cite{BondiHoyle1944} \\ \textcolor{orange}{\texttt{beta} = 0.125 $\rightarrow$ lower limit} \citep{Hurleyetal2002b}  \\ \textcolor{orange}{\texttt{epsnow} = 0.001} \citep{Hurleyetal2002b}\\ \textcolor{orange}{\texttt{eddfac} = 100 $\rightarrow$ $100 \times$ Eddington limit accretion rate} \citep{Hurleyetal2002b} \\ \texttt{gamm1} = -2.0 $\rightarrow$  Super-Eddington mass transfer rates \citep{Hurleyetal2002b} \\ \textcolor{orange}{\texttt{gamm1} = -1.0 $\rightarrow$ Lost material carries with it} \\  $\qquad\qquad$ \textcolor{orange}{the specific angular momentum of the primary} \citep{Hurleyetal2002b} \\ \texttt{gamm1} $>$ 0.0  $\rightarrow$ takes away a fraction \texttt{gamm1} of $J_{\rm orb}$ \citep{Hurleyetal2002b}\\ \textcolor{orange}{\texttt{xi} = 1.0} \citep{Hurleyetal2002b}}\\ 
				\hline \hline
			\end{tabular}
			\caption{Table showing the options in the \texttt{SSE \& BSE} stellar evolution with respect to the parameters in \texttt{Nbody6++GPU}. The parameters used in the simulations of this study (\texttt{delayedSNe-Uniform \& rapidSNe-Swana}) are shown in \textcolor{orange}{orange}. The parameters that are present in \texttt{Nbody6++GPU} version but not in \texttt{MOCCA} are shown in \textcolor{red}{red}. All the stellar evolution routines not listed here are largely identical to the original \texttt{SSE} and \texttt{BSE} \citep{Hurleyetal2000,Hurleyetal2002b}. The abbreviations are as follows: \textbf{ECSNe} - eletron capture supernova, \textbf{AIC} - accretion-induced collapse, \textbf{MIC} - merger-induced collapse, \textbf{PPISNe} - pulsating pair instability supernova, \textbf{PISNe} - pair instability supernova, \textbf{LBV} - luminous blue variable, \textbf{NS} - neutron star, \textbf{BH} - black hole.}
			\label{Nbody6++GPU_Stellar_evolution_levels_literature}
		\end{table}
	\end{landscape}
	
	\begin{landscape}
		\begin{table}
			\centering
			\begin{tabular}{|l|l|l|l|}
				\hline \textbf{Relevance} & \textbf{Routine} & \textbf{Parameter} & \textbf{Parameter option and source} \\
				\hline \hline \textbf{SSE} & \texttt{hrdiag.f} & \makecell[l]{\texttt{ecflag} - Enables or disables ECSNe \\ \texttt{compactmass} - Choices for how NS/BH masses are calculated \\ \texttt{piflag} - No, strong or Spera PPISNe \\ \texttt{wdflag} - Choices for how the white dwarfs are cooled} & \makecell[l]{\texttt{ecflag} = 0 $\rightarrow$ No ECSNe \\ \textcolor{orange}{\texttt{ecflag} = 1 $\rightarrow$ ECSNe with remnant mass from} \citep{Belczynskietal2008} \\ \texttt{compactmass} =  0 $\rightarrow$ Use original SSE NS/BH mass \citep{Hurleyetal2000} \\ \texttt{compactmass} = 1 $\rightarrow$ Use FeNi core mass from \citep{Belczynskietal2002} \\ \texttt{compactmass} = 2 $\rightarrow$ Use FeNi core mass from \citep{Belczynskietal2008} \\  \textcolor{orange}{\texttt{compactmass} = 3 $\rightarrow$ Remnant masses with \textbf{rapid} SNe} \citep{Fryeretal2012} \\   \textcolor{orange}{\texttt{compactmass} = 4 $\rightarrow$ Remnant masses with \textbf{delayed} SNe} \citep{Fryeretal2012} \\  \texttt{compactmass} = 5 $\rightarrow$ Remnant masses from \citep{EldridgeTout2004b}\\ \textcolor{orange}{\texttt{piflag} = 0 $\rightarrow$ No PPISNe/PISNe} \\ \textcolor{red}{\texttt{piflag} = 1 $\rightarrow$ PPISNe/PISNe} \citep{SperaMapelli2017} \\ \texttt{piflag} = 2 $\rightarrow$ \textbf{Strong} PPISNe/PISNe \citep{Belczynskietal2016} \\ \texttt{wdflag} = 0 $\rightarrow$ Mestel cooling \citep{Mestel1952a} \\ \textcolor{orange}{\texttt{wdflag} = 1 $\rightarrow$ Modified Mestel cooling} \citep{Toutetal1997}} \\
				\hline \textbf{SSE} & \texttt{mlwind.f} & \makecell[l]{\texttt{edd\_factor} - Sets the wind mass-loss prescription \\ \texttt{neta} - Reimers mass-loss coefficient for giant winds \\  \texttt{bwind} - Companion-enhanced mass-loss factor \\ \texttt{flbv} - Coefficient for LBV wind rate if \texttt{mdflag}$>2$} & \makecell[l]{\textcolor{orange}{\texttt{edd\_factor} = 0 $\rightarrow$ Mass loss from} \citep{Belczynskietal2010,Chenetal2015} \\ \texttt{edd\_factor} = 1 $\rightarrow$ Mass loss from \citep{Chenetal2015,Giacobboetal2018} \\ \ \textcolor{orange}{\texttt{neta} = 0.5} \citep{KudritzkiReimers1978}, \texttt{neta} = 0.477 \citep{McDonaldZijlstra2015} \\  \textcolor{red}{\texttt{neta} = -0.172} \citep{SchroederCuntz2005} \\ \textcolor{orange}{\texttt{bwind} = 0.0}, \citep{Hurleyetal2002b} \quad \textcolor{orange}{\texttt{flbv} = 1.5} \citep{Belczynskietal2010}} \\
				\hline \hline \textbf{BSE} & \texttt{comenv.f} & \makecell[l]{\texttt{LAMBDA} - Structural parameter for giant envelope \\ and control of recombination energy used \\ \texttt{ALPHA1} - factor scaling the amount of orbital energy \\ used for envelope liberation in CEE} & \makecell[l]{ \textcolor{orange}{\texttt{lambda} = 0.5, \texttt{alpha} = 3.0} \citep{GiacobboMapelli2018} ($\lambda_{\mathrm{CE}}$: depends on stellar type \\ and recombination energy; $\alpha_{\mathrm{CE}}$ = \texttt{ALPHA1}) \\ \texttt{lambda} = 0.0, \texttt{alpha} = 1.0 ($\lambda_{\mathrm{CE}}$: depends on stellar type, \\ NO recombination energy used; $\alpha_{\mathrm{CE}}$ = \texttt{ALPHA1}, see Appendix A1)}\\
				\hline \textbf{BSE} & \texttt{kick.f}  & \makecell[l]{\texttt{nsflag\_kick, bhflag\_kick} - BH kicks, same as for CC NSs \\ $\qquad\qquad$ but reduced for momentum cons. for fallback: \\ $\qquad\qquad$ standard momentum cons., convection-asymm. \\ $\qquad\qquad$ collapse-asymm. \& neutrino driven  \\ \texttt{sigmans, sigmabh} - NS/BH Maxwellian kick velocity dispersion \\ \texttt{sigmac} - ECSNe/AIC/MIC Maxwellian kick velocity dispersion \\ \texttt{vkickwd} - He/CO/ONeWDs Maxwellian kick velocity dispersion} &  \makecell[l]{\texttt{nsflag\_kick, bhflag\_kick} = 0 $\rightarrow$ no kicks at all \\ \texttt{nsflag\_kick, bhflag\_kick} = 1 $\rightarrow$ kicks for single NS formation only \\ \texttt{nsflag\_kick, bhflag\_kick} = 2 $\rightarrow$ kicks for single and binary formation NS \\ \textcolor{orange}{\texttt{nsflag\_kick, bhflag\_kick} = 3 $\rightarrow$ standard mom. conserving kick} \citep{Belczynskietal2008} \\ \texttt{nsflag\_kick, bhflag\_kick} = 4 $\rightarrow$ Convection-asymm.-driven kick \\ $\qquad\qquad$ \citep{Schecketal2004,FryerYoung2007, Banerjeeetal2020}\\ \texttt{nsflag\_kick, bhflag\_kick} = 5 $\rightarrow$ Collapse-asymm.-driven kick \\ $\qquad\qquad$ \citep{Fryer2004,MeakinArnett2006,Banerjeeetal2020}\\ \texttt{nsflag\_kick, bhflag\_kick} = 6 $\rightarrow$ Neutrino-driven kick \\ $\qquad\qquad$ \citep{Fulleretal2003,FryerKusenko2006,Banerjeeetal2020}\\ \texttt{sigmans, sigmabh} = 190.0 \citep{HansenPhinney1997}, \texttt{sigmans, sigmabh} = 30.0 \citep{Wangetal2016} \\ \textcolor{orange}{\texttt{sigmans, sigmabh} = 265.0} \citep{Hobbsetal2005} \\ \textcolor{orange}{\texttt{sigmac} = 3.0} \citep{GessnerJanka2018} \\ \textcolor{orange}{\texttt{vkickwd} = 0.0}, \quad \texttt{vkickwd} = 2.0 \citep{Fellhaueretal2003}} \\
				\hline \textbf{BSE} & \texttt{evolv2b.f} & \makecell[l]{\texttt{acc2} - Bondi-Hoyle wind accretion efficiency factor \\ \texttt{beta} - Wind velocity factor \\ \texttt{epsnow} - Fraction of material accreted onto a WD \\ $\qquad\qquad$ ejected in a nova \\ \texttt{eddfac} - Eddington limit for accretion onto \\ $\qquad\qquad$ a degenerate object \\ \texttt{gamma} - Choices for angular momentum changes \\ $\qquad\qquad$ owing to RLOF mass-loss \\ \texttt{xi} - Factor to reduce spin angular momentum \\ $\qquad\qquad$ change owing to wind accretion} & \makecell[l]{\textcolor{orange}{\texttt{acc2} = 1.5} \cite{BondiHoyle1944} \\ \textcolor{orange}{\texttt{beta} = 0.125 $\rightarrow$ lower limit} \citep{Hurleyetal2002b}  \\ \textcolor{orange}{\texttt{epsnow} = 0.0} \citep{Hurleyetal2002b}\\ \textcolor{orange}{\texttt{eddfac} = 100 $\rightarrow$ $100 \times$ Eddington limit accretion rate} \citep{Hurleyetal2002b} \\ \texttt{gamma} = -2.0 $\rightarrow$  Super-Eddington mass transfer rates \citep{Hurleyetal2002b} \\ \textcolor{orange}{\texttt{gamma} = -1.0 $\rightarrow$ Lost material carries with it} \\  $\qquad\qquad$ \textcolor{orange}{the specific angular momentum of the primary} \citep{Hurleyetal2002b} \\ \texttt{gamma} $>$ 0.0  $\rightarrow$ takes away a fraction \texttt{gamma} of $J_{\rm orb}$ \citep{Hurleyetal2002b} \\ \textcolor{red}{\texttt{gamma} = 3  $\rightarrow$ following MB prescriptions by} \citep{Rappaportetal1983,Bellonietal2018c}\\ \textcolor{orange}{\texttt{xi} = 1.0} \citep{Hurleyetal2002b}}\\ 
				\hline \hline
			\end{tabular}
			\caption{Table showing the options in the \texttt{SSE \& BSE} stellar evolution with respect to the parameters in \texttt{MOCCA}. The parameters used in the simulations of this study (\texttt{delayedSNe-Uniform \& rapidSNe-Sana}) are shown in \textcolor{orange}{orange}. The parameters that are present in \texttt{MOCCA} version but not in \texttt{Nbody6++GPU} are shown in \textcolor{red}{red} (the added CV and symbiotic star treatment by \citet{Bellonietal2018c,Belloni2020b} are not listed here). All the stellar evolution routines not listed here are largely identical to the original \texttt{SSE} and \texttt{BSE} \citep{Hurleyetal2000, Hurleyetal2002b}. The abbreviations are as follows: \textbf{ECSNe} - eletron capture supernova, \textbf{AIC} - accretion-induced collapse, \textbf{MIC} - merger-induced collapse, \textbf{PPISNe} - pulsating pair instability supernova, \textbf{PISNe} - pair instability supernova, \textbf{LBV} - luminous blue variable, \textbf{NS} - neutron star, \textbf{BH} - black hole.}
			\label{MOCCA_Stellar_evolution_levels_literature}
		\end{table}
	\end{landscape}
	\endgroup
	
	\section{McLuster}
	The original McLuster software is an open source code, which is used to either set up initial conditions for $N$-body computations or to generate artificial star clusters for direct investigation \citep{Kuepperetal2011a}.  The \texttt{McLuster} output models can be read directly into the \texttt{Nbody6++GPU} and \texttt{MOCCA} simulations as initial models. This makes \texttt{McLuster} the perfect tool to initialise realistic star cluster simulations. After choosing the initial number of objects for each sub-population and the binary content within each, we can then choose structural parameters, such as the cluster density distribution (King, Plummer, Subr, EFF, Nuker) \citep{Plummer1911,Elsonetal1987b,King1962,Kingetal1966a,Subretal2008}, mass segregation \citep{Baumgardtetal2008b}, fractal dimensions \citep{GoodwinWhitworth2004} and the virial ratio. Furthermore, we may choose from many initial mass functions (IMFs) and respective limits \citep{Kroupa2001}. For the primordial binaries, we may choose from several binary mass ratio \citep{KobulnickyFryer2007, Kouwenhovenetal2007,Kiminkietal2012,SanaEvans2011,Sanaetal2013a,Kobulnickyetal2014,MoeDiStefano2017}, semi-major axis \citep{DuquennoyMayor1991,Kroupa2008}, period \citep{Kroupa1995a,Kroupa2008,SanaEvans2011,Sanaetal2013a,Ohetal2015,MoeDiStefano2017} and eccentricity \citep{DuquennoyMayor1991,Kroupa1995b,Kroupa2008,Kroupa2009,SanaEvans2011} distributions setting minimum and maximum initial separations in the process and eigenevolution processes \citep{Kroupaetal1993,Kroupa1995b,Bellonietal2017a,Bellonietal2017b}. Lastly, we may put the star cluster model in a tidal field, such as one from a point-like MW galaxy. However, these are set in the simulations by \texttt{Nbody6++GPU} or \texttt{MOCCA} directly. In principle, there are many different options available to create star clusters with up to 10 different stellar sub-populations, each having their own distinct properties. However, for this to properly work, a large number of bugs were fixed in this version of \texttt{McLuster}. These extensive changes are reserved for a separate publication, which is in preperation currently.
	\\
	In this paper we present the updated \texttt{SSE \& BSE} routines in \texttt{McLuster}. These include all of the stellar evolution contained in the levels \texttt{A, B} and \texttt{C}. This version provides a framework, in which we can evolve the different stellar populations at the level of stellar evolution that is also discussed in this paper. This is helpful in the following way. If we want to study the evolution of clusters with multiple stellar populations as observed in \citet{Milone2012b,Grattonetal2012,Latouretal2019,Kamannetal2020b} using \texttt{Nbody6++GPU} and \texttt{MOCCA}, we can create initial models, where the first population of stars in the case of two poulations has a slight offset in \texttt{epoch} and has thus undergone stellar evolution. This stellar evolution can then be modelled with the up-to-date stellar evolution routines contained in our \texttt{SSE \& BSE} codes. In principle, however, this code may also be used as a pure population synthesis code, because by setting the \texttt{epoch} parameter we may age the population(s) up to any point in time and look at the detailed evolution of each single or binary star over the whole \texttt{epoch}. If used in this way, \texttt{McLuster} can be used for a large number of studies. It could shed light on the how stellar evolution levels affects the formation of BH-BH, BH-NS and NS-NS mergers or how they affect the development of low- and high-mass X-ray binaries (or their progenitors). Moreover, we can explore how stellar mergers would affect the overall mass function, and what the role of stellar evolution levels and orbital parameters in the determination of these are. 
	\\
	The parameters are set in \texttt{mcluster.ini} file. Here, we may switch on and off the stellar evolution by setting \texttt{BSE}=1 or \texttt{BSE}=0. Below that all the options as outlined in Tab.\ref{Nbody6++GPU_Stellar_evolution_levels_literature} are available. We note that the BHs have natal spins set by the parameter \texttt{bhspin} in the \texttt{McLuster} version, and these are set in the routines \texttt{evolv1.f} for the single stars and in \texttt{evolv2.f} for the binary stars. This is in part due to the different structure in the \texttt{SSE \& BSE} in \texttt{Nbody6++GPU}. The \texttt{McLuster} version produces next to the \texttt{dat.10}, which may be used as an input file for the \texttt{Nbody6++GPU} simulations and the \texttt{single\_nbody.dat} and \texttt{binary\_nbody.dat} for the \texttt{MOCCA} simulations (through the appropriate choice of the parameter \texttt{outputf} in \texttt{mcluster.ini}), also the following files. First of all, if \texttt{BSE=1}, we get the output file \texttt{vkick.dat}, which contains the velocity kick information for all the compact objects in the population. The files \texttt{singles.dat} and \texttt{binaries.dat} contain furthermore, the luminosities, effective temperatures, core masses and radii, stellar radii, envelope masses and radii, stellar spins and all the velocity kick information for all the stars and not just the compact objects.
	\\
	In the following two subsections, we present results from two small studies with our \texttt{McLuster} version. Future additions in this \texttt{McLuster} version may be found in the section 5.2 and are grouped together in the stellar evolution \texttt{level D}.

	\subsection{Remnant-masses of compact objects}
	\subsubsection{Delayed \& rapid SNe and metallicity dependence}
	\begin{figure*}
		\includegraphics[width=\textwidth]{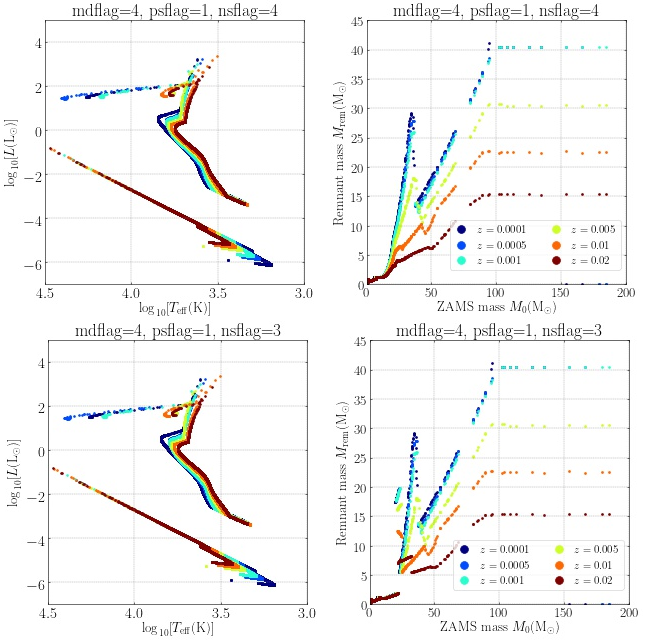}
		\caption{HRDs for the \texttt{McLuster} samples ($N=1.0\times 10^5$ single ZAMS stars) for all stars and the IFMRs of the compact objects depending on six different metallicities ranging from $Z$=0.0001 to Solar metallicity at $Z$=0.02. On top, the results for delayed SNe (\texttt{nsflag}=4) and on the bottom the results for the rapid SNe are shown (\texttt{nsflag}=3) \citep{Fryeretal2012}. The ZAMS stars suffer wind mass loss via \texttt{mdflag}=4 (no bi-stability jump) \citep{Belczynskietal2010} and the (P)PISNe are set to \texttt{psflag}=1 from \citet{Belczynskietal2016}.}
		\label{McLuster_Remnant_masses_nsflag3_nsflag4.png}
	\end{figure*}
	We simulate a star sample made up of only single ZAMS stars of size $N=1.0\times 10^5$ up to an \texttt{epoch}=12000.0, so 12 Gyr. The IMF is a \citet{Kroupa2001} IMF between (0.08-150.0)$\mathrm{M}_{\odot}$. We investigate a range of metallicities $Z$ for the two extremes of the core-collapse SNe paradigm, the rapid \texttt{nsflag}=3 and the delayed \texttt{nsflag}=4 SNe \citep{Fryeretal2012}. The ZAMS stars suffer wind mass loss via \texttt{mdflag}=4, i.e. we ignore the bi-stability jump \citep{Belczynskietal2010} (and the Reimer's mass loss coefficient set to \texttt{neta}=0.477 \citep{McDonaldZijlstra2015}), and the (P)PISNe are set to \texttt{psflag}=1 from \citet{Belczynskietal2016}. The specific time-steps \texttt{pts1, pts2, pts3} follow suggestions from \citet{Banerjeeetal2020}. The random seeds in \texttt{McLuster} are the same (\texttt{seedmc}=19640916) for all samples and therefore, we are evolving the identical ZAMS sample each time. 
	\\
	The results are shown in Fig.\ref{McLuster_Remnant_masses_nsflag3_nsflag4.png} for the delayed SNe on the top and the rapid SNe on the bottom. For both the remnant masses decrease continuously for increasing metallicity. This is mainly due to the fact that at lower metallicities the mass loss from the stars before undergoing a core-collapse SNe (or another evolutionary process that leads to a compact object) is lower than at large metallicities \citep{Vinketal2001,VinkdeKoter2005}. At metallicities as large as $Z$=0.005, the mass loss is so large and the resulting BH mass so low that the (P)PISNe are not triggered at all, see Fig.\ref{McLuster_psflag.jpg}. The results mirror those from \citet{Banerjeeetal2020}  and therefore the implementations in \texttt{Nbody7}, which confirms an accurate implementation of \texttt{Levels A, B} and \texttt{C} in \texttt{McLuster}. 
	
	\subsubsection{(P)PISNe and metallicity dependence}
	
\begin{figure}
		\includegraphics[width=\columnwidth]{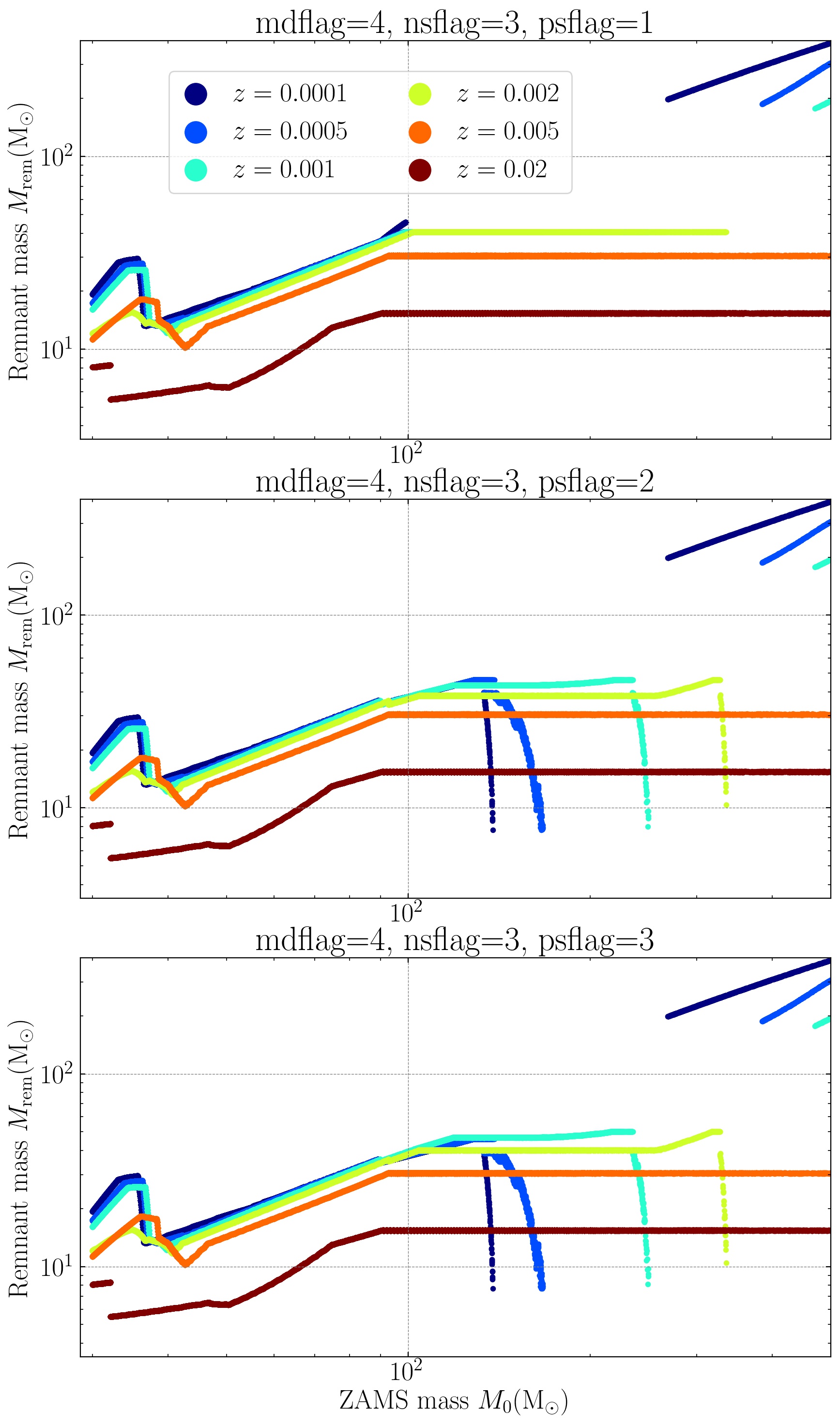}
		\caption{IFMRs of the BHs from the \texttt{McLuster} samples ($N=2.5\times 10^4$ single ZAMS stars) depending on six different metallicities ranging from $Z$=0.0001 to Solar metallicity at $Z$=0.02. Shown are the (P)PISNe recipes for \texttt{psflag}=1 on top \citep{Belczynskietal2016}, \texttt{psflag}=2 in the middle \citep{Leungetal2019c,Leungetal2020b} and \texttt{psflag}=3 on the bottom \citep{Leungetal2019c,Leungetal2020b}. The ZAMS stars suffer wind mass loss via \texttt{mdflag}=4 (no bi-stability jump) \citep{Belczynskietal2010} and the core-collapse SNe are rapid \citep{Fryeretal2012}.}
		\label{McLuster_psflag.jpg}
	\end{figure}

	We simulate a star sample made up of only single ZAMS stars of size $N=2.5\times 10^4$ up to an \texttt{epoch}=12000.0, so 12~Gyr. The IMF is a \citet{Kroupa2001} IMF between (30.0-500.0)$\mathrm{M}_{\odot}$. We note that this is a large extrapolation of what should be considered safe in the original \texttt{SSE} \& \texttt{BSE} \citep{Hurleyetal2000,Hurleyetal2002b}. But these masses are reached already in dense simulations, see \citet{DiCarloetal2021,Rizzutoetal2021a,Rizzutoetal2021b,ArcaSeddaetal2021}. We need the implementations in the \texttt{SSE\&BSE} from \citet{Tanikawaetal2020,Tanikawaetal2021a,Hijikawaetal2021} to properly models these stars in \texttt{McLuster} in the future. We investigate a range of metallicities $Z$ (0.0001-0.02). The ZAMS stars suffer wind mass loss via \texttt{mdflag}=4, i.e. we ignore the bi-stability jump \citep{Belczynskietal2010} (and the Reimer's mass loss coefficient set to \texttt{neta}=0.477 \citep{McDonaldZijlstra2015}), and we subject the stars to the rapid SNe core-collapse presciption \citep{Fryeretal2012}. The specific time-steps \texttt{pts1, pts2, pts3} follow suggestions from \citet{Banerjeeetal2020}. We investigate a range of metallicities $Z$ for the available (P)PISNe recipes: \texttt{psflag}=1 \citep{Belczynskietal2016}, \texttt{psflag}=2 \citep{Leungetal2019c,Leungetal2020b} and \texttt{psflag}=3 \citep{Leungetal2019c,Leungetal2020b}. The random seeds in \texttt{McLuster} are the same (\texttt{seedmc}=19640916) for all samples and therefore, we are evolving the identical ZAMS sample each time. 
	\\
	The results are shown in Fig.\ref{McLuster_psflag.jpg}. We see that the main difference between the three prescriptions is the onset of the (P)PISNe and the masses that result thereof. For low metallicities ($z<0.001$), the \citet{Leungetal2019c,Leungetal2020b} (P)PISNe produce high mass BHs for much larger ZAMS masses than the \citet{Belczynskietal2016} (P)PISNe. At metallicities as large as $Z$=0.005, the mass loss is so large and the resulting BH mass so low that the (P)PISNe are not triggered at all, see also Fig.\ref{McLuster_Remnant_masses_nsflag3_nsflag4.png}. Here, the remnant masses then coincide for all \texttt{psflag} ($z>0.005$). At large ZAMS and at the offset of the PISNe, the BH remnant masses are the same for \texttt{psflag}. Apart from initialising star cluster simulations with an IMF that is top-heavy and goes up to very large masses, e.g. \citet{Weatherford2021}, these BH masses may be reached through initial stellar collisions and coalescence in primordial binaries \citep{Kremeretal2020b}. Alternatively, these may be reached dynamical through BH-BH mergers \citep{Morawskietal2018,Morawskietal2019,DiCarloetal2019,Rizzutoetal2021a,Rizzutoetal2021b,ArcaSeddaetal2021}.  
	
	\bsp	
	\label{lastpage}
\end{document}